\documentclass{aa}
\usepackage[varg]{txfonts}
\usepackage{graphicx}
\usepackage{amsmath}
\usepackage{amssymb}
\usepackage{lscape}
\usepackage{color}
\usepackage{longtable}
\usepackage{natbib,twoopt}
\usepackage{enumitem}
\usepackage{mathabx}
\usepackage{booktabs}
\usepackage{footmisc}
\usepackage{multirow}
\usepackage{subfig}
\usepackage{xcolor}
\usepackage{hyperref}
\usepackage{xfrac}

\hypersetup{
    colorlinks=true,       
    linkcolor=black,          
    citecolor=teal, 
    urlcolor=teal 
}

\newcommand{\lya}{Ly$\alpha$}
\newcommand{\civ}{\ion{C}{iv}} 
\definecolor{Pink}{rgb}{1.0,0.05,0.5}

\begin{document} 

\title{Simulating MOS science on the ELT: \lya\,forest tomography}
\titlerunning{IGM tomography with the ELT}

\author{
J.~Japelj\inst{1}
\and
C.~Laigle\inst{2} 
\and 
M.~Puech\inst{3}
\and
C.~Pichon\inst{4,5}
\and
H.~Rahmani\inst{3}
\and
Y.~Dubois\inst{4}
\and
J.~E.~G.~Devriendt\inst{2}
\and
P.~Petitjean\inst{4}
\and
F.~Hammer\inst{3}
\and
E.~Gendron\inst{6}
\and
L.~Kaper\inst{1}
\and
S.~Morris\inst{7}
\and
N.~Pirzkal\inst{8}
\and
R.~S\'anchez-Janssen\inst{9}
\and
A.~Slyz\inst{2}
\and
S.~D.~Vergani\inst{3}
\and
Y.~Yang\inst{3}
}

\institute{Anton Pannekoek Institute for Astronomy, University of Amsterdam, Science Park 904, 1098 XH Amsterdam, The Netherlands; \email{j.japelj@uva.nl}
\and Sub-department of Astrophysics, University of Oxford, Keble Road, Oxford OX1 3RH, UK
\and GEPI - Observatoire de Paris Meudon. 5 Place Jules Jannsen, F-92195, Meudon, France
\and Institut d’Astrophysique de Paris, UMR 7095, CNRS-SU, 98 bis Boulevard Arago, F-75014 Paris, France
\and Korea Institute of Advanced Studies (KIAS), 85 Hoegiro, Dongdaemun-gu, Seoul 02455, Republic of Korea
\and LESIA, Observatoire de Paris, Université PSL, CNRS, Sorbonne Universit\'e, Univ. Paris Diderot, Sorbonne Paris Cit\'e, 5 place Jules Janssen, 92195 Meudon, France
\and Centre for Extragalactic Astronomy, Durham University, South Road, Durham DH1 3LE, UK
\and Space Telescope Science Institute, 3700 San Martin Drive, Baltimore, MD 21218, USA
\and UK Astronomy Technology Centre, Royal Observatory, Blackford Hill, Edinburgh, EH9 3HJ, UK
}

\date{Received DD Mmmm YYYY / Accepted DD Mmmm YYYY} 

\abstract{
Mapping the large-scale structure through cosmic time has numerous applications in studies of cosmology and galaxy evolution. At $z \gtrsim 2$, the structure can be traced by the neutral intergalactic medium (IGM) by way of observing the \lya\, forest towards densely sampled lines of sight of bright background sources, such as quasars and star-forming galaxies. We investigate the scientific potential of MOSAIC, a planned multi-object spectrograph on the {\it European Extremely Large Telescope} (ELT), for the 3D mapping of the IGM at $z \gtrsim 3$. We simulated a survey of $3 \lesssim z \lesssim 4$ galaxies down to a limiting magnitude of $m_{r}\sim 25.5$ mag in an area of 1 degree$^2$ in the sky. Galaxies and their spectra (including the line-of-sight \lya\, absorption) were taken from the lightcone extracted from the {\sc Horizon-AGN} cosmological hydrodynamical simulation. The quality of the reconstruction of the original density field was studied for different spectral resolutions ($R=1000$ and $R=2000$, corresponding to the transverse typical scales of~2.5 and 4~Mpc) and signal-to-noise ratios ($S/N$)  of the spectra. We demonstrate that the minimum $S/N$ (per resolution element) of the faintest galaxies that a survey like this has to reach is $S/N = 4$. We show that a survey with this sensitivity enables a robust extraction of cosmic filaments and the detection of the theoretically predicted galaxy stellar mass and star-formation rate gradients towards filaments. By simulating the realistic performance of MOSAIC, we obtain $S/N(T_{\rm obs}, R, m_{r})$ scaling relations. We estimate that $\lesssim 35~(65)$ nights of observation time are required to carry out the survey with the instrument's high multiplex mode and with a spectral resolution of $R=1000$ (2000). A survey with a MOSAIC-concept instrument on the ELT is found to enable the mapping of the IGM at $z > 3$ on Mpc scales, and as such will be complementary to and competitive with other planned IGM tomography surveys.
}
  
 
\authorrunning{} 
\maketitle
  
\section{Introduction}
The large-scale distribution of matter in the Universe is organized into a {\it \textup{cosmic web}} \citep{Bond1996}, which arises from the anisotropic collapse of the initial fluctuations in the matter density field \citep{Zeldovich1970}. This intricate structure consists of large void regions that are surrounded by sheet-like walls and are bordered by connected filaments \citep[e.g.][]{Springel2006}. The filaments intersect in the densest regions in the web that are associated with galaxy clusters. The structure was first observed in the Center for Astrophysics (CfA) galaxy redshift survey \citep{Lapparent1986}, has been confirmed by many subsequent surveys \citep[e.g.][]{Geller1989,Colless2001,Tegmark2004,Cole2005}, and has been successfully reproduced in cosmological {\it N}-body and hydrodynamical simulations. 

The topological analysis of the density field on large scales ($>10$ Mpc) can be used to infer details about primordial density fluctuations in the early Universe, to understand the process of the structure's growth, and the mechanisms governing the Universe's expansion, that is, to constrain the cosmology \citep[e.g.][]{park92,matsubara95,Zunckel2011,Wang2012,Codis2013,Codis2018,Appleby18}. On the other hand, the cosmic web environment on the scale of a few Mpcs and below impacts galaxy formation because galaxies accrete matter and advect angular momentum from the large-scale filaments. 
There is now ample evidence from simulations that dark matter (DM) halo and galaxy spin tend to align with the vorticity of the large-scale flows \citep[e.g.][]{libeskind13,laigle15,veena18} and therefore with large-scale filaments \citep[e.g.][]{AragonCalvo2007,codis12,Dubois2014,chen15}. These theoretical findings are now supported by a few observational measurements at low-$z$ \citep[e.g.][]{Tempel2013,zhang2015,hirv2017}, but this signal remains difficult to reliably extract at higher redshift because it requires an accurate measurement of both filament and galaxy angular momentum orientations.   
In addition to driving angular momentum acquisition, the cosmic web participates in shaping the galaxy mass assembly. Studies based on spectroscopic or photometric surveys have revealed that at low-$z$, galaxies lying closer to large-scale filaments are on average redder, more evolved, and more massive \citep{Rojas2004,Beygu2016,alpaslan16,Kuutma2017,Chen2017,Malavasi2017,Laigle2018,Kraljic2018} than those that lie farther away, an effect which seems not entirely driven by the (isotropically averaged) local density (\citealt{kraljic19}, but see \citealt{Goh2019} on the environmental dependency of DM halo properties for a different conclusion). The most direct interpretation for these results is provided by considering the impact of the tidal field, whose structure depends on the intrinsically anisotropic large-scale geometry of the matter distribution in filaments. This field shapes the accretion rate onto DM halos beyond the density \citep{Musso2018} and therefore drives an environmental assembly bias \citep[as has been described in][]{hahn2009}. However, when we consider the growth of galaxies, we also have to account for the state of the gas in filaments and the baryonic processes within the galaxies (especially feedback from active galactic nuclei, AGN). The dependency of the galaxy mass assembly on their large-scale environment and how it evolves with scale and redshift is a highly debated topic. High-redshift observations are pivotal to address this question.

However, mapping the large-scale structure with galaxy redshift surveys alone becomes progressively more difficult with increasing redshift. Securing spectroscopic redshifts for a sample of relatively faint galaxies ({\it R} $\lesssim25$ mag) in a sufficiently large volume at $z \gtrsim 2$ may require an unrealistically long exposure time per galaxy \citep[e.g.][]{Lee2014a}. A possible solution is to rely on deep multi-wavelength photometric surveys: \citet{Laigle2018} showed that the photometric redshifts from the Cosmological Evolution Survey (COSMOS) \citep{Scoville2007} are measured with a high enough precision \citep{Laigle2016} to enable the study of the 3D properties of the cosmic web at $0.5 < z < 0.9$. Alternatively, mapping large volumes at $z \gtrsim 2$ can be efficiently achieved by observations of the Lyman-$\alpha$ (\lya) forest absorption towards the bright distant background sources. The forest represents the absorption due to the Lyman transition of H\,\textsc{i} in the clouds of the highly ionized intergalactic medium (IGM) that lies in the line of sight (LOS) to the background light source \citep{Gunn1965}. H\,\textsc{i} is a good tracer of the total hydrogen density because the IGM is in a photoionization equilibrium. Furthermore, on large scales, the gas distribution in the IGM follows the DM reasonably well \citep[e.g.][]{Cen1994,Petitjean1995,Viel2004,Caucci2008,Cui2018}. In addition, this method also allows an efficient mapping at low overdensities. The \lya\, forest can therefore serve as a powerful tool for mapping the large-scale structure down to the $\sim$ Mpc scales in large volumes and/or large area in the sky. Towards that end, techniques have been developed to carry out a tomographic reconstruction of the observed \lya\, forest absorption field from a set of sight lines \citep[e.g.][]{Pichon2001,Caucci2008,cisewski14,horowitz19}. Several studies have discussed the possible applications of the technique for studying high-redshift protoclusters \citep{Stark2015b}, voids \citep{Stark2015b}, quasar light echoes \citep{Visbal2008,Schmidt2019}, and topology \citep{Caucci2008} (see also \citealt{Lee2014a} and Sect. \ref{synergy} for further discussion), among others.

The resolution of the mapping is first and foremost determined by the number density of sight lines $n_{\rm los}$; the typical distance between sight lines scales as $\approx n_{\rm los}^{-0.5}$. This method is therefore mostly limited by the number density of available bright background sources. Bright quasars are typically used as background sources to study the properties of the IGM from the local $z\sim 0$ Universe up to $z\sim 6$ \citep[e.g.][]{McQuinn2016}. However, the mean separation between sight lines even in the most ambitious quasar survey so far, the SDSS-III Baryon Oscillation Spectroscopic Survey (BOSS)  \citep{Eisenstein2011,Dawson2013} is several tens of Mpc at $z \gtrsim 2$ \citep[e.g.][]{Lee2014a,Ozbek2016}. The limiting magnitude in the BOSS survey is  $m_{r}<21.9$ \citep{Paris2012}, but because of the shallow faint end of the quasar luminosity function, even a much fainter limiting magnitude would not result in a high enough density of sources to allow the mapping of $\sim$~Mpc scales \citep{Lee2014a}. However, star-forming galaxies, also known as Lyman-break galaxies (LBG; \citealt{Steidel1996}), can also be used as background sources. The density of LBGs increases fast with decreasing luminosity at $z \sim 2 - 4$ \citep[e.g.][]{Reddy2009,Alavi2016,Bouwens2015} and enables the mapping of the large-scale structure at the scale of a few Mpc, provided that we extend spectroscopic observations down to sufficiently faint galaxies of $m_{r} \sim 25$~mag.

Spectroscopic observations of such faint galaxies in sufficient numbers may seem challenging. However, as pointed out by \citet{Lee2014a}, the signal-to-noise ratio ($S/N$) and spectral resolution required for a successful reconstruction of the \lya\ absorption field are low enough to allow the technique to be applied even in the frame of current 10-meter-class telescopes. Recently, the feasibility of a study like this has been demonstrated based on data from the COSMOS Lyman-Alpha Mapping And Tomography Observations (CLAMATO) survey \citep{Lee2014b} on the Keck~I telescope. The data were used to make a first reconstructed map of the large-scale structure at $z \sim 2 - 2.5$ using LBGs as background sources \citep{Lee2018}. This led to the detection of voids at $z\sim 2.3$ \citep{krowleski18} and a protocluster at $z \sim 2.5$ \citep{Lee2016}. IGM tomography with both the LBG galaxies  and quasars as background sources is one of the primary scientific goals for several upcoming spectrographs, such as the Prime Focus Spectrograph on Subaru \citep{Takada2014}, WEAVE on the William Herschel Telescope \citep{Pieri2016} and the multi-object spectrograph (MOSAIC) on the upcoming European Extremely Large Telescope (ELT; \citealt{Puech2018}). 

The ELT with its primary mirror of 39 meters and its advanced technology will enable us to obtain spectra of faint objects in a very short time with respect to previous facilities. This enables tomographic surveys at very high redshift  on relatively large fields. The aim of this work is to investigate the scientific potential of a multi-object spectrograph (MOS) on the ELT in the studies of the IGM, in particular, the IGM tomography and the corresponding science cases. We base our study on the MOSAIC spectrograph as envisioned at the end of the instrument's Phase A study \citep{Morris2018}. 

We begin by exploring how the quality of the reconstruction of the IGM density field changes for different spectral resolutions and $S/Ns$  (Sect.~\ref{horizon}). We use the galaxy spectral catalogue extracted from the {\sc Horizon-AGN} cosmological hydrodynamical simulation \citep{Dubois2014}, apply realistic noise to the simulated galaxy spectra, and  reconstruct the flux contrast field. We search for optimal spectral configuration. In particular, we study how well the reconstructed field can represent the actual cosmic web environment of galaxies. In Sect.~\ref{mosaic} we use the characteristics of the MOSAIC instrument to understand the performance of the instrument at blue wavelengths and to estimate the relation between S/N and several other parameters (e.g. exposure time, resolution, and brightness of the background source). Based on our results, we outline a strategy for a large survey with an MOS facility on the ELT to study the IGM (Sect.~\ref{discuss}). In this work we use cosmological parameters provided by the WMAP-7 data \citep{Komatsu2011}.

\begin{figure}
\begin{center}
\includegraphics[scale=0.52]{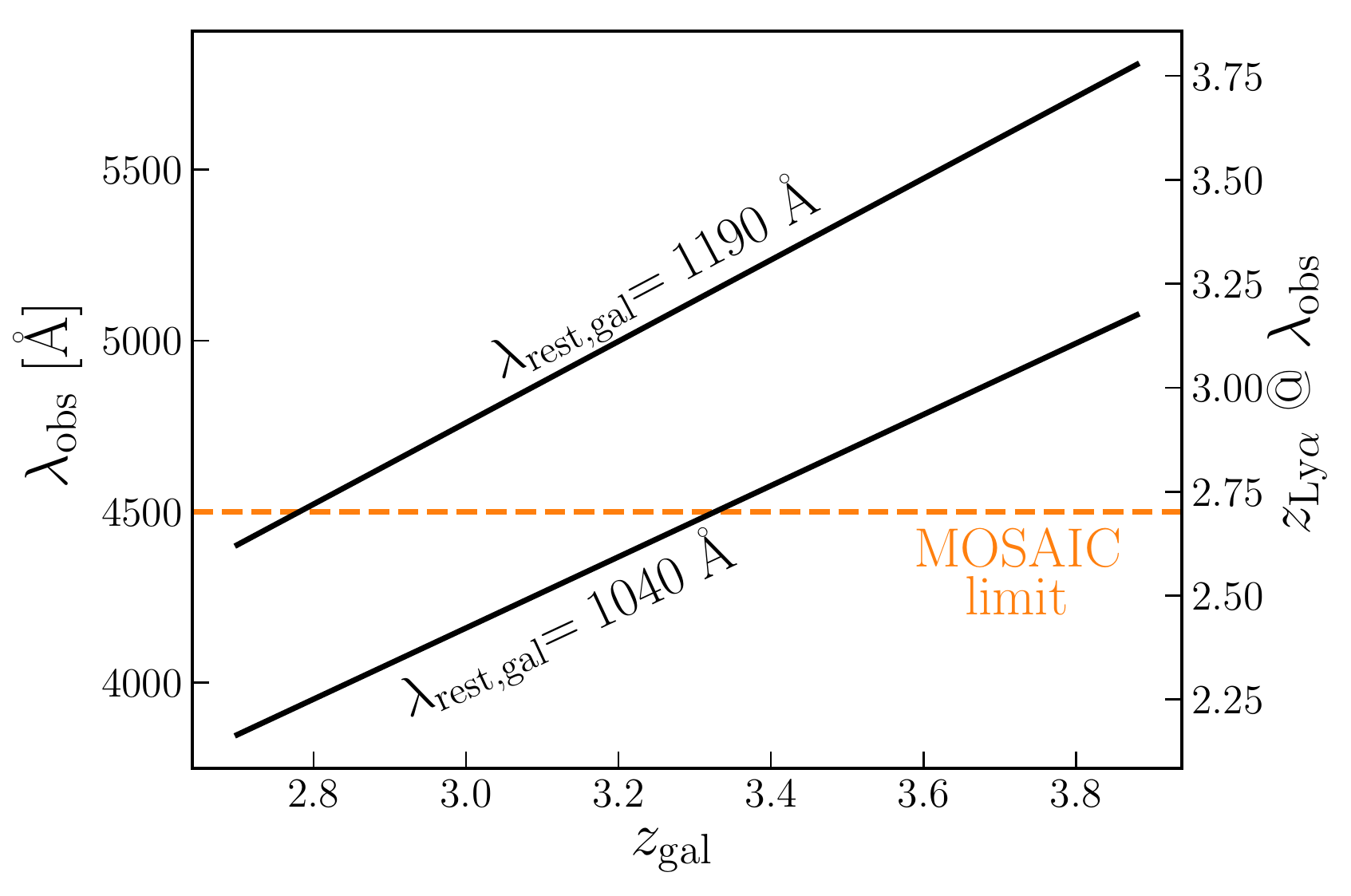}
\caption{Redshift range to be considered in our simulations. The $x$-axis shows the redshift of a background LBG $z_{\rm gal}$, the left $y$-axis shows the observed wavelength range. The two black lines show the observed wavelengths of the rest-frame $1040-1190~\mathrm{\AA}$ range for each $z_{\rm gal}$, i.e., the wavelength range of \lya\, forest. The right $y$-axis shows the redshift of a Ly$\alpha$ absorber detected at the corresponding observed wavelength. The orange dashed line indicates the currently planned blue wavelength limit of the instrument.}
\label{fig1}
\end{center}
\end{figure}

\begin{figure*}
\begin{center}
\includegraphics[scale=0.48]{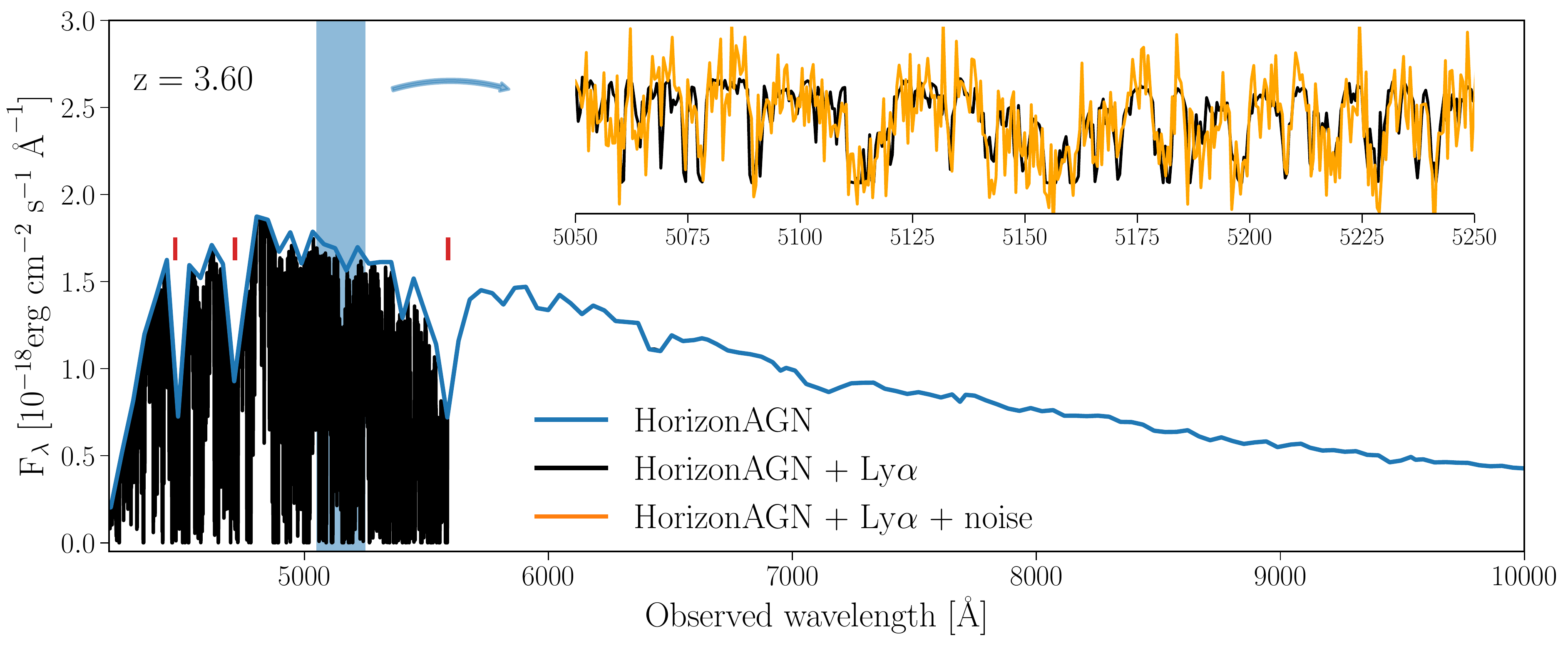}
\caption{Example of a galaxy spectrum simulated in the {\sc Horizon-AGN} simulation. The galaxy lies at $z = 3.60$ and has a magnitude of $m_{\rm rest,UV} = 25.05$ mag. Its synthetic spectrum is shown in blue. Red dashes indicate the positions of the Ly$\alpha$, Ly$\beta,$ and Ly$\gamma$ lines that originate in the galaxy. A spectrum with the added Ly$\alpha$ forest (even though we do not use it in the further analysis, we also plot the Ly$\beta$ forest etc.) is shown in black. Gaussian noise is added to this spectrum (see text for details), and the resulting spectrum is shown in orange in the zoomed plot. The noise corresponds to $S/N = 5$ and a resolution of $R=5000$ (an exposure of $\sim 1.5$ h at a fiducial resolution of MOSAIC, see Section \ref{mosaic}).}
\label{fig4}
\end{center}
\end{figure*}

\section{Simulating tomographic reconstruction}
\label{horizon}

First we explore the required parameter space (e.g. spectral resolution and $S/N$ of the background galaxy spectra) for the science case of IGM tomography. It should be emphasized that the reconstruction of the IGM is only an intermediate step leading to the actual scientific questions such as the studies of voids or the interplay between galaxies and cosmic web. It is not straightforward to understand which quality of observational data is needed to successfully carry out an investigation of a given problem. Here we use a head-on approach: we take a galaxy catalogue extracted from the {\sc Horizon-AGN} hydrodynamical simulation, and apply artificial noise to the simulated spectra. Our goal is to understand the quality of the reconstruction of the IGM, and to assess how well we can quantify the dependency of galaxy properties, such as stellar mass and star formation rate, on the distance to the closest filament. The conclusions of this section are general and overall independent of any particular instrument.

\subsection{Horizon-AGN simulation}
\subsubsection{Description of the simulation}
{\sc Horizon-AGN} is a cosmological (100 h$^{-1}$Mpc on a side) hydrodynamical simulation\footnote{\url{https://www.horizon-simulation.org/}} (see \citealt{Dubois2014} for details) that is run with the adaptive mesh refinement code RAMSES. The simulation assumes $\Lambda$CDM cosmology with cosmological parameters that are compatible with the WMAP-7 data \citep{Komatsu2011}. Heating from a uniform background starts at $z = 10$ and follows the formulation of \citet{Haardt1996}. Star formation occurs in regions of a gas hydrogen number density above $n_{0}= 0.1$ H cm$^{-3}$. Metals and energy releases from stellar winds, supernovae, and AGNs are included in the simulation.

A lightcone with an angular diameter of 1~degree (the choice for this area is justified in Sect.~\ref{survey}) was extracted from the simulation. As part of the {\sc Horizon-AGN} Virtual Observatory effort \citep[see][for details]{laigle19}, galaxies were identified from the distribution of star particles   with the halo finder {\sc AdaptaHOP}  \citep{aubert04}. A mock spectrum was assigned to each galaxy using \citet[][]{bruzual&charlot03} stellar population models, assuming that each particle behaves like a single stellar population. This computation also includes dust absorption using the gas-phase metallicity distribution as a proxy for dust distribution. Dust mass was calibrated based on the gas-phase metal mass, with a dust-to-metal mass ratio of 0.2. The redshift evolution of galaxy colours, luminosity, and mass functions of the extracted galaxies from Horizon-AGN simulation match the observational datasets well \citep[e.g.][]{Kaviraj2017}. Although the simulated galaxy counts are broadly in agreement with observations, the {\sc Horizon-AGN} simulation overestimates the abundance of low-mass blue galaxies at high redshift. This can be an issue when background sources are selected based on their photometry because the density of sight lines, and therefore the performance of the reconstruction, might be overestimated. To correct for this issue, the galaxy catalogue was pruned at the faint end. This was achieved by randomly choosing sources from the {\sc Horizon-AGN} catalogue at the faint end ($m_{r}>24$) until the number count matched the count from the COSMOS2015 catalogue \citep{Laigle2016}.

\subsubsection{Preparation of the spectra}
Along the LOS of each galaxy, the gas density, temperature, and radial velocity were extracted with a resolution of 50~kpc, and the corresponding H\,\textsc{i} optical depth and  transmitted flux distribution, including redshift space distortion, were prepared as described in \cite{laigle19}. Sight lines were drawn within the cone where the Lyman-$\alpha$ forest was implemented. The average properties of the simulated galaxy spectra, attenuated by the Lyman-$\alpha$ forest, are overall in good agreement with the observed average spectrum of LBGs at these redshifts (see Appendix \ref{stacked}).

While the added noise and the adopted spectral resolution ($R$) are not instrument-dependent (e.g. the values are generalized and do not correspond to a particular instrument), we made certain assumptions so that the redshift range of the study and resolution is the same as that of the MOSAIC instrument (the instrument design is discussed in detail in Sect.~\ref{mosaic}). Firstly, we assumed that an observation is carried out with a single spectrograph configuration in the 4500 - 6000 $\mathrm{\AA}$ wavelength range. The available redshift - wavelength space is schematically shown in Fig.~\ref{fig1}. The blue limit of 4500 $\mathrm{\AA}$ means that we can probe neutral IGM at $z_{\rm Ly\alpha} \geq 2.8$. Background LBGs therefore have to lie at $z_{\rm gal} \gtrsim 2.9$ (see Fig. \ref{fig1}). In our experiment we picked the background sources in the $3 < z < 4$ range and reconstructed the field at $3 < z < 3.5$.

We smoothed the simulated spectra to achieve the foreseen spectral resolution ($R = 5000$) taking into account the oversampling (see Section \ref{mosaic}). The throughput was fixed at $13 \%$. The brightness of the galaxies is known. Either a constant exposure time can be assumed for all galaxies and the $S/N$ estimated using the scaling relation (Equations \ref{scalingIGM} and \ref{scalingHMM}), or the opposite approach can be taken and a constant $S/N$ can be assumed for all the observed galaxy spectra. We decided to take the latter, instrument-independent, approach. Given the galaxy brightness and the intrinsic shape of the spectral energy distribution (SED), we then calculated the expected exposure time for each galaxy. For a given value of $S/N$, Gaussian noise was added to the spectra. We checked that the noise is indeed Gaussian and not e.g. Poissonian by statistically analyzing the distribution of the noise in the simulated spectra (Sec. \ref{mosaic}).~When the final resolution that we wished to achieve was lower than the fiducial resolution, then the noisy spectrum was smoothed accordingly. We emphasize that MOSAIC always observes at the fiducial resolution of $R=5000$, regardless of the resolution necessary for a particular science case. Finally, we also simulated a noise spectrum using the average properties of the sky spectrum and the adopted CCD parameters (read-out noise). An example of a part of the spectrum before and after the noise is applied is shown in Fig. \ref{fig4}. The $S/N$ throughout this paper is defined as the $S/N$ per resolution element.

The simulated galaxy spectra do not include possible intrinsic absorption from the interstellar medium (e.g. NII, NI, CIII, or SiII), which can contaminate the Lyman-$\alpha$ forest. In practice, these lines can be masked in the observed spectrum, but at the cost of missing the Lyman-$\alpha$ forest information in these regions. A portion of the LOS will be contaminated by intervening galaxies. A system like this will be impossible to identify in the absence of a rather high-resolution ($R \sim 5000$) spectroscopy (which is much higher than the resolution needed for the reconstruction, see below) and will contribute to the overall noise. In addition, errors on the continuum determination are not implemented.

\subsubsection{Tomographic reconstruction method}
The 3D distribution of the Lyman-$\alpha$ flux contrast, which is commonly used as a proxy for the inverse of the density field, was reconstructed by interpolating between the LOS using Wiener filtering in comoving space \citep[see][]{Pichon2001,Caucci2008,Lee2018}. 
We recall below the main elements of the method and the choice of the parameters. 
Let     $\mathbf{D}$ be the 1D array representing the dataset, and $\mathbf{M}$ is the 3D array  of the field estimated from the data.  Maximizing the penalized likelihood of the data given an assumed prior for the flux contrast field yields
        \begin{equation}
        \textbf{M} = \mathbf{C_{\delta^{3d} \delta }} (\mathbf{C_{\delta \delta}} +\mathbf{N})^{-1} \mathbf{D}\,,
        \end{equation}
where $\mathbf{C_{\delta^{3d} \delta}}$ is the mixed parameter-data covariance matrix, and $\mathbf{C_{\delta \delta}}$ is the data covariance matrix. We assume that the noise is uncorrelated, therefore the noise covariance matrix can be expressed as $\mathbf{N}=n^{2}\mathbf{I}$, $n$ being set by the S/N on the spectra. In addition, we assume a priori normal covariance matrices:
\begin{equation}
\mathbf{C_{\delta \delta}}(x_{1},x_{2},\mathbf{x_{1\rm T}},\mathbf{x_{2\rm T}})={\sigma^{2}}e^{-\dfrac{\vert x_{1}-x_{2} \vert ^{2}}{2 L_{x}^2}}\!\!e^{-\dfrac{\vert \mathbf{x_{1 T}}-\mathbf{x_{2 T}} \vert^{2}}{2 L_{\rm T}^{2}}}
,\end{equation}
where ($x_{i}$,$\mathbf{x_{iT}}$) are the coordinates of the point along and perpendicular to the LOS.

The reconstruction depends on $\sigma^{2}$, $n$, and the correlation lengths  $L_{x}$  and $L_{\rm T}$, along and perpendicular to the LOS, respectively. The transverse correlation length $L_{\rm T}$ is set by the mean inter-LOS distance, the longitudinal correlation length $L_{x}$ by the comoving scale corresponding roughly to our spectral resolution, and $\sigma^{2}$ quantifies the fluctuations of the field in a volume $L_{\rm T}^{2}\times L_{\rm x}$. The light-cone volume was partitioned into sub-boxes and we reconstructed each block individually in order to speed up the computation. Blocks overlapped (over $3 \times L_{\rm T}$) to avoid edge effects. 
Tomography was performed on the flux  contrast, $\delta = F/\langle F\rangle_{z} - 1$, where $F = \exp\left(-\tau\right)$ is the \lya\, transmitted flux and $\langle F\rangle_{z}$ is the mean value at a given $z$. Practically, this imposes assuming or precisely determining the mean flux in the Lyman-$\alpha$ forest at a given redshift. For the purpose of this work, we used the exact value from the simulation ($\langle F \rangle \sim$0.7 at $z=3$). An error in determining the mean flux will induce systematics in the reconstruction.

\begin{figure*}
\begin{center}
\includegraphics[scale=0.8, angle=270
]{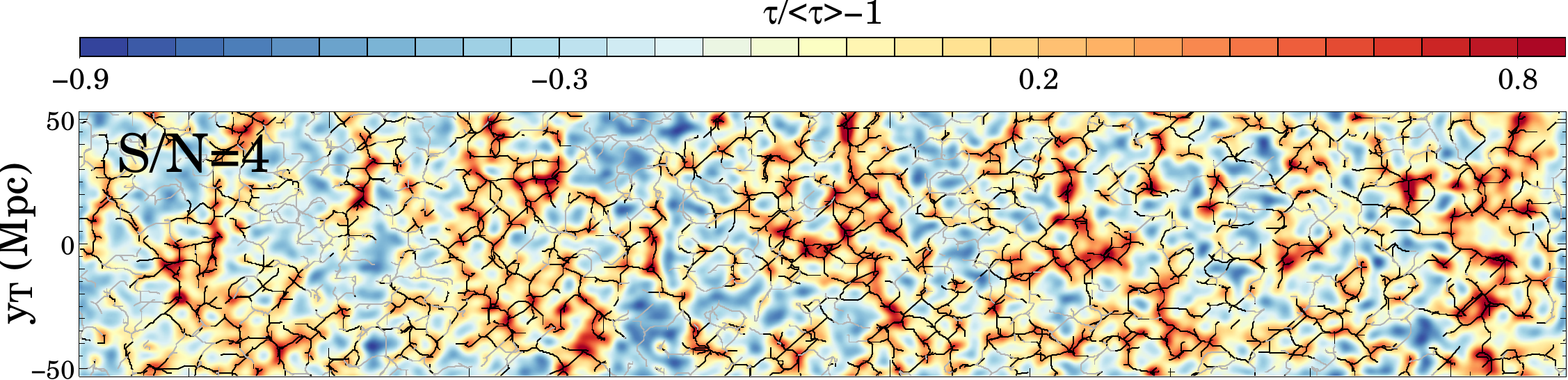} \\
\vspace{0.2cm}
\includegraphics[scale=0.8, angle=270
]{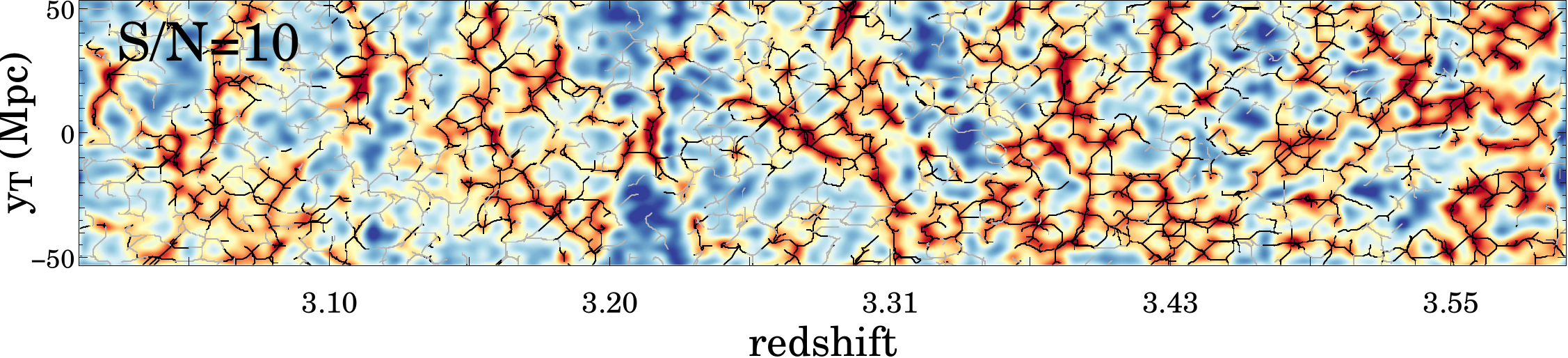} \\
\vspace{0.2cm}
\includegraphics[scale=0.8, angle=270
]{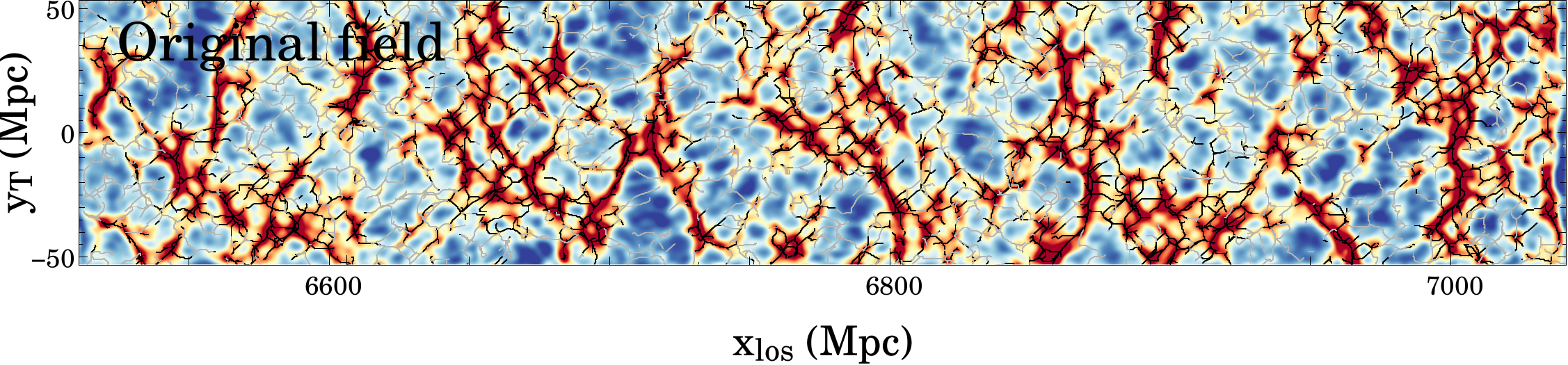} 
\caption{Longitudinal slices of  thickness $\sim 10$~Mpc (i.e. four times the smoothing length) of the reconstructed (\textit{top} for $S/N=4$ and \textit{middle} for $S/N=10$) and original (\textit{bottom}) optical depth contrast field $\tau/\langle \tau \rangle -1$, where $\tau = -\log F$ and $F$ is transmitted flux. $\tau$ is taken as a proxy for the HI density.
The original field is smoothed with a Gaussian kernel over 2.5 Mpc. Filaments extracted with {\sc DisPerSE} are overplotted in grey and black. The black lines correspond to the 50\% densest filaments. As noise in the spectra increases, the reconstruction shows lower contrast, and filaments are more randomly located. }
\label{fig09}
\end{center}
\end{figure*}

\subsubsection{Configurations}
\label{configuration}
As mentioned before, we discuss here a volume-limited survey, where all spectra have the same $S/N$, and we focus on how the reconstruction varies as a function of $S/N$ in order to determine a sensible $S/N$ that can be aimed for in an actual survey. In theory, the reconstruction can be performed anisotropically with different values for $L_{\rm x}$ and $L_{\rm T}$. However, for the purpose of studying galaxy evolution in the cosmic web, we require an isotropic smoothing to extract the large-scale structure. We therefore chose here to perform an isotropic reconstruction. We assumed that we are able to observe all the sources brighter than 25.5 in the $r$ band, which corresponds to a mean inter-LOS distance of $\sim 2$~Mpc at redshift $z = 3.4$. We tested the quality of the reconstruction for two spectral resolutions ($R=1000$ and $R=2000$). The spectral resolution was here the limiting length, and therefore we adopted two configurations: $L_{\rm T}=L_{\rm x}=4$~Mpc ($R=1000$) and $L_{\rm T}=L_{\rm x}=2.5$~Mpc ($R=2000$). We measured $\sigma= 0.14$ and 0.19 for the two configurations, respectively, and we set $n$ as a function of the $S/N$ on the spectra. We set $0$ and $1$ as the lower and upper limit for the transmitted flux, that is, the pixels that are brought below 0 or above 1 due to the noise were set to 0 and 1, respectively. The two sets of configurations, called C1 and C2 in the following, are summarized in Table~\ref{Tab:conf}.

\begin{table}
\small 
\begin{center}
  \caption{Summary of the parameters for the two main configurations, {\cal C}$_1$ and {\cal C}$_2$.  \label{Tab:conf}}
\begin{tabular}{c c c c} \hline \hline
                 &  & {\cal C}$_1$ & {\cal C}$_2$  \\ \hline
    Characteristics    &  $m_r$ &25.5 & 25.5 \\
    of the survey    &  R & 1000 & 2000 \\ \hline
     Parameters for   &  $L_{\rm T}$ & 4 Mpc & 2.5 Mpc \\
    the   tomographic  &   $L_{\rm x}$ & 4 Mpc & 2.5 Mpc \\
     reconstruction   &   $\sigma$ & 0.14 & 0.19 \\ \hline
    Persistence cut of the skeleton   &  c & 0.005 & 0.03 \\
  \hline
\end{tabular}
\end{center}
\end{table}

In addition to the presented main two sets of reconstructions, we also inspected several other scenarios (but with the same basic C1 and C2 configurations) that we detail below. All these cases are also summarized in Table~\ref{tab:perf}, and are discussed and presented alongside the results of the main study throughout this section.
\begin{itemize}[noitemsep,nolistsep]
\item We tested the effect of noise on the tomographic reconstruction using in addition to the various constant $S/N$ cases an additional configuration with spectra that were not perturbed by noise (i.e. with an infinite $S/N$, labelled {\it \textup{'no noise'}} in the following).\item In order to probe the convergence of the Wiener filtering method we performed a reconstruction on a uniform distribution of sight lines (labelled {\it \textup{'no noise and uniform'}} in the following) at very high spectral resolution, and un-perturbed by noise. For this case, the same density of background sources was adopted, but uniformly distributed on the grid. This configuration helps us to understand in particular the effect of the clustering of background sources on degrading the quality of the reconstruction; 
\item It is unreasonable to expect that the observed galaxy spectra will all have the same $S/N$, as assumed in the main part of the study. For the C2 configuration we built an additional sample (called {\it SN=4, CT} in the following) by setting a constant observation time to be $T_{\rm obs} = 3.7$ h, which corresponds to the $S/N = 4$ for the faintest $m_{r} = 25.5$~mag sources (see equation \ref{scalingIGM}). By performing the reconstruction with this sample, we tested the improvement in the reconstruction compared to the case when $S/N=4$ for all sources, regardless of the magnitude. This is the only case that depends on instrument specifications.
\item For the C1 configuration the number density of sources brighter than $\sim 25$ mag is already sufficient for the performance of the reconstruction, while we also included the fainter sources in the main analysis. We performed a set of reconstructions by taking only the sources with $m_{r} \lesssim 25$ mag. This allowed us to test in particular whether adding more sources at fainter magnitudes without changing the transverse length allows a better reconstruction of the field. 
\end{itemize}

\subsection{Performance of the reconstruction: galaxies in  the cosmic web}
\label{galaxies}

We estimated the performance of the tomographic reconstruction for the purpose of the cosmic web analysis in the different configurations detailed above. 
As a first illustration, Fig.~\ref{fig09} presents three $\sim 10$~Mpc-thick longitudinal slices, built from the original and two reconstructed ($R=2000$ and $S/N=4$ and 10) optical depth contrast fields. We display here $\tau/\langle \tau \rangle -1$, where the optical depth $\tau=-\log F$ is taken as a proxy for the density. In order to  compare to the reconstructed field, the original field was smoothed with a Gaussian kernel over 2.5~Mpc. By eye, the reconstructed and original fields agree relatively well. The reconstructed field seems to show less contrast as noise increases, and filaments are more randomly distributed. 

\begin{figure*}[!t]
\begin{center}
\includegraphics[width=\textwidth]{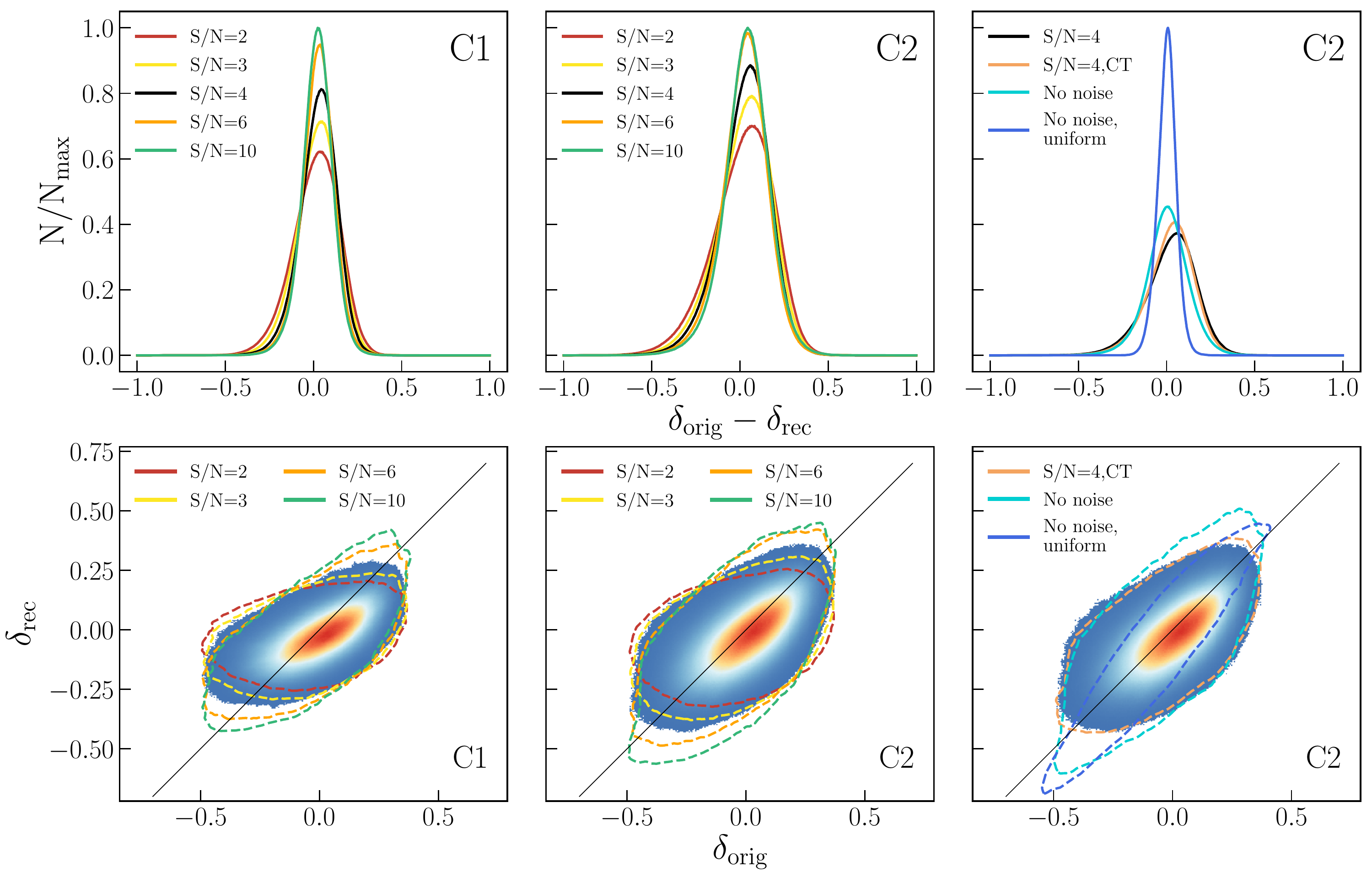}
\caption{Assessing the quality of the reconstructed flux contrast field. The quality is checked for the different values of $S/N$ and two different configurations, corresponding to the resolution of $R=1000$ and the reconstruction scale of $L_{\rm T}=4$ Mpc (configuration C1, left column) and the resolution of $R=2000$ and the reconstruction scale of $L_{\rm T}=2.5$ Mpc (configuration C2, middle column). The right column shows 3 additional flavors of the C2 configuration (see Section~\ref{configuration} for details), with a constant exposure time and at least $S/N=4$ on all spectra (S/N=4,CT), with no noise on spectra (No noise) and with a spatially uniform background sources distribution (No noise, uniform). The upper row shows the distributions of the pixel-by-pixel differences between the smoothed original and reconstructed flux contrast fields. Note that the distributions are normalized in a way that the peak of the narrowest distribution equals 1. The bottom row shows scatter plots of the original flux contrast field (for $S/N = 4$) against the reconstructed one. The original field has been smoothed with the same $L_{\rm T}$ scale as the corresponding reconstructed one. Black line shows $\delta_{rec} = \delta_{orig}$ relation. Overplotted contours of number counts (computed at a level of 10 counts and smoothed afterwards) show the change of the 2D distribution for reconstruction for different $S/N$ values of the spectra.}
\label{fig10}
\end{center}
\end{figure*}

\begin{figure}
\begin{center}
\includegraphics[scale=0.63]{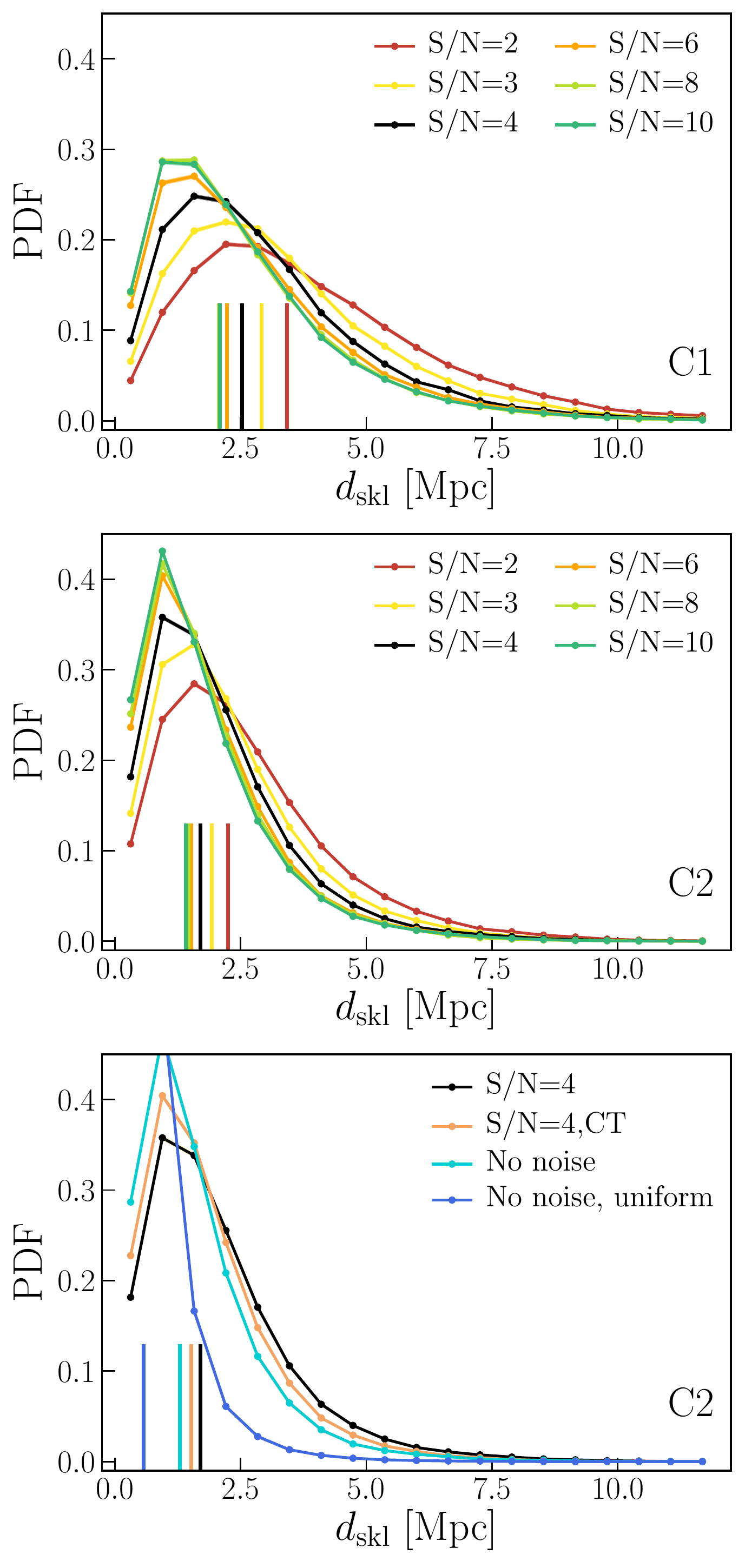}
\caption{Probability density function of the pseudo-distance between the skeletons computed on the original and reconstructed fields in the C1 (top) and C2 (middle and bottom) configurations for various $S/N$ (see
the caption of Fig.~\ref{fig10} and Section~\ref{configuration} for details). Vertical lines indicate the median values of the distributions.}
\label{fig11a}
\end{center}
\end{figure}

\begin{table*}
\small
\begin{center}
\caption{Quality of the reconstructed field and cosmic filaments as a function of $S/N$. The quality of the  reconstruction is estimated from the standard deviation $\sigma_{\rm f}$ of $\delta_{\rm orig}-\delta_{\rm rec}$ (see Fig. \ref{fig10}) and the median distance $d_{\rm skl}$ between the skeleton extracted from the original (smoothed) and the reconstructed field (see Fig. \ref{fig11a}). We also provide the Spearman correlation coefficients between the closest distances of the galaxies to filaments and nodes ($\rho_{\rm orig,rec}$(fil) and $\rho_{\rm orig,rec}$(nod)) as measured from the skeletons computed on the original and reconstructed fields (see Figs. \ref{figC1} and \ref{figC2}). All the values are computed for two configurations: the low-resolution case with $R=1000$ (and $L_{\rm T} = 4$ Mpc) and the high-resolution case with $R=2000$ (and $L_{\rm T} = 2.5$ Mpc). We also show the results of the reconstruction for several additional configurations (see Section~\ref{configuration} for the details of these configurations). \label{tab:perf}}
\renewcommand{\arraystretch}{1.4}
\begin{tabular}{ c c c c c c c c c} \hline \hline
                     &\multicolumn{4}{c}{R=1000, $L_{\mathrm{T}}=4~\mathrm{Mpc}$} & \multicolumn{4}{c}{R=2000, $L_{\mathrm{T}}=2.5~\mathrm{Mpc}$}\\ 
             \cmidrule(lr{.75em}){2-5}\cmidrule(lr{.75em}){6-9}
             $S/N$           &  $\sigma_{f}$  &  $d_{\rm skl}$ [Mpc] & $\rho_{\rm orig,rec}$(fil) & $\rho_{\rm orig,rec}$(nod) & $\sigma_{f}$ &  $d_{\rm skl}$ [Mpc] & $\rho_{\rm orig,rec}$(fil) & $\rho_{\rm orig,rec}$(nod)\\ 
             \hline
             2  & -0.14,+0.12 & 3.2 & 0.07 & 0.03 & -0.19,+0.15 & 2.3 & 0.10 & 0.06\\
             3  & -0.13,+0.12 & 2.8 & 0.14 & 0.06 & -0.17,+0.14 & 2.0 & 0.15 & 0.09\\
             4  & -0.11,+0.10 & 2.5 & 0.18 & 0.13 & -0.15,+0.12 & 1.8 & 0.20 & 0.10\\
             6  & -0.09,+0.08 & 2.2 & 0.24 & 0.17 & -0.13,+0.12 & 1.6 & 0.24 & 0.14\\
             8  & -0.09,+0.08 & 2.0 & 0.27 & 0.13 & -0.12,+0.11 & 1.5 & 0.25 & 0.16\\
             10 & -0.08,+0.08 & 2.0 & 0.28 & 0.25 & -0.12,+0.12 & 1.5 & 0.26 & 0.16\\
  \hline
          4, 25 & -0.12,+0.11 &  2.8   &0.12 &  0.08 & --          & --& --& --\\
          4, CT &      --     & --  & --   & --   & -0.13,+0.12 & 1.5 & 0.22 & 0.14  \\
          No noise &      --     & --  & --   & --   & -0.11,+0.11 & 1.3 & 0.27 & 0.21  \\
          No noise + uniform & --     & --  & --   & --   & -0.05,+0.05 & 0.6 & 0.51 & 0.40\\
   \hline
\end{tabular}
\end{center}
\end{table*}

\subsubsection{Quality of the reconstructed flux field}
\label{density_comp}
Figure~\ref{fig10} compares the original flux contrast field ($\delta_{\rm orig}$) with the reconstructed field ($\delta_{\rm rec}$) in both the C1 (left) and C2 (middle) configurations. For comparison, the original flux contrast field was also smoothed with a Gaussian kernel at the same scale (2.5~or 4~Mpc). Both the scatter of the reconstruction (bottom) and its 1D-equivalent probability density function (PDF; i.e. the distribution of $\delta_{\rm orig}-\delta_{\rm rec}$; top) are shown. From the 1D distributions we learn not only how scattered the data are (quantified by the standard deviation $\sigma_{f}$, see Table~\ref{tab:perf}), but also whether the distribution is skewed. For low $S/N$ the distributions are skewed towards lower values. It is indeed easier to overestimate the flux  contrast by creating false voids (due to the non-uniform distribution of sight lines because of the clustering of background sources) than underestimating it (we recall that the logarithm of the flux scales as the opposite of the density). The same effect is illustrated by the change with $S/N$ of the contours in Fig. \ref{fig10} (bottom): with higher $S/N$ the scatter is less skewed away from the $\delta_{\rm rec}$=$\delta_{\rm orig}$ relation. The 1$\sigma$ equivalent values of dispersion are provided in Table~\ref{tab:perf}. As shown by our simulations, the quality of the reconstruction does not improve drastically above $S/N\approx$ 4 in either the C1 or C2 configuration. 

In order to estimate how the reconstruction is improved in the more realistic case where the exposure time is similar for all sources, we compare in the right panel of Figure~\ref{fig10} the performance of the $S/N=4$ reconstruction (black line) with the performance of a reconstruction where the $S/N$ is variable but equals 4 for the \textit{\textup{faintest}} sources (orange curve labelled `CT'). The performance of the reconstruction on ideal spectra, that is, spectra that are not perturbed by noise (cyan line) is also shown. Finally, we present the reconstruction in the case of extremely high-resolution spectra that are not perturbed by noise and are spatially uniformly distributed (ocean blue line). Although there is a real improvement in decreasing the noise on spectra (compare orange, cyan, and ocean blue lines), the main gain comes from distributing the background sources more uniformly. Targeting spectra not only based on their magnitude distribution but also on their spatial distribution in order to make it more uniform might help in improving the quality of a matter density field reconstruction based on Lyman-$\alpha$ tomography. 

\begin{figure*}
\begin{center}
\includegraphics[width=\textwidth]{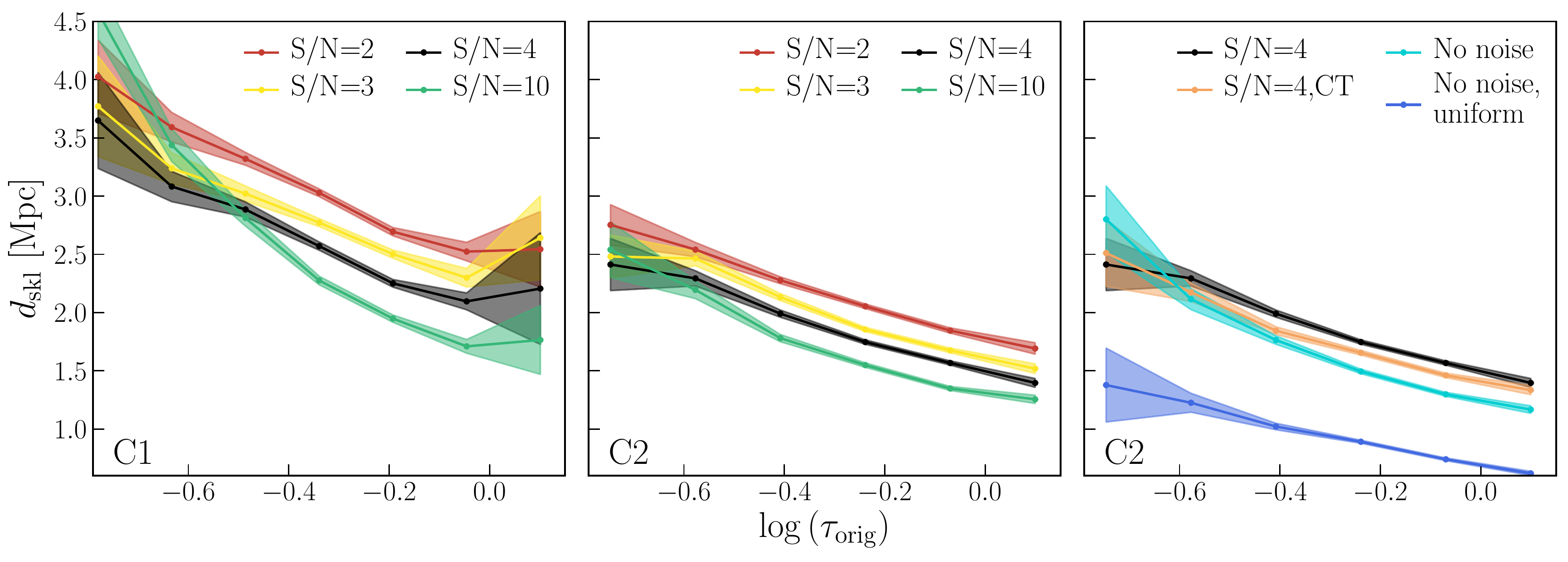}
\caption{Pseudo-distance between the skeletons on the original and reconstructed fields as a function of the density of the original field, where the optical depth $\tau_{\rm orig}$ is taken as a proxy for the density, in the C1 (left panel) and different flavours of the C2 configurations (middle and right panels). See the caption of Fig.~\ref{fig10} and Section~\ref{configuration} for details.}
\label{fig11}
\end{center}
\end{figure*}

\subsubsection{Cosmic web extraction method}

After the flux contrast field was reconstructed in a specific configuration, the filamentary cosmic web was extracted using {\sc DisPerSE} \citep{Sousbie2011,sousbieetal2011}. {\sc DisPerSE} usually identifies the filament from the density field (either a Cartesian grid or  a Delaunay tesselation computed from the distribution of the particles). In our case, filaments were identified on the optical depth field $\tau=-\log F$, which is a better proxy for the density field than the flux contrast itself. 
Filaments were extracted as the ridges connecting the filament-like saddle points to peaks. Each filament was  defined to be a set of connected small segments. The set of filaments is called the skeleton. 

To each pair of critical points, that is, points with a vanishing gradient of the $\tau$ field (maxima, minima, and saddle points), we assigned a persistence, defined as the difference between the underlying $\tau$-value at each critical point. Persistence quantifies the robustness of the underlying topological features that are characterized by this pair.  Filtering low-persistence pairs is a way to filter topologically weak filaments. The ideal persistence threshold needs to be calibrated, and the final distribution of filaments will depend upon this choice. The motivation for the choice of our persistence cut ($c$) is described in Appendix~\ref{App:SkelExt}. The chosen persistence cuts (which vary with the smoothing scale) are displayed in Table~\ref{Tab:conf} (in units of $\tau$). At a given smoothing scale, these cuts yield similar numbers of extracted filaments regardless of the underlying $S/N$. Fewer filaments are extracted in the $C_1$ configuration ($L_{T}=4$ Mpc)  than in $C_2$ because smoothing the  field decreases the number of structures and preserves only the most prominent structures.  

Examples of the  skeletons extracted from the smoothed original and reconstructed $\mathbf{\tau}$-fields are shown in Fig.~\ref{fig09}. Although topologically robust, all the extracted filaments are not equally important for the formation of galaxies, depending on their matter content and the depth of their gravitational potential well. As shown in Fig.~\ref{fig09}, some filaments are extremely dense (black lines), while some can be qualified as `tendrils' (grey lines) that connect galaxies in low-density environments. We might expect that filaments of different densities drive different environmental effects, depending on the relative mass of the halos with respect to the filaments. Therefore a comprehensive analysis should bin filaments depending on their underlying density \citep[as done in e.g.][]{katz2019}, for instance. In the following analysis, we focus on the densest filaments of the cosmic web, which roughly corresponds to the structures studied in the low-$z$ analyses \citep[e.g.][]{Kraljic2018,Laigle2018}. All segments in a same filament (running from a saddle point to a node) are assigned a single density value, corresponding to the average of the underlying optical depth $\tau$ at each segment location. In our analysis, only the 50\% densest filaments are kept in each sample. We briefly discuss in the following how the results change when this selection is varied. 

\subsubsection{Analysis of the quality of the reconstructed web}
Following the method introduced by \cite{Sousbie2011}, the quality of the skeleton extraction was first quantified by comparing the pseudo-distance $d_{\rm skl}$ between the skeletons on the smoothed original $\tau$-field (SKL$_{\rm orig}$) and the reconstructed field (SKL$_{\rm rec}$). The PDFs of $d_{\rm skl}$ are shown in Fig.~\ref{fig11a} and the median values of the distributions are provided in Table \ref{tab:perf}. The median distance decreases with $S/N$, although the improvement after the $S/N\approx$4 case is again limited. As expected, the distances are smaller for the $L_{\rm T}=2.5$~Mpc and $R=2000$ scenario. In any case, the distributions peak at much shorter distances than the mean separation between filaments ($\langle d_{\rm sep} \rangle$ is  $\sim 6.2$ Mpc for $C_{1}$ and 4.5 Mpc for $C_{2}$). For each saddle point in  SKL$_{\rm orig}$ , the distance $d_{\rm sep}$ to its closest neighbour was measured.  

The pseudo-distance between skeletons is found to be dependent on the density, as shown in Fig.~\ref{fig11}. In this figure, pseudo-distances are plotted in bins of $\tau_{\rm orig}$, which is the optical depth in the smoothed original $\tau$-field. The 1$\sigma$ error regions were obtained by bootstrapping the distribution of distances at each density  bin. The distance is most uncertain in the lowest-density regions. The pseudo-distance between the skeletons becomes smaller with increasing $S/N$ and saturates at $\sim L_{\rm T}/2$. The pseudo-distance does not fully encode how similar two skeletons are because they can be close to each other without being aligned. Therefore we also measured the alignment between the reconstructed and the original skeletons. We measured the angle $\theta$ between each segment in SKL$_{\rm rec}$ and the closest neighbour in SKL$_{\rm orig}$. The probability distribution function of $\cos \theta$ is displayed in Fig.~\ref{figC6}. We measure an excess of probability for $\cos \theta  > 0.5$, that is, for the segments to be aligned. This signal increases with increasing $S/N$. 

The comparisons of the reconstructed $\tau$-field to the original field and the value of the pseudo-distance between the computed skeletons  suggest that the reconstruction performs already well with a constant $S/N\sim4$. A minimum $S/N\sim4$ is therefore a reasonable value to aim for in an IGM tomography survey (for the purpose of studying the cosmic web). In the following we use these reconstructions and compute skeletons in combination with the galaxy catalogue of the {\sc Horizon-AGN} simulation in order to asses the level at which the large-scale environments of galaxies can be studied at $3 \lesssim z \lesssim 3.5$.

\subsubsection{Mapping galaxies in filaments}
\label{sec:mapgal}
Galaxy mass assembly is the result of  accretion from both their surrounding and their secular evolution. The galaxy properties are therefore expected to be partly dependent on their host halo mass, and in addition to the effect of the halo, on the large-scale environment in which they are embedded, the geometry of which shapes the tidal field. The mass gradient towards filaments is therefore one of the  expected imprints of the cosmic web on galaxy properties: at low-$z$ ($z \lesssim 1$) more massive galaxies have been found closer to the geometrical centre of the filament than less massive galaxies \citep{ Malavasi2017,Kraljic2018,Laigle2018}, a measurement in agreement with  theoretical prediction \citep{Musso2018}. This trend is expected to be stronger at high redshift, when the  effect of the initial large-scale tides has not yet been perturbed by non-linear processes such as mergers or AGN feedback. At $z \gtrsim 2$, Lyman-$\alpha$ tomographic reconstruction can be used to test the importance of the large-scale environment in shaping the galaxy growth. The purpose of our present analysis is not to  measure and interpret this signal, but rather to quantify to which extent an existing trend in the original field can be measured in the noisy reconstruction. 

All  galaxies more massive than $10^{9} {\rm M}_{\odot}$ from the {\sc Horizon-AGN} catalogue in the redshift range of interest ($3<z<3.5$) were used in the analysis.  We measured their distances to the closest filaments ($d_{\rm fil}$) and nodes ($d_{\rm nod}$) for skeletons computed on all configurations. Ideally, the distances from the reconstructed fields should within a certain scatter correspond to those measured from the original field. In practice, however, some of the information is lost when  realistic noisy spectra are used. Because the filament positions are  scattered in the reconstructed field, galaxies tend to be less clustered around them. 
This results in larger distances on average of galaxies to the closest filaments with respect to the original field. This is shown in Figs. \ref{figC1} and \ref{figC2} (top panels) and quantified with the low correlation coefficient in Table \ref{tab:perf}. This has a noticeable effect on the measured environmental trends. 

\begin{figure*}[!t]
\begin{center}
\includegraphics[width=\textwidth]{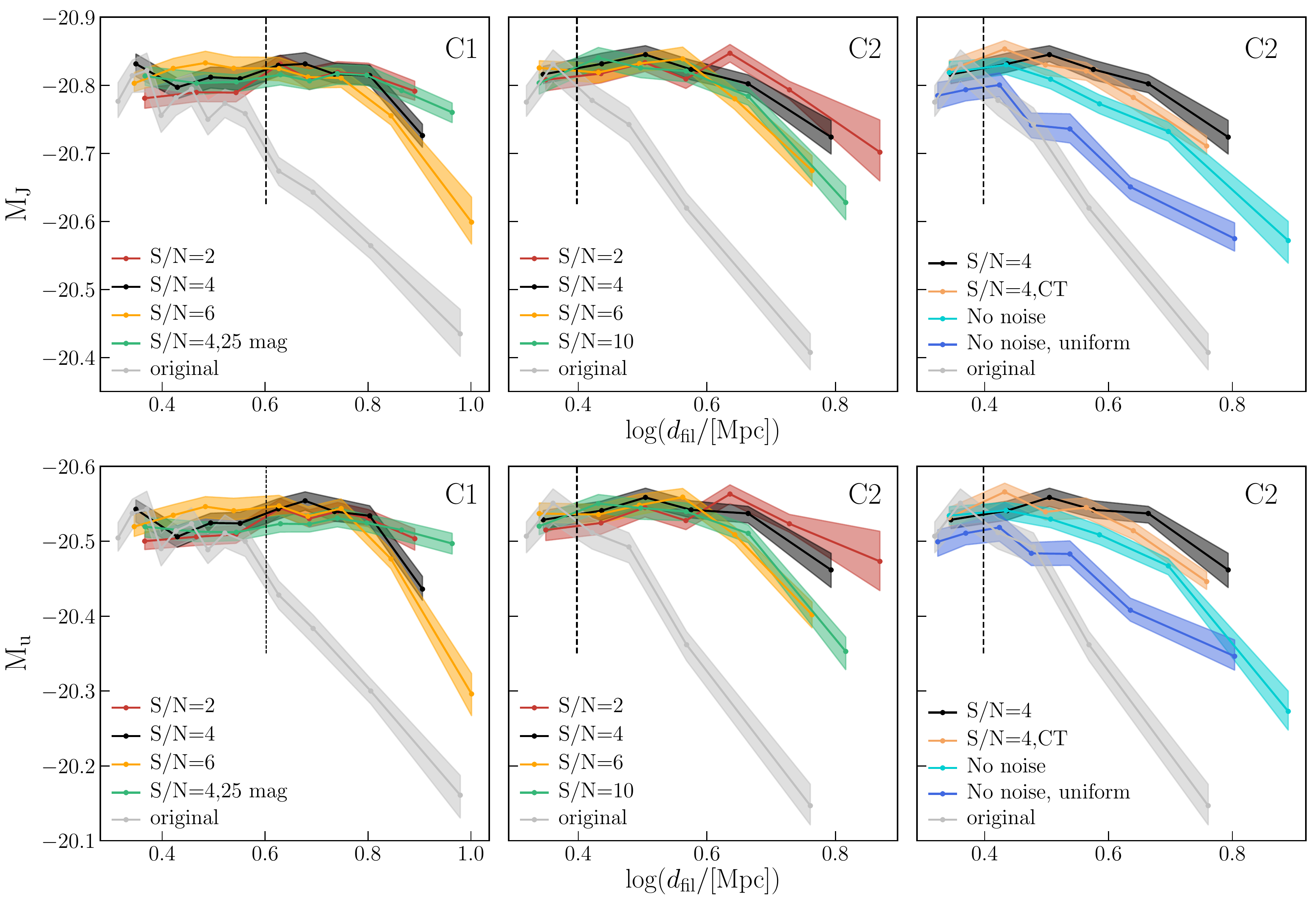}
\caption{Distributions of the mean $M_{\mathrm{J}}$ (top) and $M_{\rm u}$  (bottom) galaxy rest-frame magnitudes as a function of the distance to the filaments ($d_{\mathrm{fil}}$) for galaxies with $\log M_{\star}/{\rm M}_{\odot} > 9.0$. Results are shown for the C1 (left column) and the C2 (middle column) configurations. The right column shows the additional flavours of the C2 configuration (see Section~\ref{configuration} and the caption of Fig.~\ref{fig10} for details). The vertical dashed line indicates the reconstruction scales $L_{\rm T}$ of 4 and 2.5 Mpc for the C1 and C2 configurations, respectively. Below this scale, no trend is expected towards the reconstructed filaments. }
\label{fig12}
\end{center}
\end{figure*}

We then studied the environmental dependency of four galaxy properties: rest-frame ${\it M}_{\rm u}$ and ${\it M}_{\rm J}$ magnitudes, stellar mass M$_{\star}$, and star formation rate (SFR). ${\it M}_{\rm u}$ traces young star populations and can therefore be a proxy for the SFR, while ${\it M}_{\rm J}$ is a better proxy for the total stellar mass. We indeed show that using ${\it M}_{\rm u}$ and ${\it M}_{\rm J}$  yields a similar environmental trend as the SFR and M$_{\star}$, but is easier to measure (see Section \ref{synergy}). We  present only the results of the analysis of magnitudes in this section and show similar plots for M$_{\star}$ and SFR in Fig.~\ref{figC3} in Appendix.

\begin{figure*}
\begin{center}
\includegraphics[scale=0.58]{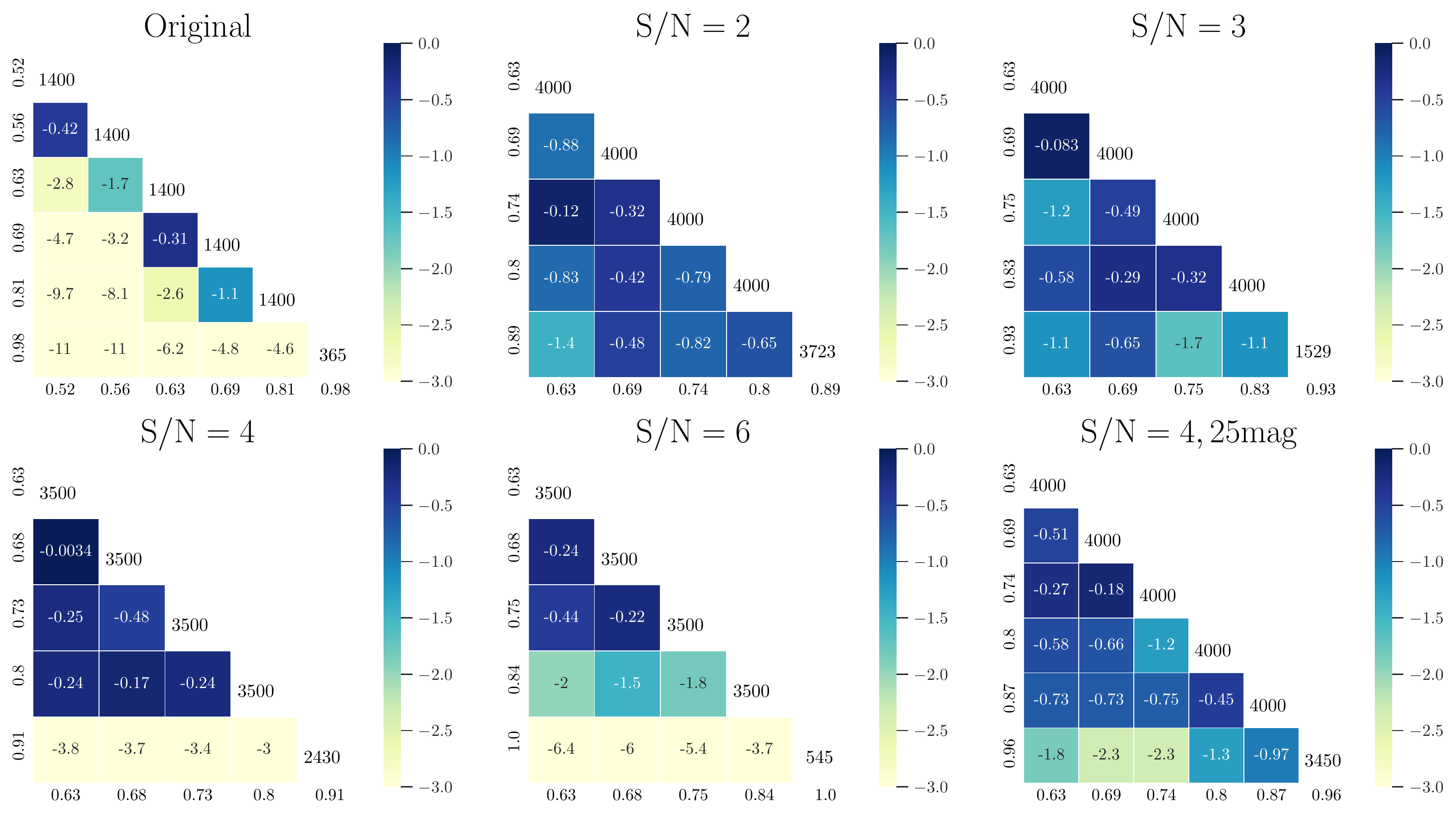}
\caption{Measuring the significance of galaxy {\it M}$_{\rm J}$ magnitude gradient towards filaments. A KS test is performed in order to asses the difference between galaxy mass distributions in different (logarithmic) $d_{\rm fil}$ bins. Colours indicate the logarithm of the $p$-value, i.e. the probability that the two compared samples are drawn from the same distribution (the actual value is also indicated in each square). Numbers on the diagonal indicate the number of galaxies in each corresponding bin. Plots are shown for the C1 configuration and for different $S/N$. The KS test is performed only for  $d_{\rm fil} > 4$ Mpc  (except for the original field case).}
\label{fig12a}
\end{center}
\end{figure*}

\begin{figure*}
\begin{center}
\includegraphics[scale=0.58]{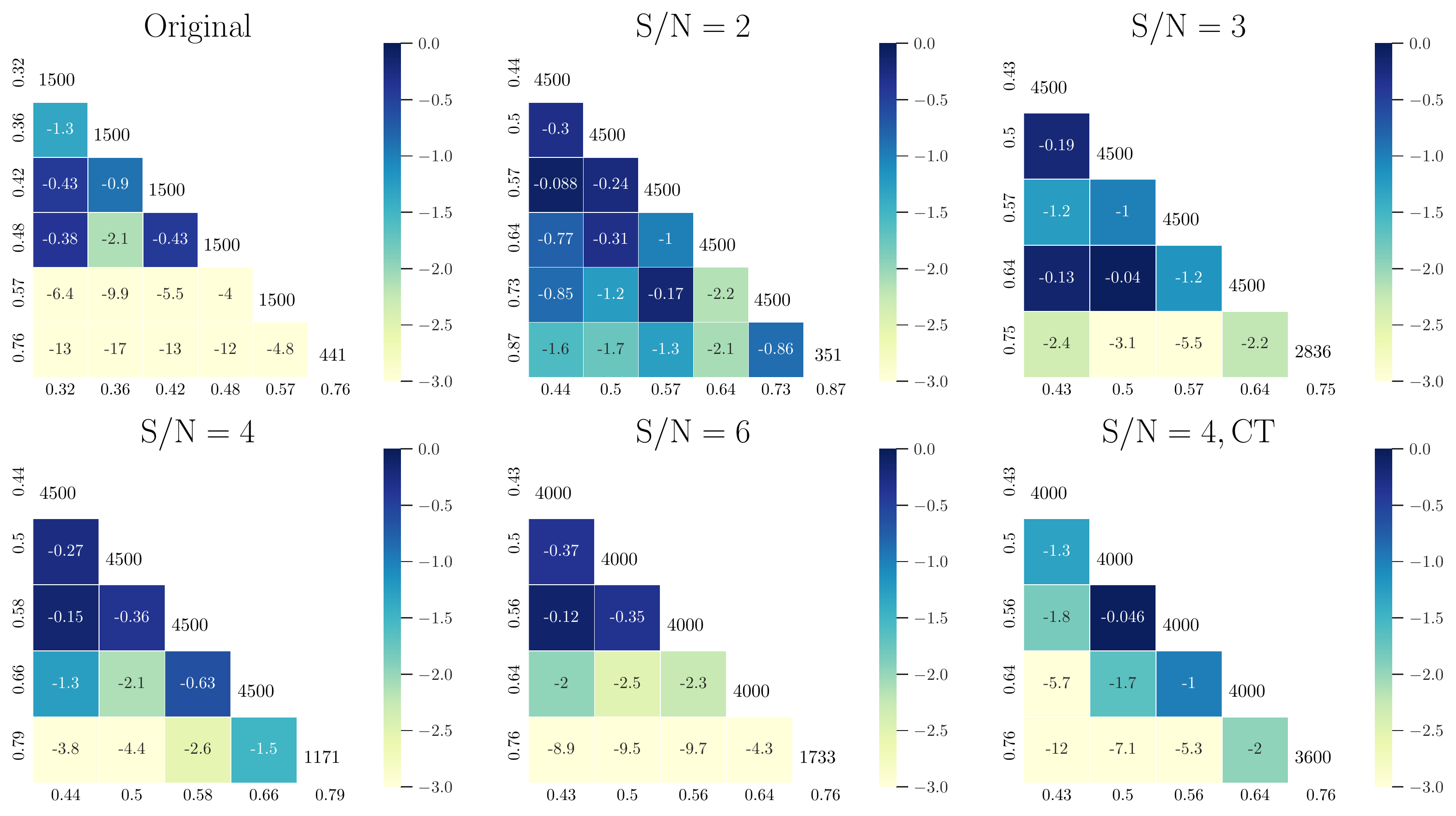}
\caption{Same as Fig. \ref{fig12a}, but for the C2 configuration. The KS test is performed only for  $d_{\rm fil} > 2.5$~Mpc (except for the original field case).}
\label{fig12b}
\end{center}
\end{figure*}

Galaxies were binned by their measured distance to the closest filament (keeping only the 50\% densest filaments), and the distributions of their properties in each bin were measured. The mean values of ${\it M}_{\rm u}$ and ${\it M}_{\rm J}$ in each bin and $\pm 1\sigma$ dispersion are plotted in Fig.~\ref{fig12}. In order to probe an effect that is only induced by filaments, galaxies closer than 2~Mpc from a node were removed from the analysis. Several features are readily noticeable. In the original field,  ${\it M}_{\rm J}$ and ${\it M}_{\rm u}$ increase towards filaments because higher mass and a higher SFR is expected closer to filaments. In this measurement, we did not distinguish the density effect from the proximity effect to filaments. Because the filament positions are scattered in the reconstructed field, the gradient strength decreases with increasing noise. With lower resolution (and therefore larger smoothing of the  field), fewer filaments are present in the density field, and therefore the slope of the gradients is less pronounced because smoothing tends to mix environments. The gradient is indeed more pronounced when the $S/N$ is increased, but the real limitation of detecting the gradient is the clustering of sources and therefore the non-uniform distribution of the LOS, as illustrated in the right plots of Fig.~\ref{fig12} and as we described in the mere analysis of the density field (see Fig.~\ref{fig10}).

In Figure~\ref{figC3} in the appendix we show what happens when all filaments (not only the 50\% densest) for the original field are kept in the sample (dotted black line). The trend is slightly weaker but similar, although the distance range probed by galaxies is slightly smaller (because more filaments are included in the skeleton). It is notable that the gradients tend to completely disappear for the reconstructed fields in this case (not shown). This is a consequence of both the fact that gradients towards low-density filaments are intrinsically weaker, and that the low-density filaments are not very well recovered in the reconstructed fields (as shown in Fig.~\ref{fig11}).

To quantify the required $S/N$  for a statistically significant detection of this environmental trend, we performed the following analysis for each $S/N$ case. Firstly, to ensure that we had a similar number of galaxies in each $d_{\rm fil}$ bin, we binned the galaxies into logarithmic bins. The width of the bins varied according to a predefined number of galaxies per bin (typically several thousand, see Figs. \ref{fig12a} and \ref{fig12b}). A two-sample Kolmogorov-Smirnov (KS) test was then performed between the distributions in all the bins. The test was only performed for $d_{\rm fil}$ that are larger than the transverse reconstruction scale $L_{\rm T}$. The resulting {\it p}-values are shown in Figs~\ref{fig12a} and~\ref{fig12b}. The distribution of a given galaxy property (here only $M_{\rm J}$ is shown, but the results are similar for $M_{\rm u}$) was considered to be significantly different within two  bins when the associated p-value was $<0.05$. A dependency of the galaxy property on the distance to filaments was measured when at least two of four density bins presented significantly different distributions from the other ones (i.e. a p-value $<0.05$).

\begin{figure*}
\begin{center}
\includegraphics[width=\textwidth]{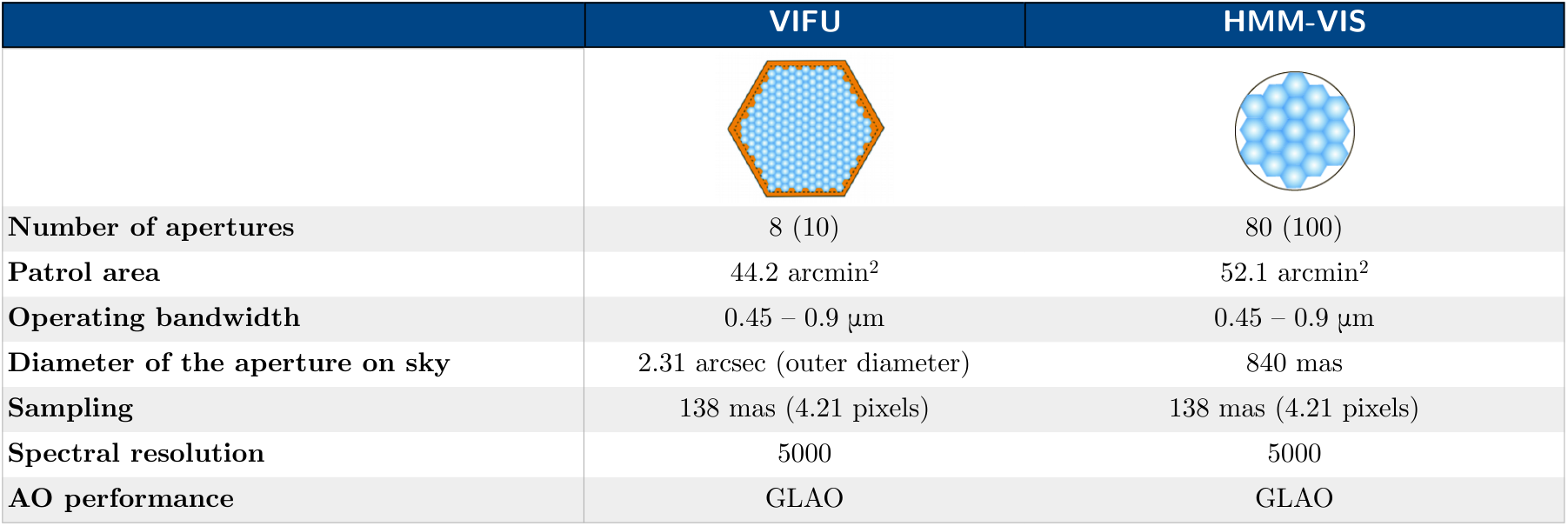}
\caption{Summary of the basic characteristics of the VIFU and HMM-VIS modes according to the Phase A design (based on \citealt{Jagourel2018}). The number of apertures reported in brackets is the goal that is not the baseline of the Phase A design, however. For the purpose of this paper we consider only the low-resolution $R=5000$ modes.}
\label{fig0}
\end{center}
\end{figure*}

Our analysis demonstrates that $S/N \sim$~4 is sufficient to detect gradients of $M_{\rm J}$ towards filaments in the C2 configuration, but the $S/N$ might need to be pushed to~6 in C1. This value is in line with the analysis of the flux contrast field reconstruction and the computation of the skeleton. We therefore conclude that there will be not much gain by increasing the $S/N$ above this threshold for the purpose of studying the relation between galaxy properties and their large-scale environments. This is the lowest $S/N$ we need for our spectra. In reality, the $S/N$ of all but the faintest target galaxies in a survey will therefore be higher. For example, if the galaxies were all observed with a fixed exposure time, which is necessary to reach $S/N=4$ for the faintest sources, then the reconstruction of such a sample would give better results and gradients would be detected with higher significance (compare the results of the KS test in Fig.\ref{fig12b} for SN=4 and SN=4, CT). The limiting factor then is the finite number of sight lines and their non-uniform spatial distribution.

A detailed analysis of the mass or SFR environmental dependency at $z \sim 3$ is beyond the scope of this paper. The connection between the filaments and other properties, such as AGN activity and redshift evolution, will also be discussed in a future work.

\section{Simulations of MOSAIC observations}
\label{mosaic}

In the previous section we have studied the spectral resolution and S/N that are required to successfully reconstruct the IGM. In this section we study a performance of the MOSAIC instrument on the ELT in detail, which will allow us to make a connection between the resolution $R$, $S/N,$ and the actual observing time.

MOSAIC is the proposed multi-object spectrograph for the Extremely Large Telescope \citep{Morris2018}. This fiber-based spectrograph will have several observing modes to tackle different science cases \citep{Evans2015,Puech2018,Jagourel2018}. The modes that will be used for observations in the blue and visible, the range of interest of this study, are the high multiplex mode (HMM-VIS mode) and the visible integral field unit mode (VIFU mode). Each unit of HMM-VIS and VIFU mode will have an aperture with a diameter of 840 mas and 2.3$\arcsec$, respectively. The size of the patrol field of view for VIFU (HMM-VIS) will be $\sim 44 ~(52)$ arcmin$^{2}$. The current conceptual design baseline predicts at least 80 (8) units for the HMM-VIS (VIFU) mode, which means that about $\sim 40-60$ (8) objects can be observed simultaneously. The maximum number of observed objects in HMM-VIS depends on the number of fibers that are dedicated to the observations of sky. The blue spectral band will cover a wavelength range of $\lambda = 450 - 600$ nm with a spectral resolution of $R = 5000$. The PSF on the detector will be oversampled with 4.21 pixels per spatial and spectral element. The summary of basic characteristics of the VIFU and HMM-VIS modes is provided in Fig. \ref{fig0}. We note that the current specifications of the MOSAIC instrument will undoubtedly change to some extent. This will affect the estimates of observation time of a survey in Sect. \ref{discuss}. The analysis of this paper is performed and presented in such a way that the envisioned possible changes can be used together with the presented results to easily obtain new estimates of the observation times.

In order to understand the performance of the instrument at blue wavelengths, we performed a set of realistic simulations (a preliminary overview of the performance of the MOSAIC in the near-infrared is presented in \citet{Disseau2014}). Simulations were made with the WEBSIM-COMPASS simulator that is dedicated to ELT simulations \citep{Puech2010,Puech2016}. The astrophysical target in these simulations is first modelled as a high-resolution data cube, where the spatial dimensions sample the light distribution at the resolution of the telescope ($\sim \lambda/2D$, where $D$ is the diameter of the telescope). This data cube is then convolved with the PSF that is representative of the optical path through the telescope and the atmosphere, including the adaptive optics (AO) system. Realistic sky background, photon noise, and detector noise are added. The pipeline produces simulations in FITS format that mimic the result of a data reduction pipeline with perfectly extracted and reduced data (although the actual background subtraction and extraction of spectra can be carried out by hand from "raw" frames). In the following we provide the detailed input of our simulations.

\subsection{Input considerations}
\label{inputs}

\subsubsection{Instrumental and atmospheric parameters}
\label{atmospheric}

\begin{figure}
\begin{center}
\includegraphics[scale=0.55]{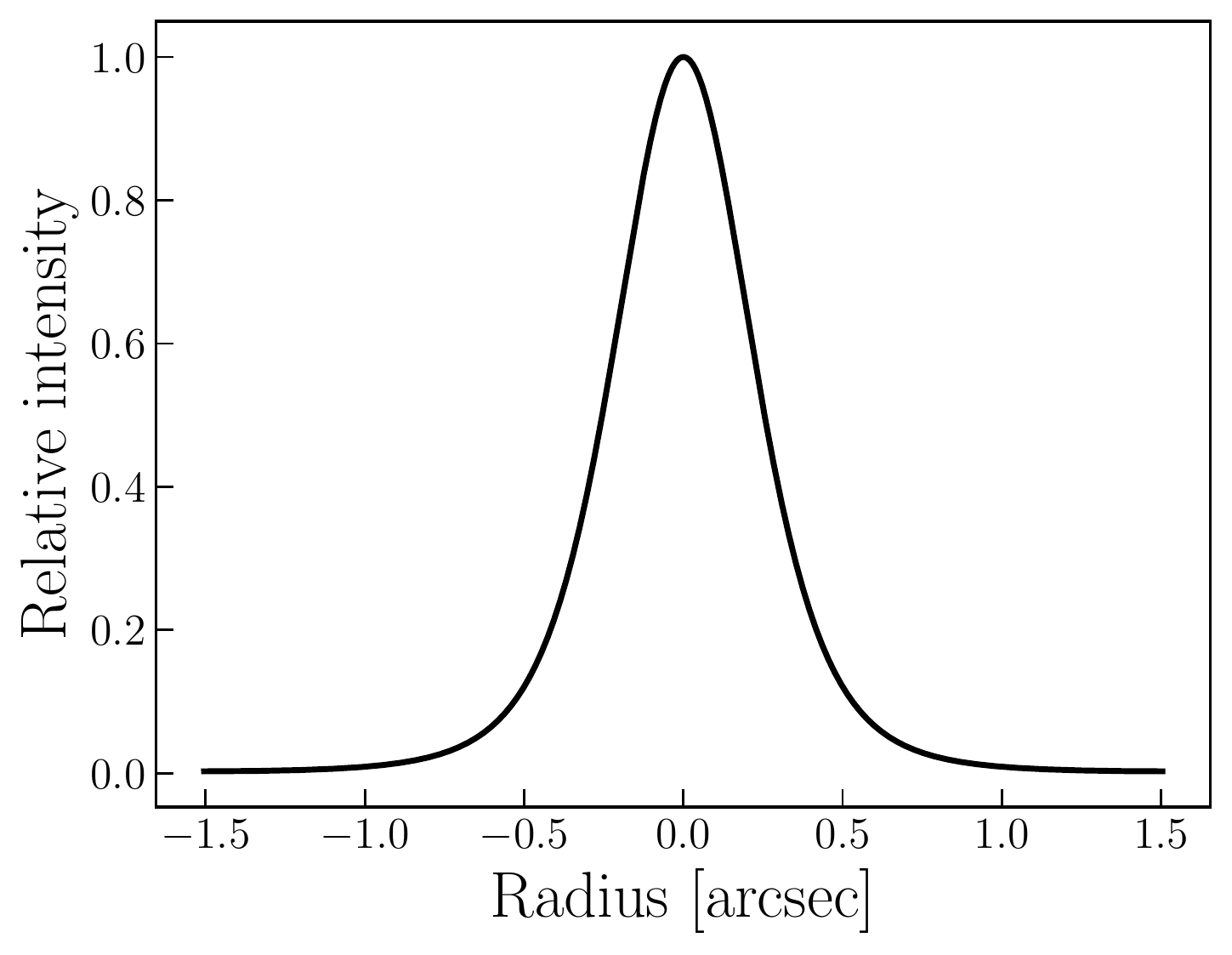}
\caption{Transverse cut of the simulated MOSAIC GLAO PSF at $\lambda = 400 ~\mathrm{nm}$, calculated for an input seeing of 0.7 arcsec.}
\label{fig01}
\end{center}
\end{figure}

We assumed realistic CCD parameters (dark current, read-out noise, charge transfer efficiency, and quantum efficiency) and telescope parameters (pupil diameter, effective central obscuration, typical temperature, and emissivity in the blue). We adopted the calculated total throughput of the atmosphere, telescope, and the instrument that were calculated in the throughput analysis. We assume a conservative total throughput of $T = 13\%$ throughout this work: we did not take the wavelength dependency of the throughput  into account (the assumed value was estimated at the blue spectral edge), but we investigated the dependence between $S/N$ and throughput. The spectral resolving power was $R = 5000$. The default spectral and spatial sampling will be 4.21; in order to be conservative, we rounded the sampling to 5 (e.g. 5 pixels per element of spectral or spatial resolution). This configuration results in a significant read-out noise and is hardly suitable for observations of (very) faint sources. Simulations at different sampling rates (Appendix \ref{noiseregime}) led us to finally assume a (on-the-ccd) binning that resulted in an effective spatial sampling of $S_{\rm s} = 1$ and a spectral sampling of $S_{\lambda} = 2$. The PSF is the ground-layer adaptive optics (GLAO) PSF calculated at $\lambda = 400$ nm: the PSF has an ensquared energy of $5.4 \%$, calculated in an aperture of 150 mas, and a Strehl ratio of 0.0017$\%$. The transverse cut of the PSF is shown in Fig. \ref{fig01}. Because MOSAIC will operate at $\lambda > 450$ nm and the quality of the AO correction increases towards longer wavelengths, this is a conservative choice. The expected performance at the blue wavelengths ($<6000$ $\mathrm{\AA}$) is nearly equivalent to seeing-limited.

All simulations were carried out under dark-sky conditions, at airmass of 1.15 and seeing at zenith of 0.7$\arcsec$. An observation was separated into $N$ one-hour exposures (in the following, the number of integration times, or NDIT), which is the current baseline for operations at the VLT. We neglected the effect of overheads so that in the following, all observing times are to be interpreted as integration times. Sky emission and transmission were adopted assuming the parametrization of \citet{Noll2012} and \citet{Jones2013} as implemented into the Paranal sky advanced model\footnote{https://www.eso.org/observing/etc/doc/skycalc/helpskycalc.html}. The background model accounts for airglow and residual continuum contributions as well as molecular emission and absorption lines.

The apertures of the two modes are shown in Fig. \ref{fig0}. In the case of the simulation of the VIFU mode, we assumed the aperture to be a square with a side of 2$\arcsec$, instead of the hexagonal shape. In case of the HMM-VIS mode, we performed the simulations in the so-called "simple aperture" mode \citep{Puech2016}. The full aperture of a fiber is assumed to be fragmented into 19 hexagonal-shaped microlenses. While the HMM-VIS field has been segmented into 19 fibers due to technical constraints (i.e. the plate scale on the ELT), this should not be considered as a small IFU unit, however.

\subsubsection{Physical input}

In Sect.~\ref{horizon} we established that we need to observe galaxies as faint as $m_{\rm UV,rest} = 25.5$ mag. This choice allows us to have a sufficient number of targets per FOV (see Section \ref{discuss} for a detailed discussion) and to reach a sufficient transverse tomographic map resolution scale at redshifts $\sim 3 - 3.5$. We note that, in our calculations, we applied basic corrections to take into account the fact that galaxies are at different redshifts and therefore observations with a certain filter probe somewhat different rest-frame spectral ranges of these galaxies. Nevertheless, to avoid confusion, we use a neutral $m_{\rm UV,rest}$ nomenclature corresponding to $\lambda_{\rm rest} = 150$ nm.

Because galaxies are not point sources, we adopted the relation between the apparent size of a galaxy (characterized by its half-light radius $r_{\rm half}$), its brightness, and redshift. These three quantities are required as input of the simulation. We determined the mean relation between the three properties in the following way. We calculated the typical size ($r_{\rm half}$) of a galaxy of luminosity $L$ using the relation from the study of \citet{Shibuya2015}, which  covers the redshift range and galaxy brightness considered in our work:

\begin{equation}
\label{size}
r_{\rm half} = 6.9\left( 1 + z\right)^{-1.2} \left( \frac{L}{L^{\star}}\right)^{0.27} ~\mathrm{kpc,}
\end{equation}

\noindent where $L^{\star} = L^{\star}({\rm z=3})$ is a characteristic luminosity in the Schechter function \citep{Schechter1976} for the star-forming galaxy population at $z \sim 3$.  \citet{Shibuya2015} did not find a strong dependence between the size of the galaxy and the rest-frame wavelength at which the size is measured. We therefore assumed that the sizes all correspond to the rest-frame UV band corresponding to the wavelength band in which $m_{\rm lim}$ is measured. The luminosity as a function of apparent magnitude was furthermore calculated as (neglecting the second term in the colour correction)

\begin{equation}
L(m_{\rm lim},z) = L_{0}10^{0.4\left[ M^{\star} - m_{\rm lim} + 5\log d_{\rm L}(z) + 25 - 2.5\log \left(1 + z\right)\right]},
\end{equation}

\noindent where $M^{\star}$ corresponds to $L^{\star}$ and $d_{\rm L}$ is the luminosity distance. To be consistent with the work of \citet{Shibuya2015}, we assumed $M^{\star} = -21$ mag, which may be somewhat bright compared to the values in\,\citet{Bouwens2015}, for instance, but the overall effect does not have strong implications on our conclusions (i.e. using the value of $M^{\star} = -20.8$ mag \citep{Bouwens2015} gives $\sim 5 \%$ larger radii). The resulting predictions are given in Table \ref{tab1}. The diameter of a galaxy was assumed to be equal to 4$r_{\rm half}$ in the simulation. Because the apparent size of a galaxy of a certain brightness does not change appreciably in the considered redshift range, we ran all our simulations assuming $z = 3.3$ because in the first part of our work (Sect. \ref{horizon}), we performed reconstruction in the $z = 3 - 3.5$ range.

\begin{table}
\centering
\small
\renewcommand{\arraystretch}{1.5}
\begin{tabular}{lcccccc}
\hline
\hline
\vspace{-0.2cm}
\multirow{2}{*}{$m_{\rm lim}$} & \multicolumn{6}{c}{Redshift}\\
& 2.5 & 2.8 & 3.0 & 3.3 & 3.5 & 3.7\\
\hline
23   & 0.25 & 0.25 & 0.24 & 0.24 & 0.24 & 0.24\\
24   & 0.20 & 0.19 & 0.19 & 0.19 & 0.19 & 0.18\\
24.5 & 0.17 & 0.17 & 0.17 & 0.17 & 0.17 & 0.16\\
25   & 0.15 & 0.15 & 0.15 & 0.15 & 0.15 & 0.14\\
25.5 & 0.14 & 0.13 & 0.13 & 0.13 & 0.13 & 0.12\\
\hline
\end{tabular}
\caption{Predicted typical half-light radius (in arcseconds) of a galaxy of brightness $m_{\rm lim}$ lying at redshift $z$.}
\label{tab1}
\end{table}

Galaxy morphology templates, representing different galaxy types, were taken from the simulator's library \citep[see][]{Puech2010}. We used nine templates obtained either from real observations \citep{Fuentes2004,Garrido2002,Garrido2004} or simulations \citep{Cox2006,Bournaud2008}. The sample of galaxies represents diverse morpho-kinematic properties, including galaxy mergers and clumpy discs. We assumed that the galaxy itself has no internal kinematics, that is, except for the absolute scale, each part of the galaxy has exactly the same spectrum. We only provide the simulator with flat spectra (see Sec. \ref{additional}), therefore this assumption does not affect our conclusions.

\begin{figure}
\begin{center}
\includegraphics[scale=0.57]{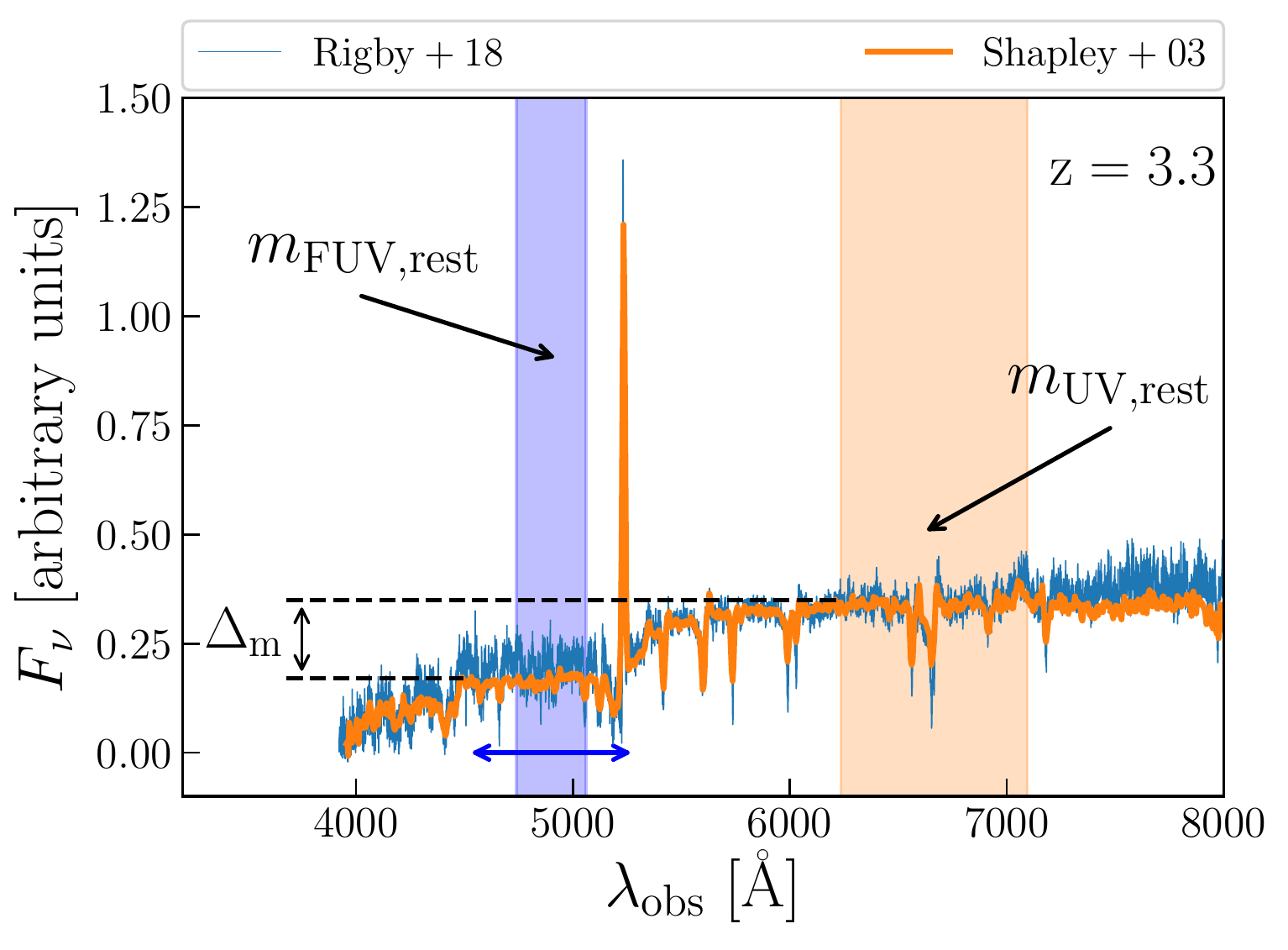}
\caption{Illustration of the $\Delta_{\rm m}$ correction (see Section \ref{additional} for details) for a galaxy at $z = 3.3$. For a galaxy of certain brightness $m_{\rm UV,rest}$, we use this spectrum \citep{Shapley2003} to estimate the expected brightness in the far UV part of the spectrum. The blue and orange shaded regions indicate spectral ranges for which $m_{\rm FUV,rest}$ and $m_{\rm UV,rest}$ are measured. The blue arrow indicates the spectral range that is simulated in the simulation when $m_{\rm FUV,rest}$ is estimated. For comparison, we also show the stacked spectrum of intermediate resolution by \citet{Rigby18}.}
\label{fig1a}
\end{center}
\end{figure}

\subsubsection{Simulation method}
\label{additional}

The main purpose of the simulation is to understand the performance of the instrument at the operating (blue) wavelengths. For example, we wish to understand how the $S/N$ depends on the brightness of the target. For most purposes, our spectral input was assumed to be a simple flat spectrum, for which the $S/N$ can be easily measured. In the following, the $S/N$ therefore refers to the $S/N$ in the case of an ideal spectrum, for example, an intrinsic spectrum without additional absorption from intervening IGM absorbing systems (or equivalently, an average $S/N$ over the absorbed region of a given brightness). This approach is assumed because the $S/N$ in the strongly absorbed \lya\, forest region is difficult to quantify.

\begin{figure*}
\begin{center}
\begin{tabular}{cc}
\includegraphics[scale=0.55]{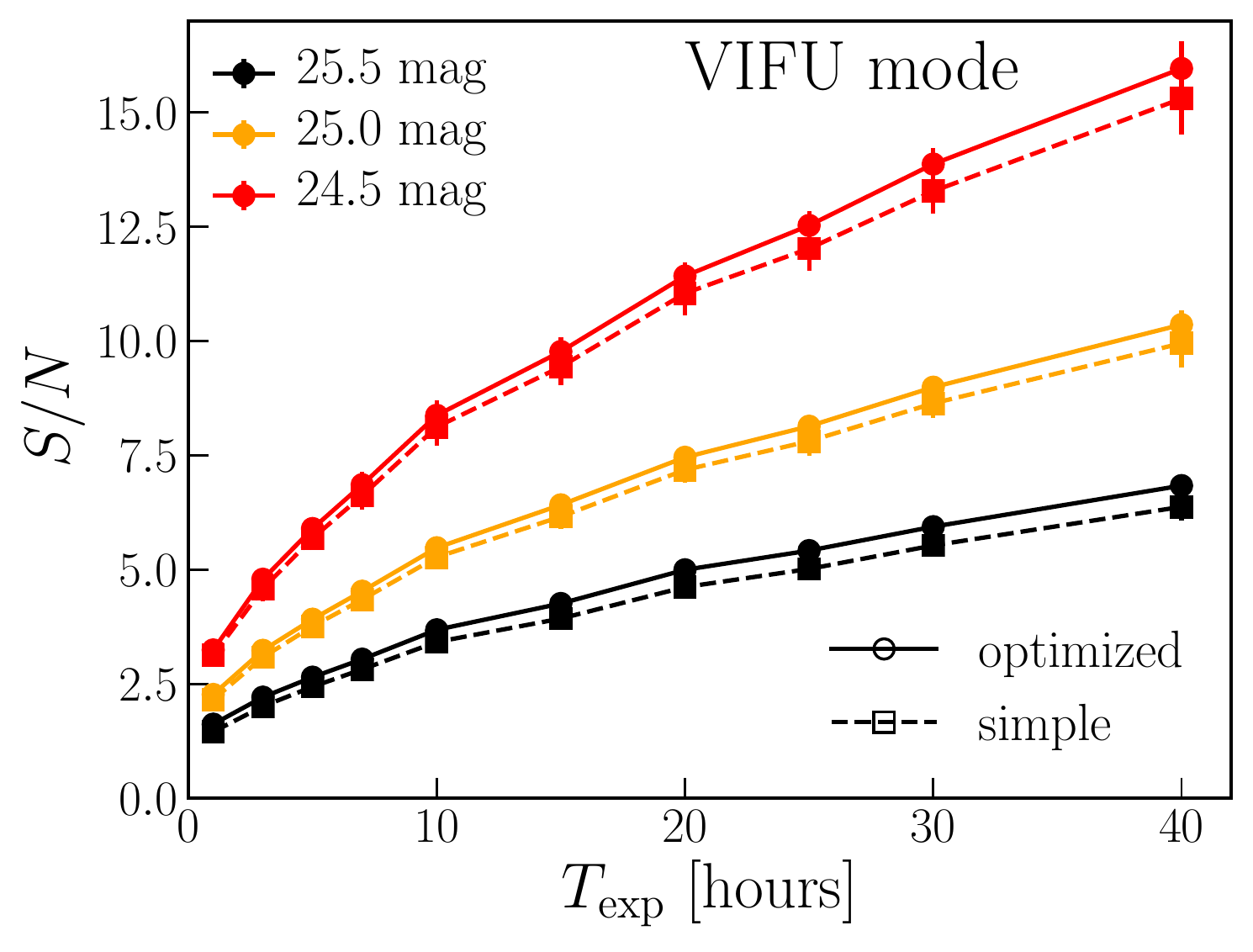}&
\includegraphics[scale=0.55]{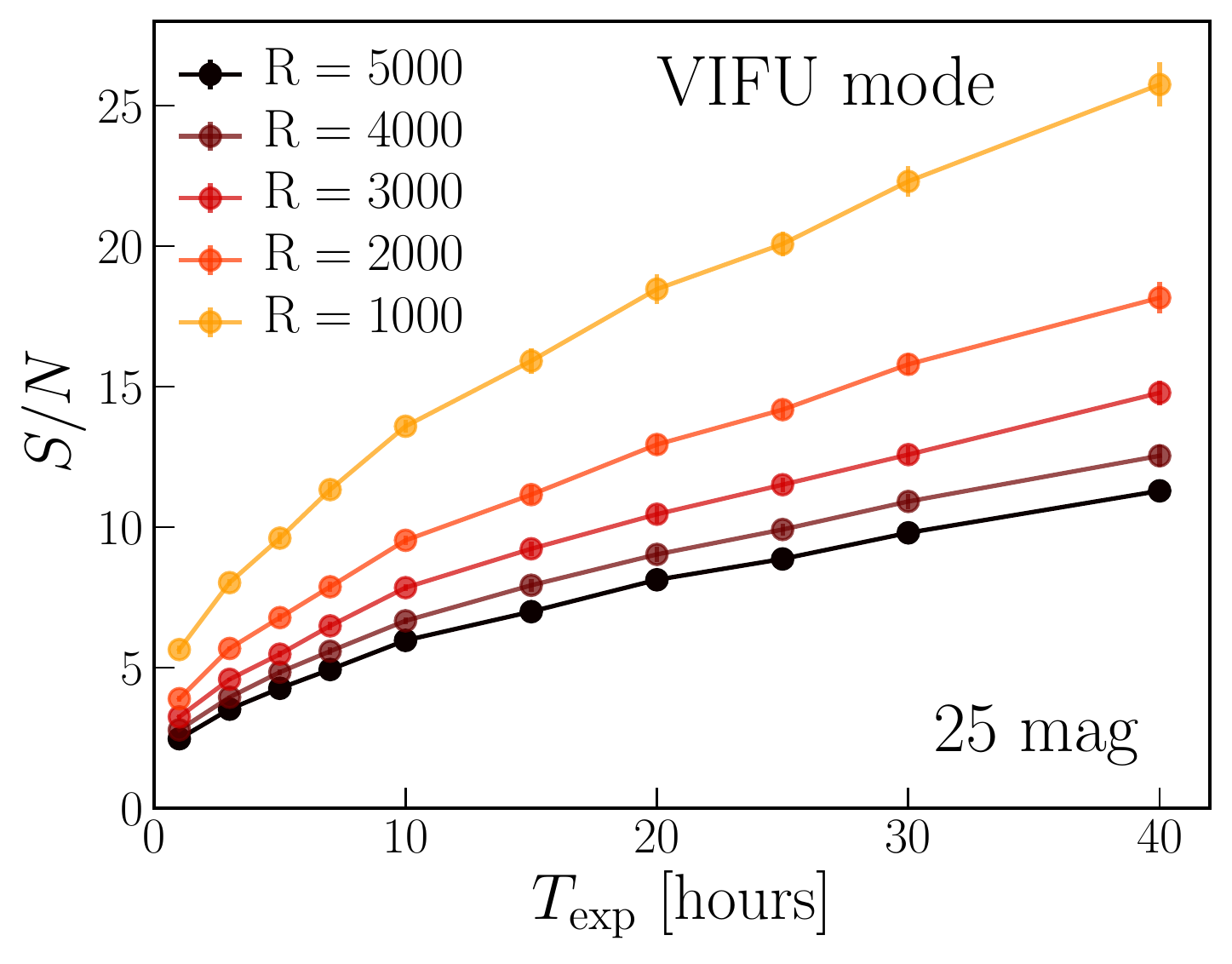}
\end{tabular}
\caption{Left: Signal-to-noise ratio per resolution element as a function of exposure time for galaxies of three different brightnesses at $z=3.3$. $S/N$ is estimated from integrated spectra obtained with two different methods; see the text for details. Right: Illustration of the change in $S/N$ if the original $R=5000$ spectrum (for a $m_{\rm UV,rest} = 25$ mag) is smoothed down to a lower resolution.}
\label{fig2}
\end{center}
\end{figure*}

A spectral template is needed to understand the connection between the rest-frame UV magnitude of a galaxy (all magnitudes in this report are quoted as rest-frame UV at $\lambda \sim 1500~\mathrm{\AA}$) and the brightness in the far UV region, for example, in the \lya\, forest region. We used the average spectral template of an LBG population at $z \sim 3$ by \citet{Shapley2003} to estimate the correction from $m_{\rm UV,rest}$ to $m_{\rm FUV,rest}$, the latter being the input of the simulation. This is illustrated in Fig.~\ref{fig1a}. For the assumed \citet{Shapley2003} template, the difference in the rest UV and far-UV (FUV) magnitude is $\Delta_{\rm m} \sim 0.7$. In Fig.~\ref{fig2} the $S/N$ plots should be understood with this correction in mind: an $S/N$ result for a quoted $m_{\rm UV,rest}=25$ mag is calculated for a source of a magnitude $m_{\rm FUV,rest} = m_{\rm UV,rest} + \Delta_{\rm m} = 25.7$ mag. The spectral template used by \citet{Shapley2003} includes the averaged contribution from the Ly$\alpha$ forest. Our final $S/N$ scaling relations (Section \ref{scaling}) do not lose any generality because of this as they can be easily modified for a different choice of magnitude correction (i.e. galaxy spectrum).

In summary, for each simulation, we chose a galaxy morphology template, galaxy redshift, and $m_{\rm UV,rest}$ (which in turn determine its apparent size in the sky according to Eq. \ref{size}), applied the $\Delta_{\rm m}$ correction, and ran the simulation for a given set of instrumental and observational parameters (e.g. exposure time). The simulated data were analysed as discussed in the following, and the $S/N$ was measured for each simulation.

\subsection{VIFU-mode performance}

The advantage of an IFU-mode observation is that the sky background can be estimated directly from the IFU in which the object of interest lies. The size of an individual IFU is projected to be large enough that the majority of the galaxies at $z \sim 3$ will be significantly smaller than the size of the IFU: the size ($\sim 4r_{\rm half}$) of the brightest galaxies at $z \sim 3$ is $\sim 1\arcsec$ (Table \ref{tab1}), while the outer diameter of the VIFU aperture is $2.3\arcsec$. The outer spaxels of the IFU are therefore not contaminated by the source. The median of the signal from these spaxels is combined into a sky spectrum that is then subtracted from all the spaxels of an IFU.

\subsubsection{Estimation of the S/N}
\label{snsn}

We estimated the S/N as a function of exposure time by assuming a flat spectrum as an input. The mean value was subtracted from the spectrum to obtain the RMS, and this was used to calculate the $S/N$. All $S/N$ values in the following are reported as $S/N$ per spectral resolution element. First, we checked how different extraction methods affect the quality of the extracted spectra. We considered two extraction methods. The first method relies on the size of the object. By collapsing a data cube, we checked that the size of a galaxy in the simulated data was the same as the input size, that is, its diameter is $\sim 4r_{\rm half}$. We therefore considered all the spaxels lying within a circle of $r = 2r_{\rm half}$ whose center was determined by a simple Sersic profile fit to a collapsed spectrum (i.e. galaxy image). All these spaxels were summed to create an integrated galaxy spectrum.

This method is fairly naive because it does not take the intrinsic surface brightness distribution of a galaxy into account, and by doing so, we added much noise to the integrated spectrum. An alternative approach is to optimize the extraction in the following way \citep{RosalesOrtega2012}: First, the $S/N$ is calculated for each individual spaxel of a data cube. Spaxels are then ordered according to their $S/N$ in a decremental order. Then we examine the changes in the $S/N$  when we gradually add increasingly more spaxels together (starting from the spaxel with the highest $S/N$). The $S/N$ curve calculated in this way first increases, reaches its peak, and then drops as increasingly more noisy spaxels are added to the extracted spectrum. In this way, we can determine which spaxels should be combined in order to reach an optimal $S/N$. 

\subsubsection{Effect of exposure time and spectral resolution}
\label{scaling}

The S/N that is estimated with the two methods as a function of exposure time is shown in Fig. \ref{fig2} (left). Simulations were run for nine different morphological templates and for galaxies of three different magnitudes. For all cases, the assumed redshift was $z = 3.3$. The error bars in the $S/N$ show the dispersion of the results of different models. The dispersion increases with exposure time. The optimized extraction results in a slightly improved quality of the spectra. The improvement is not larger because the AO correction in the blue is significantly worse than in the near-infrared (NIR) and is similar to seeing-limited observations. Given the typical sizes of galaxies at considered redshifts (Table \ref{tab1}), we therefore cannot resolve individual morphological features, and in the optimized case, we more or less summed the signal from the same spaxels as if we simply considered the galaxy size (simple method). For comparison, the situation would be completely different when the quality of the AO correction in the blue were the same as in H band: in this case, the $S/N$ of the spectra obtained from the optimized method would improve by $\sim 25 \%$ with respect to the spectra from simple extraction. In the following, we only use the results from the optimal extraction method.

\begin{figure*}
\begin{center}
\includegraphics[scale=0.45]{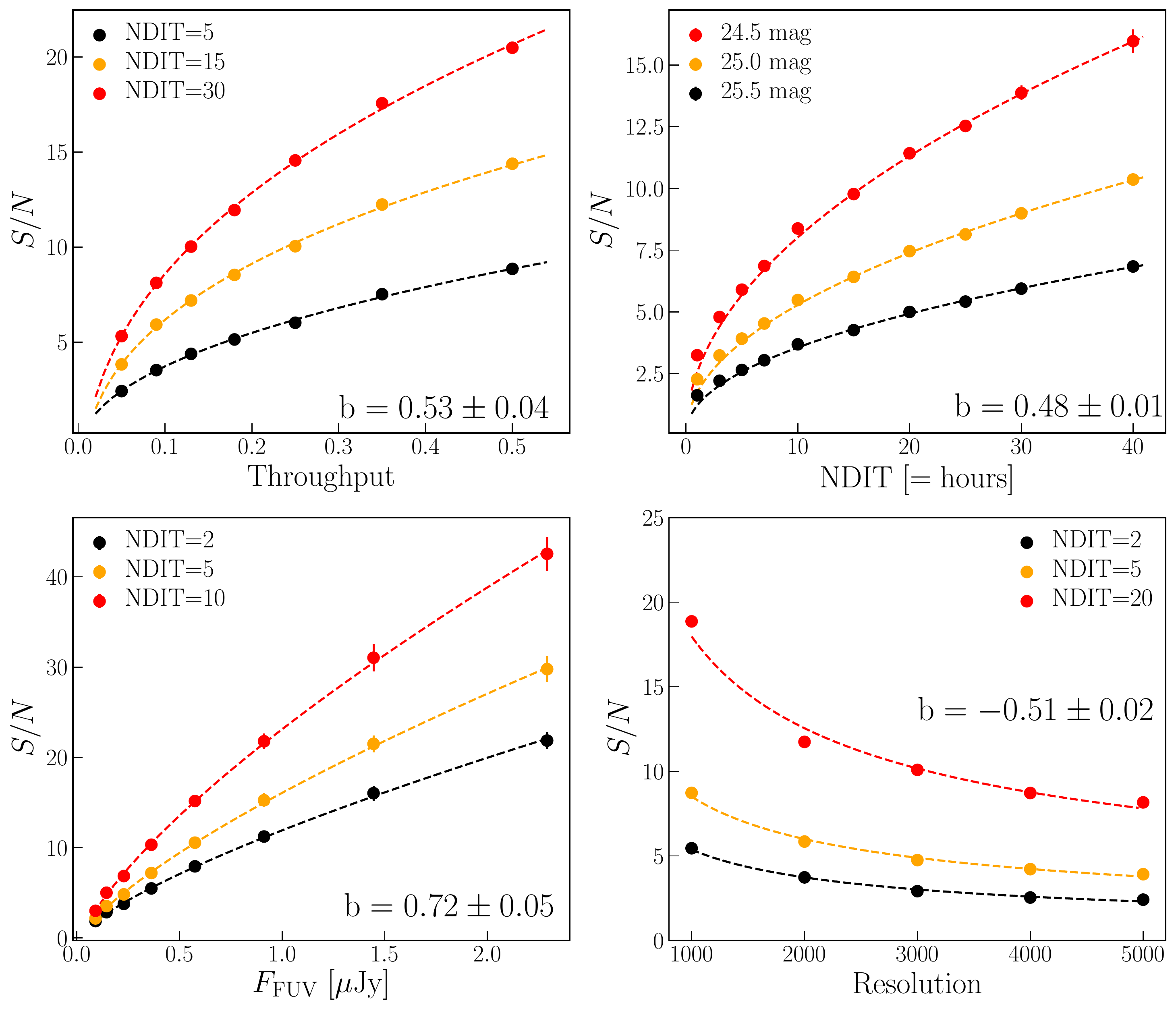}
\caption{Representative plots to show how the $S/N$ for the VIFU mode scales with the number of exposures (exposure time), resolution, total throughput, and brightness of the source. Relationships were obtained by fitting a power-law function $\left( x/x_{0}\right)^{b}$ to each $S/N$ curve, where $x$ represents one of the four quantities, $x_{0}$ is the  normalization (see text and Equation \ref{scalingIGM}), and $b$ is a power-law index. An equivalent plot for the HMM-VIS mode is shown in the appendix (Fig. \ref{figA0}).}
\label{fig3}
\end{center}
\end{figure*}

The effect of the spectral resolution on the $S/N$ is shown in Fig. \ref{fig2} (right). The fiducial resolution equals $R=5000$. However, as we showed in Sect.~\ref{horizon}, this high resolution is not required. We show the effect of the resolution on the $S/N$ by smoothing the spectra to lower resolution, i.e. by convolving the spectrum with a Gaussian of appropriate FWHM. For example, a source with $m_{\rm UV,rest} = 25$ mag would have to be observed for $\sim 7.5~(2.8)$ hours to achieve an $S/N\sim5$ at $R = 5000~(2000)$ (i.e. $m_{\rm FUV,rest} = 25.7$ mag is the actual magnitude of the observation). The observations will be performed at $R=5000$, which means that the resolution can only be degraded afterwards if required.

In order to make our results easily applicable, we generalized the performance of the VIFU mode in the blue by providing scaling relations between $S/N$ and several other quantities, in particular, the number of exposures NDIT ($T_{\rm exp}$ = NDIT h), resolution ($R$), total throughput ($T_{\%}$), and galaxy magnitude ($m_{\rm UV} = m_{\rm rest,UV}$). As a reference set of parameters, we used a case with NDIT = 20, $R = 5000$, $T_{\%}=13 \%,$ and $m_{\rm rest,UV} = 25.0$ mag. Then we investigated how each of the quantities scales with $S/N$. The results are parametrized in the following equation (see Fig. \ref{fig3}):

\begin{multline}
\label{scalingIGM}
S/N_{\rm VIFU} \approx 8.2\left(\frac{NDIT}{20}\right)^{0.48\pm 0.01}\left(\frac{R}{5000} \right)^{-0.51 \pm 0.02}\times \\ \times\left( \frac{T_{\%}}{13}\right)^{0.53\pm0.03}10^{-0.4(m_{\rm UV} - 25.0)(0.72 \pm 0.05)}\,.
\end{multline}

\noindent The exponents in the scaling relation (for NDIT, $R,$ and $T_{\rm \%}$) are those that are expected in the shot-noise (background limited) regime. The relation with brightness is more complicated due to the relation between the size and brightness of a galaxy that we used in the simulation. Many uncertainties affect the simulated $S/N$. However, our choice of input ingredients (e.g. type of PSF) reflects a very conservative approach: by relaxing some of the constraints, the resulting simulated $S/N$ could be higher. In particular, we used $\Delta_{\rm m}$ = 0.7 mag as a correction to the normalization of the flux in the FUV (see discussion in Section \ref{additional}). This value is the average expected value, while an individual galaxy will have a bluer or redder spectrum. The scaling relation can be easily modified to account for a different correction, for instance, Equation \ref{scalingIGM} can be used with the exponent in the last term set as $-0.4\left( m_{\rm UV} - 25.0 + \Delta_{\rm m} - 0.7~\mathrm{mag}\right)$.

\subsection{HMM-VIS mode performance}
\label{hmmvis}

In this section we compare the expected performance of the VIFU and HMM-VIS mode observations in the blue. The HMM-VIS apertures (Fig. \ref{fig0}) are still larger than the average size of all but the brightest galaxies at $z \gtrsim 3$ (Table \ref{tab1}), therefore we do not expect substantial aperture losses (but see Sect.~\ref{reqtime} for a discussion about the effect of different seeing conditions on the performance of the HMM-VIS mode). We assumed the same spectral and spatial binning as in the VIFU mode. The aperture in the HMM-VIS mode is a bundle of 19 fibers, and each observation therefore results in 19 spectra (see Fig. \ref{fig0}).

Following the steps in the previous analysis, we started by computing the $S/N$ as a function of exposure time using the flat spectra as input. For each exposure time we first combined the number of DITs corresponding to the exposure time (i.e. spectra for each fiber were simply summed), and calculated the $S/N$ for each fiber individually. Then we ordered the fibers according to their $S/N$ and repeated the extraction with the optimal $S/N$ as presented in Section \ref{snsn}.

The scaling relations for the case of HMM-VIS mode are shown in Fig. \ref{figA0} and are measured to be

\begin{multline}
\label{scalingHMM}
S/N_{\rm HMM-VIS} \approx 6.7\left(\frac{NDIT}{20}\right)^{0.48\pm 0.01}\left(\frac{R}{5000} \right)^{-0.49 \pm 0.03}\times \\ \times \left( \frac{T_{\%}}{13}\right)^{0.48\pm0.04}10^{-0.4(m_{\rm UV} - 25.0)(0.76 \pm 0.07)}\,.
\end{multline}

\noindent Again the scaling relations indicate the shot-noise regime. For the same exposure time, the $S/N$ in the HMM-VIS mode observations is $\sim 20 \%$ lower than the $S/N$ for the VIFU mode. Equivalently, under the same observational parameters, the difference in magnitudes to reach the same $S/N$ with the two modes is only $\sim 0.25$ mag. 

The advantage of using an IFU over a simple aperture is that the sky can be estimated from the same data cube in which the science source is located. In the case of the HMM-VIS mode, however, we have to rely on a secondary fiber observing a part of the sky near the source. We performed a detailed analysis of the effect that the sky variability has on the $S/N$ (see Appendix \ref{skyvar}) and conclude that under normal circumstances, and assuming that the differential response between fibers can be measured and corrected for to a good accuracy, the effect of sky variability on the performance of the instrument in the blue band is negligible when sources down to 26 mag are observed.

\section{Discussion}
\label{discuss}

We have demonstrated the quality of spectra necessary to perform a successful 3D reconstruction of an IGM at $3 \lesssim z \lesssim 3.5$ and to study the properties of galaxies in relation to their large-scale environment at these redshifts. Using these results, combined with the simulated MOSAIC performance, we can now discuss how such a survey would be carried out. We first determine the number of galaxies we expect to have in the instrument FOV, after which we estimate the time it would take to complete the survey. We conclude with a brief discussion of the synergy programs on the ELT and of the complementary IGM tomography programs on other facilities.

\subsection{Background galaxy counts}

We first estimate the number of galaxies that we expect to have in a FOV at each pointing by considering real fields. We chose to consider the GOODS-South field \citep{Giavalisco2004} because it is one of the best-observed parts of the sky. For our purposes, we need photometric observations of galaxies in the field complete down to $m_{\rm rest,UV} < 25.5$ mag and information on the redshifts. We based our analysis on the data collected by the 3D-HST survey \citep{Skelton2014}. The survey collected photometric data of many previous deep surveys and through an SED analysis computed photometric redshifts. In addition, for the bright part of the sample, Skelton and collaborators provide grism-based redshifts \citep{Momcheva2016}. They compiled a final catalogue of sources with measured redshifts, where the redshifts were taken from three different ways of measurement: ground spectroscopy, grism redshifts, and photometry. While photometric redshifts are not completely reliable, we still took them at face value because the completeness of the grism-based redshifts does not reach sufficiently faint sources.

The rest-frame UV spectrum of galaxies lying at $z \sim 3.0 - 4$ is covered by the HST/F606W filter. We therefore selected all the galaxies with $m_{\rm F606V} < 25.5$ mag. Several sources lack reported measurements with this filter. We checked the literature to determine whether any of the sources without $m_{\rm F606V}$ measurement have been observed, but we found that the catalogue we used was complete. The sources that are not found in these catalogue are probably very red (stars or high-$z$ galaxies). The number of sources in \citet{Guo2013}, for example, is approximately the same as in the \citet{Momcheva2016} catalogue (down to $m_{\rm F606V} < 25.5$), therefore we assumed for our purposes that the number of sources (down to our limiting magnitude) represents the whole galaxy population in the field.

\begin{figure*}
\begin{center}
\begin{tabular}{cc}
\includegraphics[scale=0.58]{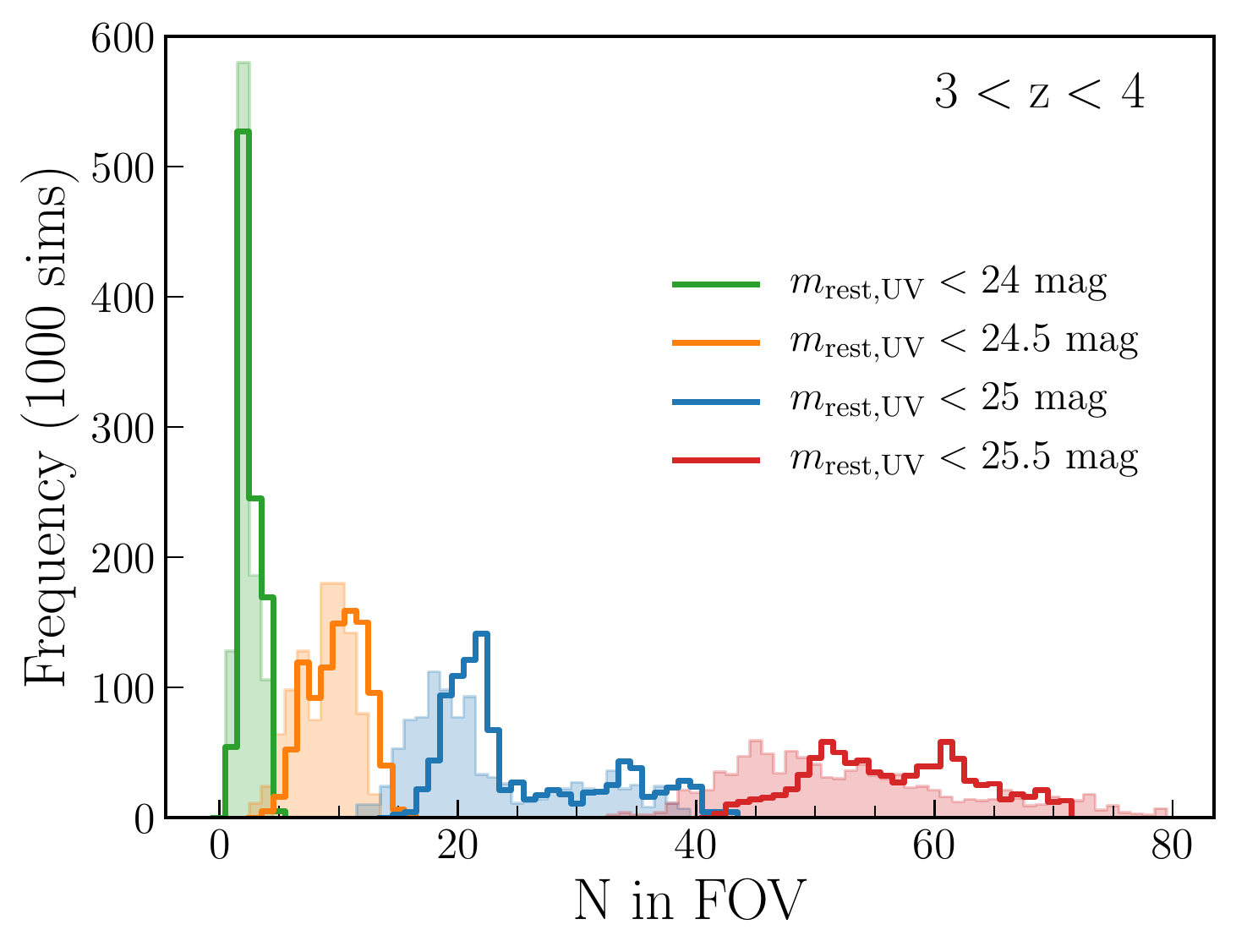}&
\includegraphics[scale=0.58]{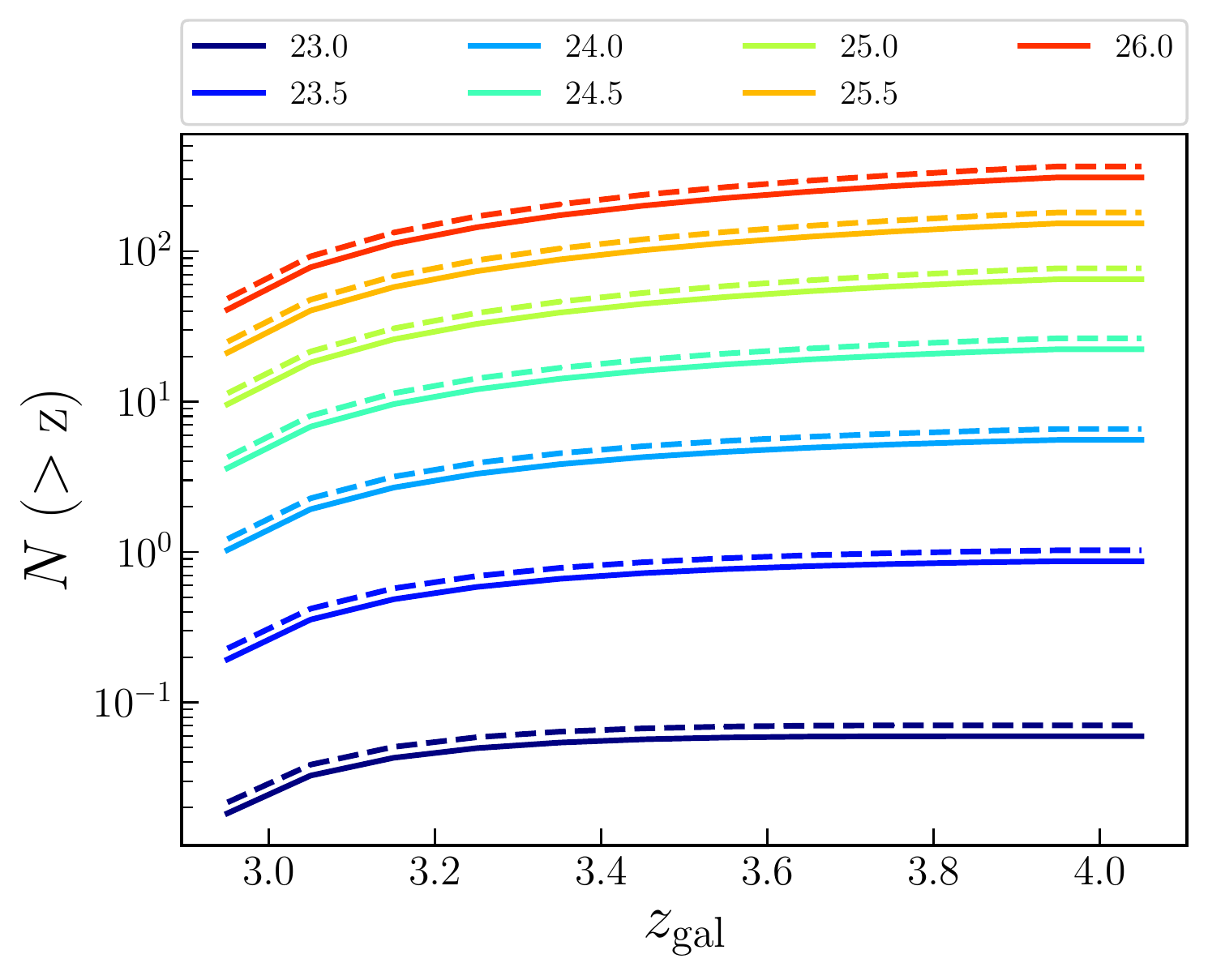}
\end{tabular}
\caption{{\it Left}: Number of galaxies in a randomly placed FOV in the 3D-HST GOODS-South field. The filled and empty histograms are computed for the FOV = 44 (52) arcmin$^2$, corresponding to the VIFU and HMM-VIS modes, respectively. The FOV is repositioned 1000 times. The results are given for different magnitude cuts, and a redshift range of $3 < z < 4$ is considered. {\it Right:} Predicted cumulative numbers of galaxies in a 44 (solid) and 52 (dashed) arcmin$^2$ FOV obtained by integrating the measured rest-UV luminosity function of LBG galaxies \citep{Reddy2009,Bouwens2015}. Different colours correspond to different limiting magnitudes.}
\label{fig13}
\end{center}
\end{figure*}

The scientific FOV of MOSAIC is projected to be 44 and  52 arcmin$^2$ for the VIFU and HMM-VIS mode, respectively (Figure \ref{fig0}). We randomly placed a FOV like this (assumed to be circular for simplicity) on the 3D-HST GOODS-South field. Then we counted the sources within the FOV. We repeated this for 1000 random positions. The numbers of detected galaxies for different magnitude cuts in the $3 < z < 4$ redshift range are shown in Fig. \ref{fig13} (left). On average, going down to $\sim$25.5 mag, we expect to have $\sim 30-40$ background galaxies in the FOV, depending on the redshift range that is considered. Similar numbers can be estimated by integrating the observed luminosity function of LBGs at $z \sim 3 - 4$ \citep{Reddy2009,Bouwens2015}, as illustrated in Fig. \ref{fig13} (right).

\subsection{Survey}
\label{survey}

\subsubsection{Survey area}
The first step towards designing a tomographic survey is to determine the dimension of the field in which the reconstruction is to be performed. A sufficient volume is required in order to properly define the large-scale structure as a whole. In this sense, we can argue that each dimension of the field should encompass several times the typical size of the large-scale voids at this redshift \citep[i.e. roughly between $\sim 10$ and $\sim 20$~Mpc in diameter for the largest ones, see e.g.][]{Arbabi02}. 

\begin{figure}
\begin{center}
\includegraphics[scale=0.58]{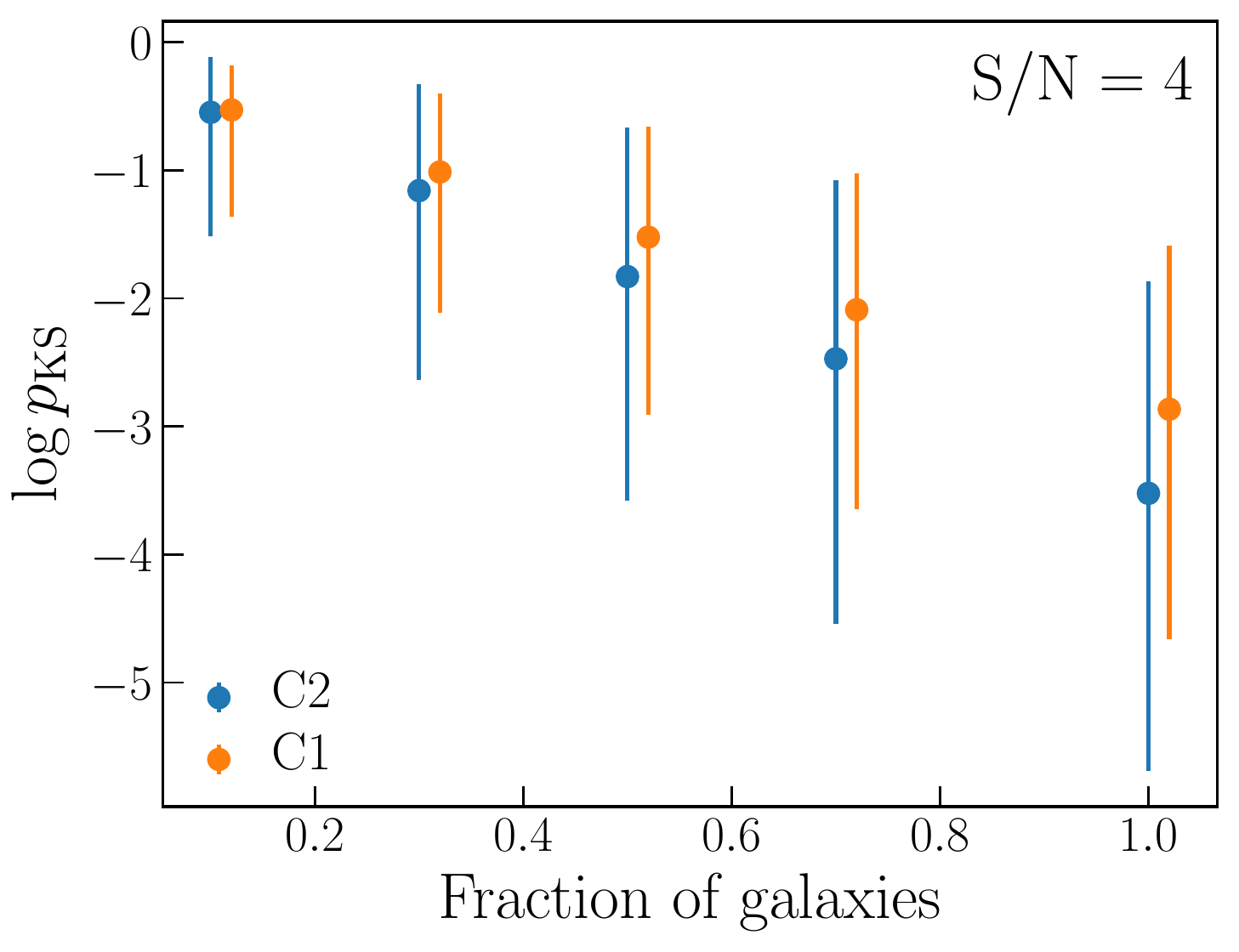}
\caption{Significance of the detected galaxy $M_{\rm J}$ gradient towards the filaments as a function of galaxy sample size. The latter is measured relative to the fiducial size (Section \ref{horizon}) and is used throughout the paper.}
\label{fig14}
\end{center}
\end{figure}

In addition, the environmental dependency of galaxy properties on the distance to the filaments is a subtle effect (second order with respect to the halo mass, which is the dominant driver), which, in order to be measured, requires us to decrease the statistical errors as much as possible. To explore this effect, we carried out a simple test to measure the decrease in significance of the signal when we pruned our initial sample. Focusing on the S/N=4 case, we made a series of galaxy samples with a decreasing number of galaxies, where the number was defined as a fraction of the parent sample. Each sample was N$\times$fraction of randomly selected (with replacement) galaxies from the parent sample. The sample was then divided into two subsamples, based on its median $d_{\rm fil}$ (e.g. we obtain two data points in Figure \ref{fig12}). A KS test was used to calculate the significance of the difference between the distributions of $M_{\rm J}$ of two subsamples. For each fraction of galaxies we made a Monte Carlo (MC) simulation. The result is shown in Figure \ref{fig14} for both configurations. As expected, the signal becomes less pronounced with decreasing galaxy numbers. In order to detect the signal at a level of $p_{\rm KS} \lesssim 0.05$ within errors, we cannot prune the parent sample much. The fraction of galaxies can be interpreted as a fraction of the area in the sky as long as the galaxies are uniformly distributed inside the volume. From this test, we conclude that a one-degree$^{2}$ field offers a good compromise between the required observational time and the statistical significance of the signal. 

Finally, we emphasize that we here focused on a single science case, related to the effect of the large-scale cosmic web on galaxy properties. However, this 3D reconstruction of the matter distribution down to the scale of Mpc opens up the door to a wealth of additional measurements relying on the 3D topology of the field to probe the cosmology (void statistics, peak counts, connectivity of the cosmic web, etc.). All these science cases would strongly benefit from fields as large as possible in order to be competitive. 

\subsubsection{Required time}
\label{reqtime}

Ideally, we would like to carry out the survey so that it would be consistent with the parameters of the C2 configuration studied in Section \ref{horizon}. In this case, we need to target the galaxy population down to the limiting magnitude of $25.5$ mag, and the spectra should have at least an $S/N\sim$4 after they are smoothed down to $R=2000$. In order to have a representative volume, the area that is covered in the survey should be of $\sim 1$ degree$^2$. The total amount of the observational time in the number of nights can be summarized as

\begin{equation}
\label{surveyEq}
N_{\rm night} = N_{\rm point} \frac{T_{\rm exp,point}}{T_{\rm night}},
\end{equation}

\begin{equation}
\label{nights}
N_{\rm point} = N_{\rm sub} \frac{{\rm area}}{{\rm FOV}},
\end{equation}

\noindent where $T_{\rm night}$ is the observing time per night (hereafter $T_{\rm night} = 8$ h), $T_{\rm exp,point}$ is the minimum required exposure time per pointing, and $N_{\rm sub}$ takes into account that a certain field may have to be observed several times, depending on the adopted observation mode. As discussed in Section \ref{mosaic}, this could be either the multiplex (HMM-VIS) or the IFU (VIFU) mode. The VIFU and HMM-VIS modes will have a FOV = 44 (52) arcmin$^2$, respectively. As shown in the left panel of Fig.\,\ref{fig13}, we expect on average $\sim50$ or 60 galaxies (with $m_{\rm R}< 25.5$ mag) in the FOV of the two modes. This means that we could observe all the galaxies with one pointing using the multiplex mode, and several pointings would be necessary using the IFU mode (about five pointings, based on the current design specifications). $T_{\rm exp,point}$ is determined by the required $S/N$ and can be computed by the appropriate scaling relations (Equations \ref{scalingIGM} or \ref{scalingHMM}). For example, for the C2 configuration, $T_{\rm exp,point} = 3.7$ (5.2) h is required to obtain spectra with $S/N = 4$ for the faintest of sources with the VIFU (HMM-VIS) mode. Similar estimates can also be made for the C1 configuration. We note that, when we applied the noise to the simulated spectra, we measured the correcting factor for each galaxy instead of assuming the average $\Delta m = 0.7$ mag value. Even though the correction factor changes from galaxy to galaxy as a result of the differences in the shape of the galaxy SED, the minimum required exposure, calculated from our scaling relations used at face value, is correct, that is, the spectra of $\sim99\%$ of the simulation galaxies indeed reach an $S/N=4$ within this calculated minimum exposure time.

\begin{table}
\begin{center}
\caption{Estimates of the number of nights in which the IGM tomography survey with the MOS instrument on the ELT needs to be carried out in order to achieve the scientific goals outlined in this paper. The estimates are given for the two observation modes (VIFU and HMM-VIS) and the two configurations studied here. The numbers in parentheses indicate the estimates if the observations are conducted under poorer seeing conditions (see the main text in Section \ref{reqtime}). The third line gives an independent estimate of the survey speed ratio of the two modes.}
\label{Tab:survey}
\renewcommand{\arraystretch}{1.3}
\begin{tabular}{ c c c} \hline \hline
             mode & $R=1000$ & $R=2000$  \\ \hline
          VIFU   & 92     &  180\\
         HMM-VIS & 24 (34) &  45 (63)\\
         \hline
         Survey speed ratio & 0.3 & 0.3\\
  \hline
\end{tabular}
\end{center}
\end{table}

Using the above considerations, we can estimate the number of nights that are required to carry out the survey when it covers an area of 1 degree$^2$. The numbers are provided in Table \ref{Tab:survey} for the two configurations and the two observational modes. The VIFU mode is not competitive because too few units are available. The same conclusion can be reached for the {\it \textup{survey speed}} (SS), expressed as SS = $S/N^{2}\times multiplex$ \citep{Puech2018}. The ratio SSR = SS$_{\rm VIFU}$/SS$_{\rm HMM-VIS}$ shows approximately how efficient the two modes are for a particular science case. In our case, the ratio is SSR $\approx \left( 1.2\right)^2 \times (8/40) = 0.3$, which is about the same as the ratio of the required nights estimated from Eq. \ref{nights}.

In the case of the C1 configuration, it might be considered to target only $m_{\rm R} < 25$ mag galaxies, which would effectively halve the number of targets per field. In this case, we estimate a similar number of nights for the two modes. However, we have shown throughout the Section \ref{horizon} that the results of the reconstruction for the limit of $m_{\rm R}< 25$ mag are slightly worse than for the fiducial case. It would take twice the time to carry out the second configuration with respect to the first if we were to limit ourselves to the HMM-VIS mode. While that is a significant difference, considering the higher resolution would be beneficial not just for the science case of IGM tomography, but also for the synergy programs (see Section \ref{synergy}).

The estimates are realistic, but there are several factors that could affect the real observation times. In the case of multiplex (HMM-VIS) observations, our results are affected by the uncertainty in the atmospheric conditions. The simulations in Section \ref{mosaic} were carried out under the assumption that the seeing FWHM equals 0.7 arcsec. According to the data of the atmospheric monitoring at the Paranal site in Chile\footnote{$\mathrm{http://archive.eso.org/wdb/wdb/asm/dimm\_paranal/form}$}, a seeing of 0.7 arcsec is about the median value in the period of 2017-2019. Because of the long time that such a survey would take, observations also at worse seeing conditions might be considered, which would result in a loss of received light and therefore poorer performance. For example, according to the data, a seeing of $\sim 0.9~(1.2)$ arcsec corresponds to the 75$\%$ (90$\%$) in the seeing distribution. When the same observing parameters (e.g. typical galaxy sizes) as in Section \ref{mosaic} are assumed, observations at this seeing would result in a loss of light of $\sim 20\%$ (50$\%$) with respect to the fiducial 0.7 arcsec case. This implies that in order to reach the same $S/N$, the observing time would have to be increased by a factor of $\sim1.4~(3)$ at the corresponding seeing conditions, according to Eq. 6. To illustrate this effect, we include in Table \ref{Tab:survey} (in parentheses) the estimates for the case of a seeing at the time of observation (for all targets) of 0.9 arcsec.

In addition, about $\sim 10\%$ of the LOS are expected to be contaminated by damped Lyman-alpha absorbers \citep{Wolfe2005} and will therefore not be used for the IGM tomography. Finally, more accurate times are difficult to predict without the actual fields in the sky that would allow us to optimize the survey. However, especially in the HMM-VIS case, the numbers in Table \ref{Tab:survey}, which were calculated for two different seeing conditions, provide a  reasonable confidence interval.

\subsection{Synergy with other surveys and science cases}
\label{synergy}
\subsubsection{Accurate redshift and mass measurements}
So far, we have assumed that the redshifts and physical properties (such as stellar mass and SFR) of the targeted galaxies are known. In reality, a multi-band photometric survey of a sky with an area of $\sim1$ degree$^2$ will have to be carried out down to the required limiting magnitude before the gradients of galaxy population at $z\sim 3$ can be addressed.
First of all, such a survey is required to obtain preliminary information on the source redshifts. Furthermore, because the galaxy evolution in the frame of the large-scale structure is one of the fundamental questions that we wish to address, galaxy characteristics have to be known, such as stellar mass and SFR. Because mass and SFR are not easy to measure accurately, reliable tracers of these properties can be used, such as the rest-frame UV and NIR brightness. In Section \ref{sec:mapgal} we showed that $M_{\mathrm{J}}$ and $M_{\rm u}$ are good tracers of stellar mass and SFR and can be used to measure gradients. In our targeted redshift range ($z \gtrsim 3$), the rest-frame UV galaxy spectrum can be accessed from the ground. On the other hand, the rest-frame NIR brightness would require  IR observations from space. Our imposed galaxy brightness limit does not require extremely deep observations of the targeted area in principle. However, the resources required for multi-band photometry observations of a $\sim 1~{\rm degree}^2$ field will not be negligible. This suggests that this and other surveys should be coordinated. 

For example, the southern sky will be simultaneously targeted by the European Space Agency's {\it Euclid} mission \citep{Laureijs2011} and by the Large Synoptic Survey Telescope \citep{LSST2009} mission. These two surveys will provide {\it u} - {\it H} band SEDs with completeness in brightness as required by our science case \citep{Rhodes2017} and will allow, if combined, accurate redshift measurements \citep{laigle19}. Obtaining reliable stellar mass measurements at $z>3$ might be more problematic at the depth of the \emph{Euclid} core programs ($H\sim 24$ at 5$\sigma$ for extended sources), and we might need to either rely on the NIR photometry of the \emph{Euclid} Deep fields (which should reach two magnitudes deeper) or on additional rest-frame NIR data, which could be obtained by observing the same field with the MIRI instrument \citep{Rieke2015} on the forthcoming James Webb Space Telescope \citep{Gardner2006}. We used the JWST Exposure Time Calculator\footnote{https://jwst.etc.stsci.edu/} \citep{Pontoppidan2016} to estimate the minimum observation time required to obtain the rest-frame $M_{\rm J}$
magnitudes for our targeted galaxies. To estimate the typical brightness of galaxies in the MIRI F560W band, we adopted the galaxy spectral templates of \citet{bruzual&charlot03}, solar metallicity, and a single stellar population. We find that a galaxy at $z = 3.5$ is typically  $\gtrsim 2$ mag brighter in the MIRI F560W band than in the HST WFC/F606W band ($\sim 0.5$ Gyr after starburst). According to the ECT, it would take $\sim 4$ min of exposure time to detect a source with a brightness of F560W = 23.5 mag (AB) with an $S/N \sim 10$. This means that $\sim 90$ h of observing time on the JWST would be required to cover $1~{\rm degree}^2$ down to the stated limits.

\subsubsection{CGM studies}
We here only focused on the Ly$\alpha$ forest. However, the spectra will also contain the absorption from various metals in the IGM and in the circum-galactic medium (CGM). One of the best candidates for probing the metal budget in the IGM is triply ionized carbon (\civ), a species observed to be abundant not only in the CGM, but also in the diffuse IGM \citep[e.g.][]{DOdorico2010,DOdorico2016}. The \civ\,transition is known for its characteristic doublet ($1548.20~\mathrm{\AA}$ and $1550.78~\mathrm{\AA}$). The large-scale clustering of \civ\, has been studied in several surveys that probed the CGM \citep{Adelberger2003,Turner2014} and larger Mpc scales \citep{Tytler2009,Blomqvist2018}. 

With the IGM tomography survey, as outlined in the previous sections, information on the \civ$\lambda\lambda1548,1550$ over a wide range of redshifts $z\sim1.9-2.9$ is obtained automatically. \civ\ is known to be associated with the photoionized gas of $\sim10^5$\, K that is seen to have a high covering fraction in the CGM at low redshifts \citep[e.g., ][]{Bordoloi2014ApJ,Burchett2016ApJ}. This makes  it a useful tracer for CGM studies. With its rest-frame wavelength of $\sim1550~\mathrm{\AA}$, it lies outside the region that is polluted by the Ly-$\alpha$ forest absorption over a large redshift path. The survey will also result in an observation of the host galaxies for a large fraction of \civ\ absorbers. This survey will therefore provide a unique dataset for CGM studies at high redshifts. 

Here we estimate the expected number of \civ\ systems for the IGM tomography survey redwards of the Ly-$\alpha$ forest. To do so, we used the $dN$/$dz$ obtained from XQ-100 survey \citep{Perrotta2016MNRAS}. This must provide a reasonable prediction because it is obtained using X-Shooter \citep{Vernet2011}, which has a spectral resolution of $\sim5000$ in the blue arm; this is close to the resolution of MOSAIC. To obtain this estimate, we considered the redshift paths redwards of the Ly-$\alpha$ of the galaxies. We then used the $dN$/$dz$ from \citet{Perrotta2016MNRAS} and counted the expected number of \civ\ for the available redshift path. For a minimum S/N of 3 (and 5), we estimate that $\sim9500$ (and $\sim1700$) systems will be detected. This will be one of the largest \civ\ samples in which we are likely to observe for a reasonable fraction of the absorbers the host galaxy as well.

\subsubsection{Void and wall analysis}
As outlined above, although we have focused on the study of galaxy evolution in filaments, a 3D reconstruction of the matter distribution like this will also enable more cosmology-oriented studies. In particular, the probed scale range should allow reliably extracting voids in 3D, as postulated in \cite{krowleski18}. Voids and walls are very interesting laboratories for galaxy evolution and cosmology. On the one hand, voids represent a somewhat primitive environment for galaxies where the density is low and the matter flow is still laminar and curl free. Void galaxies are more numerous at higher than at lower redshift, although they are usually fainter and therefore more difficult to observe than their high-density counterparts. They are of prime interest for testing the theories of galaxy formation \citep[e.g.][]{lindner1996,hoeft06, Moorman16}. Walls are the loci of galaxy formation and affect early spin acquisition \citep{codis15}. On the other hand, voids and walls are the tool of choice for probing the cosmology, for instance to constrain the dark energy equation of state or to test the theory of modified gravity \citep[e.g.][inter alia]{Gay12,lavaux12,Cai15,hamaus16,falck18}. Their large sizes and tri-dimensional volumes (in contrast to filaments, which are almost uni-dimensional) mean that voids will naturally be better reconstructed than filaments for a given density of background sources because on average, each void will be sampled by many more sightlines than each filament. Conversely, these sighlines would also naturally cross their surrounding walls. A survey  that would enable a reconstruction of filaments as proposed here should therefore also enable a robust reconstruction of voids and walls. Fortunately, {\sc DisPerSE}  self-consistently provides the distribution of walls and voids at no additional computational cost. 
As mentioned above, however, void statistics potentially require a larger volume than one square degree in order to place stringent constraints on cosmology. Exploring the prospects for void analysis from Lyman-$\alpha$ tomography will be the focus of a future work.

\subsection{Synergies with other tomographic Lyman-$\alpha$ surveys}

The interest of mapping the large-scale structure in~3D through Lyman-$\alpha$ tomography has grown strongly over the past years, and this science case is now one of the primary scientific goals of several current or future surveys. The design of these surveys has to compromise between the size of the field and the resolution to stay within reach in terms of observational time. 

At large scale, the quasar survey of the William Herschel Telescope Enhanced Area Velocity Explorer \citep[WEAVE; Jin et al. in prep,][]{Pieri2016} will be a sample of choice to perform a reconstruction over $\sim 6000$ deg$^{2}$. However, as WEAVE will primarily target quasi-stellar objects (QSO), or quasars, as a part of WEAVE-QSO survey, the resolution that is reached will be limited to $15-20$ cMpc/$h$ (although higher resolution might be reachable in some specific regions). 
The PFS \citep{Takada2014} will enable a smaller scale exploration, down to some~Mpc.  Its observing program will target the range $2.2\lesssim z \lesssim 2.8$, which will extend the pilot study performed on the COSMOS field over 0.8 deg$^{2}$ on a slightly tighter redshift range \citep[CLAMATO,][]{Lee2018} to several square degrees. In a more distant future, the MaunaKea Spectroscopic Explorer (MSE) aims for a reconstruction like this up to $z\sim 3$ over several tens of degree$^{2}$ (see ``The Detailed Science Case for the Maunakea Spectroscopic Explorer, 2019 edition"\footnote{\url{https://arxiv.org/abs/1904.04907}}). The Giant Magellan Telescope (GMT) has also identified it as one of its science focuses (see the GMT Science Book 2018\footnote{\url{https://www.gmto.org/wp-content/uploads/GMTScienceBook2018.pdf}}). Although the design of this survey is barely defined as yet, it is expected that through its $\sim$30-meter diameter, the GMT will enable a very high-redshift exploration of the IGM at reasonable cost, potentially comparable to what we plan with MOSAIC on the ELT. Finally, the BlueMUSE integral field spectrograph proposed for the VLT (BlueMUSE: Project Overview and Science Cases\footnote{\url{https://arxiv.org/pdf/1906.01657}}) could offer an interesting counterpart to these reconstructions because it proposes to image the IGM\textup{ in emission} at $z\sim 2-3$.

Interestingly, MOSAIC is currently the only facility (to the best of our knowledge) that may enable a Lyman-$\alpha$ tomographic reconstruction at $z>3$ down to the Mpc scale. It would therefore advantageously complement the other planned surveys, and offer the possibility of probing the full redshift evolution of galaxy growth in the cosmic web throughout a cosmic period of intense star formation in galaxies.

\section{Conclusions}

One of the most ambitious programs envisioned for the multi-object spectrograph on the European Extremely Large Telescope is to directly conduct the full inventory of matter at $z \sim 3$ in galaxies, their circumgalactic medium, and in the IGM \citep{Puech2018}. This will be achieved by complementary observations of galaxies in the visible and NIR wavelengths: the MOSAIC instrument concept \citep{Morris2018} has been designed with this scientific goal in mind. In this work we investigated the potential of such an instrument for the IGM tomography at $z \gtrsim 3$, that is, the 3D mapping of the IGM. The idea is to observe \lya\, forest at $z\sim 3$ that is imprinted in the spectra of background LBGs. If the density of the sight lines is high enough, the set of spectra can be used to reconstruct the 3D matter distribution.

We made use of the galaxy spectral catalogue extracted from the {\sc Horizon-AGN} cosmological hydrodynamical simulation (Sect.~\ref{horizon}) to determine which spectral configuration and quality of galaxy spectra are required to enable a reconstruction of the IGM that is good enough to successfully address various science cases. Noise was applied to the simulated galaxy spectra, and the resulting realistic set of spectra was used to reconstruct the density field from the {\sc Horizon-AGN} simulation. Afterwards, the filamentary cosmic web, also known as the skeleton, was extracted from the reconstructed field. We reconstructed two different configurations, corresponding to two different spectral resolutions ($R=1000$ and $R=2000$) and therefore two reconstruction scales.

We investigated the quality of the reconstruction and that of the skeleton extraction as a function of the assumed $S/N$ of the spectra. We found that the optimal S/N (per resolution element) of the faintest objects in the survey is $S/N = 4$, that is, no substantial gain is achieved for an $S/N$ higher than this. We also showed that this reconstruction allows us to observe and study the galaxy stellar mass gradient towards the filaments at $3 \lesssim z \lesssim 3.5$. We found that the less pronounced filaments, when the spectral noise is taken into account, are lost in the process of the reconstruction and skeleton extraction. Following the analysis of the data from a real survey, the details of this introduced bias will therefore have to be understood and taken into account to interpret the results. We stress that this part of our analysis is independent of the characteristics of a specific instrument.

In order to estimate the time that such a survey would take on the ELT, we used the MOSAIC concept design and simulated the performance of the instrument at visible ($\lambda \gtrsim 450$) wavelengths (Sect.~\ref{mosaic}). In particular, we provided the scaling relations between the S/N and other relevant observational (exposure time and source brightness) and instrumental (resolution and total throughput) quantities for the VIFU and HMM-Vis observing modes (equations \ref{scalingIGM} and \ref{scalingHMM}). By combining the scaling relations with the results of the density field reconstruction analysis (Sect.~\ref{reqtime}), we estimated that conservatively, a survey as envisioned in this paper would take $\lesssim 35~(65)$ nights on the ELT. This estimate corresponds to the observations being carried out by the high-multiplex mode. We found that for this science case, the VIFU mode is not competitive because there are too few IFU units. The specifications of the MOS instrument on the ELT will be changed to some extent in the next design phase. Unless the changes are very significant, the results of this paper can be used to easily obtain new estimates of the observation times that are required to carry out such a survey. 

A great scientific potential, combined with technological advancement, has led to a steady increase of interest in the mapping of the large-scale structure at high $z \gtrsim 2$ redshifts over the past years. We look forward to the many planned IGM tomography surveys in the next decade. With the combined power of the shallower large-field surveys (e.g. with the PSF instrument on Subaru) and the deeper surveys of the smaller fields (as offered by the MOSAIC), we will gain an unprecedented insight into the galaxy evolution on different scales and over a long interval of cosmic time.  

\begin{acknowledgements}

We wish to thank C. J. Evans and C. Peroux for fruitful discussions. JJ and LK acknowledge support from NOVA and NWO-FAPESP grant for advanced instrumentation in astronomy. CL acknowledges funding support from Adrian Beecroft and would like to thank Stephane Arnouts for discussions and advice. This work is partially supported by the Spin(e) grants ANR-13-BS05-0005 (\url{http://cosmicorigin.org}) of the French Agence Nationale de la Recherche, and ERC grant 670193. We thank S. Rouberol for smoothly running the Horizon cluster, hosted by the Institut d’Astrophysique de Paris, where some of the postprocessing was carried out, and T. Sousbie for distributing {\sc DisPerSE}. This research made use of Astropy, a community-developed core Python package for Astronomy (Astropy Collaboration, 2013). We thank D.~Munro for freely distributing his {\sc \small  Yorick} programming language and opengl interface (available at \url{http://yorick.sourceforge.net/}

\end{acknowledgements}

\bibliographystyle{aa}
\bibliography{ms_mosaic_bib}

\begin{thebibliography}{133}
\expandafter\ifx\csname natexlab\endcsname\relax\def\natexlab#1{#1}\fi

\bibitem[{{Adelberger} {et~al.}(2003){Adelberger}, {Steidel}, {Shapley}, \&
  {Pettini}}]{Adelberger2003}
{Adelberger}, K.~L., {Steidel}, C.~C., {Shapley}, A.~E., \& {Pettini}, M. 2003,
  \apj, 584, 45

\bibitem[{{Alavi} {et~al.}(2016){Alavi}, {Siana}, {Richard}, {Rafelski},
  {Jauzac}, {Limousin}, {Freeman}, {Scarlata}, {Robertson}, {Stark}, {Teplitz},
  \& {Desai}}]{Alavi2016}
{Alavi}, A., {Siana}, B., {Richard}, J., {et~al.} 2016, \apj, 832, 56

\bibitem[{{Alpaslan} {et~al.}(2016){Alpaslan}, {Grootes}, {Marcum}, {Popescu},
  {Tuffs}, {Bland-Hawthorn}, {Brough}, {Brown}, {Davies}, {Driver}, {Holwerda},
  {Kelvin}, {Lara-L{\'o}pez}, {L{\'o}pez-S{\'a}nchez}, {Loveday}, {Moffett},
  {Taylor}, {Owers}, \& {Robotham}}]{alpaslan16}
{Alpaslan}, M., {Grootes}, M., {Marcum}, P.~M., {et~al.} 2016, \mnras, 457,
  2287

\bibitem[{{Appleby} {et~al.}(2018){Appleby}, {Chingangbam}, {Park}, {Hong},
  {Kim}, \& {Ganesan}}]{Appleby18}
{Appleby}, S., {Chingangbam}, P., {Park}, C., {et~al.} 2018, \apj, 858, 87

\bibitem[{{Arag{\'o}n-Calvo} {et~al.}(2007){Arag{\'o}n-Calvo}, {van de
  Weygaert}, {Jones}, \& {van der Hulst}}]{AragonCalvo2007}
{Arag{\'o}n-Calvo}, M.~A., {van de Weygaert}, R., {Jones}, B.~J.~T., \& {van
  der Hulst}, J.~M. 2007, \apjl, 655, L5

\bibitem[{{Arbabi-Bidgoli} \& {M{\"u}ller}(2002)}]{Arbabi02}
{Arbabi-Bidgoli}, S. \& {M{\"u}ller}, V. 2002, \mnras, 332, 205

\bibitem[{{Aubert} {et~al.}(2004){Aubert}, {Pichon}, \& {Colombi}}]{aubert04}
{Aubert}, D., {Pichon}, C., \& {Colombi}, S. 2004, \mnras, 352, 376

\bibitem[{{Beygu} {et~al.}(2016){Beygu}, {Kreckel}, {van der Hulst}, {Jarrett},
  {Peletier}, {van de Weygaert}, {van Gorkom}, \& {Aragon-Calvo}}]{Beygu2016}
{Beygu}, B., {Kreckel}, K., {van der Hulst}, J.~M., {et~al.} 2016, \mnras, 458,
  394

\bibitem[{{Blomqvist} {et~al.}(2018){Blomqvist}, {Pieri}, {du Mas des
  Bourboux}, {Busca}, {Slosar}, {Bautista}, {Brinkmann}, {Brownstein},
  {Dawson}, {de Sainte Agathe}, {Guy}, {Percival}, {P{\'e}rez-R{\`a}fols},
  {Rich}, \& {Schneider}}]{Blomqvist2018}
{Blomqvist}, M., {Pieri}, M.~M., {du Mas des Bourboux}, H., {et~al.} 2018,
  \jcap, 5, 029

\bibitem[{{Bond} {et~al.}(1996){Bond}, {Kofman}, \& {Pogosyan}}]{Bond1996}
{Bond}, J.~R., {Kofman}, L., \& {Pogosyan}, D. 1996, \nat, 380, 603

\bibitem[{{Bordoloi} {et~al.}(2014){Bordoloi}, {Tumlinson}, {Werk},
  {Oppenheimer}, {Peeples}, {Prochaska}, {Tripp}, {Katz}, {Dav{\'e}}, {Fox},
  {Thom}, {Ford}, {Weinberg}, {Burchett}, \& {Kollmeier}}]{Bordoloi2014ApJ}
{Bordoloi}, R., {Tumlinson}, J., {Werk}, J.~K., {et~al.} 2014, \apj, 796, 136

\bibitem[{{Bournaud} {et~al.}(2008){Bournaud}, {Daddi}, {Elmegreen},
  {Elmegreen}, {Nesvadba}, {Vanzella}, {Di Matteo}, {Le Tiran}, {Lehnert}, \&
  {Elbaz}}]{Bournaud2008}
{Bournaud}, F., {Daddi}, E., {Elmegreen}, B.~G., {et~al.} 2008, \aap, 486, 741

\bibitem[{{Bouwens} {et~al.}(2015){Bouwens}, {Illingworth}, {Oesch}, {Trenti},
  {Labb{\'e}}, {Bradley}, {Carollo}, {van Dokkum}, {Gonzalez}, {Holwerda},
  {Franx}, {Spitler}, {Smit}, \& {Magee}}]{Bouwens2015}
{Bouwens}, R.~J., {Illingworth}, G.~D., {Oesch}, P.~A., {et~al.} 2015, \apj,
  803, 34

\bibitem[{{Bruzual} \& {Charlot}(2003)}]{bruzual&charlot03}
{Bruzual}, G. \& {Charlot}, S. 2003, \mnras, 344, 1000

\bibitem[{{Burchett} {et~al.}(2016){Burchett}, {Tripp}, {Bordoloi}, {Werk},
  {Prochaska}, {Tumlinson}, {Willmer}, {O'Meara}, \& {Katz}}]{Burchett2016ApJ}
{Burchett}, J.~N., {Tripp}, T.~M., {Bordoloi}, R., {et~al.} 2016, \apj, 832,
  124

\bibitem[{{Cai} {et~al.}(2015){Cai}, {Padilla}, \& {Li}}]{Cai15}
{Cai}, Y.-C., {Padilla}, N., \& {Li}, B. 2015, \mnras, 451, 1036

\bibitem[{{Caucci} {et~al.}(2008){Caucci}, {Colombi}, {Pichon}, {Rollinde},
  {Petitjean}, \& {Sousbie}}]{Caucci2008}
{Caucci}, S., {Colombi}, S., {Pichon}, C., {et~al.} 2008, \mnras, 386, 211

\bibitem[{{Cen} {et~al.}(1994){Cen}, {Miralda-Escud{\'e}}, {Ostriker}, \&
  {Rauch}}]{Cen1994}
{Cen}, R., {Miralda-Escud{\'e}}, J., {Ostriker}, J.~P., \& {Rauch}, M. 1994,
  \apjl, 437, L9

\bibitem[{{Chen} {et~al.}(2017){Chen}, {Ho}, {Mandelbaum}, {Bahcall},
  {Brownstein}, {Freeman}, {Genovese}, {Schneider}, \& {Wasserman}}]{Chen2017}
{Chen}, Y.-C., {Ho}, S., {Mandelbaum}, R., {et~al.} 2017, \mnras, 466, 1880

\bibitem[{{Chen} {et~al.}(2015){Chen}, {Ho}, {Tenneti}, {Mandelbaum}, {Croft},
  {DiMatteo}, {Freeman}, {Genovese}, \& {Wasserman}}]{chen15}
{Chen}, Y.-C., {Ho}, S., {Tenneti}, A., {et~al.} 2015, \mnras, 454, 3341

\bibitem[{{Cisewski} {et~al.}(2014){Cisewski}, {Croft}, {Freeman}, {Genovese},
  {Khandai}, {Ozbek}, \& {Wasserman}}]{cisewski14}
{Cisewski}, J., {Croft}, R. A.~C., {Freeman}, P.~E., {et~al.} 2014, \mnras,
  440, 2599

\bibitem[{{Codis} {et~al.}(2012){Codis}, {Pichon}, {Devriendt}, {Slyz},
  {Pogosyan}, {Dubois}, \& {Sousbie}}]{codis12}
{Codis}, S., {Pichon}, C., {Devriendt}, J., {et~al.} 2012, \mnras, 427, 3320

\bibitem[{{Codis} {et~al.}(2015){Codis}, {Pichon}, \& {Pogosyan}}]{codis15}
{Codis}, S., {Pichon}, C., \& {Pogosyan}, D. 2015, \mnras, 452, 3369

\bibitem[{{Codis} {et~al.}(2013){Codis}, {Pichon}, {Pogosyan}, {Bernardeau}, \&
  {Matsubara}}]{Codis2013}
{Codis}, S., {Pichon}, C., {Pogosyan}, D., {Bernardeau}, F., \& {Matsubara}, T.
  2013, \mnras, 435, 531

\bibitem[{{Codis} {et~al.}(2018){Codis}, {Pogosyan}, \& {Pichon}}]{Codis2018}
{Codis}, S., {Pogosyan}, D., \& {Pichon}, C. 2018, \mnras, 479, 973

\bibitem[{{Cole} {et~al.}(2005){Cole}, {Percival}, {Peacock}, {Norberg},
  {Baugh}, {Frenk}, {Baldry}, {Bland-Hawthorn}, {Bridges}, {Cannon}, {Colless},
  {Collins}, {Couch}, {Cross}, {Dalton}, {Eke}, {De Propris}, {Driver},
  {Efstathiou}, {Ellis}, {Glazebrook}, {Jackson}, {Jenkins}, {Lahav}, {Lewis},
  {Lumsden}, {Maddox}, {Madgwick}, {Peterson}, {Sutherland}, \&
  {Taylor}}]{Cole2005}
{Cole}, S., {Percival}, W.~J., {Peacock}, J.~A., {et~al.} 2005, \mnras, 362,
  505

\bibitem[{{Colless} {et~al.}(2001){Colless}, {Dalton}, {Maddox}, {Sutherland},
  {Norberg}, {Cole}, {Bland-Hawthorn}, {Bridges}, {Cannon}, {Collins}, {Couch},
  {Cross}, {Deeley}, {De Propris}, {Driver}, {Efstathiou}, {Ellis}, {Frenk},
  {Glazebrook}, {Jackson}, {Lahav}, {Lewis}, {Lumsden}, {Madgwick}, {Peacock},
  {Peterson}, {Price}, {Seaborne}, \& {Taylor}}]{Colless2001}
{Colless}, M., {Dalton}, G., {Maddox}, S., {et~al.} 2001, \mnras, 328, 1039

\bibitem[{{Cox} {et~al.}(2006){Cox}, {Jonsson}, {Primack}, \&
  {Somerville}}]{Cox2006}
{Cox}, T.~J., {Jonsson}, P., {Primack}, J.~R., \& {Somerville}, R.~S. 2006,
  \mnras, 373, 1013

\bibitem[{{Cui} {et~al.}(2018){Cui}, {Knebe}, {Yepes}, {Yang}, {Borgani},
  {Kang}, {Power}, \& {Staveley-Smith}}]{Cui2018}
{Cui}, W., {Knebe}, A., {Yepes}, G., {et~al.} 2018, \mnras, 473, 68

\bibitem[{{Dawson} {et~al.}(2013){Dawson}, {Schlegel}, {Ahn}, {Anderson},
  {Aubourg}, {Bailey}, {Barkhouser}, {Bautista}, {Beifiori}, {Berlind},
  {Bhardwaj}, {Bizyaev}, {Blake}, {Blanton}, {Blomqvist}, {Bolton}, {Borde},
  {Bovy}, {Brandt}, {Brewington}, {Brinkmann}, {Brown}, {Brownstein}, {Bundy},
  {Busca}, {Carithers}, {Carnero}, {Carr}, {Chen}, {Comparat}, {Connolly},
  {Cope}, {Croft}, {Cuesta}, {da Costa}, {Davenport}, {Delubac}, {de Putter},
  {Dhital}, {Ealet}, {Ebelke}, {Eisenstein}, {Escoffier}, {Fan}, {Filiz Ak},
  {Finley}, {Font-Ribera}, {G{\'e}nova-Santos}, {Gunn}, {Guo}, {Haggard},
  {Hall}, {Hamilton}, {Harris}, {Harris}, {Ho}, {Hogg}, {Holder}, {Honscheid},
  {Huehnerhoff}, {Jordan}, {Jordan}, {Kauffmann}, {Kazin}, {Kirkby}, {Klaene},
  {Kneib}, {Le Goff}, {Lee}, {Long}, {Loomis}, {Lundgren}, {Lupton}, {Maia},
  {Makler}, {Malanushenko}, {Malanushenko}, {Mandelbaum}, {Manera}, {Maraston},
  {Margala}, {Masters}, {McBride}, {McDonald}, {McGreer}, {McMahon}, {Mena},
  {Miralda-Escud{\'e}}, {Montero-Dorta}, {Montesano}, {Muna}, {Myers},
  {Naugle}, {Nichol}, {Noterdaeme}, {Nuza}, {Olmstead}, {Oravetz}, {Oravetz},
  {Owen}, {Padmanabhan}, {Palanque-Delabrouille}, {Pan}, {Parejko},
  {P{\^a}ris}, {Percival}, {P{\'e}rez-Fournon}, {P{\'e}rez-R{\`a}fols},
  {Petitjean}, {Pfaffenberger}, {Pforr}, {Pieri}, {Prada}, {Price-Whelan},
  {Raddick}, {Rebolo}, {Rich}, {Richards}, {Rockosi}, {Roe}, {Ross}, {Ross},
  {Rossi}, {Rubi{\~n}o-Martin}, {Samushia}, {S{\'a}nchez}, {Sayres}, {Schmidt},
  {Schneider}, {Sc{\'o}ccola}, {Seo}, {Shelden}, {Sheldon}, {Shen}, {Shu},
  {Slosar}, {Smee}, {Snedden}, {Stauffer}, {Steele}, {Strauss}, {Streblyanska},
  {Suzuki}, {Swanson}, {Tal}, {Tanaka}, {Thomas}, {Tinker}, {Tojeiro},
  {Tremonti}, {Vargas Maga{\~n}a}, {Verde}, {Viel}, {Wake}, {Watson}, {Weaver},
  {Weinberg}, {Weiner}, {West}, {White}, {Wood-Vasey}, {Yeche}, {Zehavi},
  {Zhao}, \& {Zheng}}]{Dawson2013}
{Dawson}, K.~S., {Schlegel}, D.~J., {Ahn}, C.~P., {et~al.} 2013, \aj, 145, 10

\bibitem[{{de Lapparent} {et~al.}(1986){de Lapparent}, {Geller}, \&
  {Huchra}}]{Lapparent1986}
{de Lapparent}, V., {Geller}, M.~J., \& {Huchra}, J.~P. 1986, \apjl, 302, L1

\bibitem[{{Disseau} {et~al.}(2014){Disseau}, {Puech}, {Flores}, {Hammer},
  {Yang}, \& {Pentericci}}]{Disseau2014}
{Disseau}, K., {Puech}, M., {Flores}, H., {et~al.} 2014, in \procspie, Vol.
  9147, Ground-based and Airborne Instrumentation for Astronomy V, 914791

\bibitem[{{D'Odorico} {et~al.}(2010){D'Odorico}, {Calura}, {Cristiani}, \&
  {Viel}}]{DOdorico2010}
{D'Odorico}, V., {Calura}, F., {Cristiani}, S., \& {Viel}, M. 2010, \mnras,
  401, 2715

\bibitem[{{D'Odorico} {et~al.}(2016){D'Odorico}, {Cristiani}, {Pomante},
  {Carswell}, {Viel}, {Barai}, {Becker}, {Calura}, {Cupani}, {Fontanot},
  {Haehnelt}, {Kim}, {Miralda-Escud{\'e}}, {Rorai}, {Tescari}, \&
  {Vanzella}}]{DOdorico2016}
{D'Odorico}, V., {Cristiani}, S., {Pomante}, E., {et~al.} 2016, \mnras, 463,
  2690

\bibitem[{{Dubois} {et~al.}(2014){Dubois}, {Pichon}, {Welker}, {Le Borgne},
  {Devriendt}, {Laigle}, {Codis}, {Pogosyan}, {Arnouts}, {Benabed}, {Bertin},
  {Blaizot}, {Bouchet}, {Cardoso}, {Colombi}, {de Lapparent}, {Desjacques},
  {Gavazzi}, {Kassin}, {Kimm}, {McCracken}, {Milliard}, {Peirani}, {Prunet},
  {Rouberol}, {Silk}, {Slyz}, {Sousbie}, {Teyssier}, {Tresse}, {Treyer},
  {Vibert}, \& {Volonteri}}]{Dubois2014}
{Dubois}, Y., {Pichon}, C., {Welker}, C., {et~al.} 2014, \mnras, 444, 1453

\bibitem[{{Eisenstein} {et~al.}(2011){Eisenstein}, {Weinberg}, {Agol},
  {Aihara}, {Allende Prieto}, {Anderson}, {Arns}, {Aubourg}, {Bailey},
  {Balbinot}, \& et~al.}]{Eisenstein2011}
{Eisenstein}, D.~J., {Weinberg}, D.~H., {Agol}, E., {et~al.} 2011, \aj, 142, 72

\bibitem[{{Evans} {et~al.}(2015){Evans}, {Puech}, {Afonso}, {Almaini}, {Amram},
  {Aussel}, {Barbuy}, {Basden}, {Bastian}, {Battaglia}, {Biller}, {Bonifacio},
  {Bouch{\'e}}, {Bunker}, {Caffau}, {Charlot}, {Cirasuolo}, {Clenet}, {Combes},
  {Conselice}, {Contini}, {Cuby}, {Dalton}, {Davies}, {de Koter}, {Disseau},
  {Dunlop}, {Epinat}, {Fiore}, {Feltzing}, {Ferguson}, {Flores}, {Fontana},
  {Fusco}, {Gadotti}, {Gallazzi}, {Gallego}, {Giallongo}, {Gon{\c c}alves},
  {Gratadour}, {Guenther}, {Hammer}, {Hill}, {Huertas-Company}, {Ibata},
  {Kaper}, {Korn}, {Larsen}, {Le F{\`e}vre}, {Lemasle}, {Maraston}, {Mei},
  {Mellier}, {Morris}, {{\"O}stlin}, {Paumard}, {Pello}, {Pentericci},
  {Peroux}, {Petitjean}, {Rodrigues}, {Rodr{\'{\i}}guez-Mu{\~n}oz}, {Rouan},
  {Sana}, {Schaerer}, {Telles}, {Trager}, {Tresse}, {Welikala}, {Zibetti}, \&
  {Ziegler}}]{Evans2015}
{Evans}, C., {Puech}, M., {Afonso}, J., {et~al.} 2015, ArXiv e-prints

\bibitem[{{Falck} {et~al.}(2018){Falck}, {Koyama}, {Zhao}, \&
  {Cautun}}]{falck18}
{Falck}, B., {Koyama}, K., {Zhao}, G.-B., \& {Cautun}, M. 2018, \mnras, 475,
  3262

\bibitem[{{Fuentes-Carrera} {et~al.}(2004){Fuentes-Carrera}, {Rosado}, {Amram},
  {Dultzin-Hacyan}, {Cruz-Gonz{\'a}lez}, {Salo}, {Laurikainen}, {Bernal},
  {Ambrocio-Cruz}, \& {Le Coarer}}]{Fuentes2004}
{Fuentes-Carrera}, I., {Rosado}, M., {Amram}, P., {et~al.} 2004, \aap, 415, 451

\bibitem[{{Ganeshaiah Veena} {et~al.}(2018){Ganeshaiah Veena}, {Cautun}, {van
  de Weygaert}, {Tempel}, {Jones}, {Rieder}, \& {Frenk}}]{veena18}
{Ganeshaiah Veena}, P., {Cautun}, M., {van de Weygaert}, R., {et~al.} 2018,
  \mnras, 481, 414

\bibitem[{{Gardner} {et~al.}(2006){Gardner}, {Mather}, {Clampin}, {Doyon},
  {Greenhouse}, {Hammel}, {Hutchings}, {Jakobsen}, {Lilly}, {Long}, {Lunine},
  {McCaughrean}, {Mountain}, {Nella}, {Rieke}, {Rieke}, {Rix}, {Smith},
  {Sonneborn}, {Stiavelli}, {Stockman}, {Windhorst}, \& {Wright}}]{Gardner2006}
{Gardner}, J.~P., {Mather}, J.~C., {Clampin}, M., {et~al.} 2006, \ssr, 123, 485

\bibitem[{{Garrido} {et~al.}(2004){Garrido}, {Marcelin}, \&
  {Amram}}]{Garrido2004}
{Garrido}, O., {Marcelin}, M., \& {Amram}, P. 2004, \mnras, 349, 225

\bibitem[{{Garrido} {et~al.}(2002){Garrido}, {Marcelin}, {Amram}, \&
  {Boulesteix}}]{Garrido2002}
{Garrido}, O., {Marcelin}, M., {Amram}, P., \& {Boulesteix}, J. 2002, \aap,
  387, 821

\bibitem[{{Gay} {et~al.}(2012){Gay}, {Pichon}, \& {Pogosyan}}]{Gay12}
{Gay}, C., {Pichon}, C., \& {Pogosyan}, D. 2012, \prd, 85, 023011

\bibitem[{{Geller} \& {Huchra}(1989)}]{Geller1989}
{Geller}, M.~J. \& {Huchra}, J.~P. 1989, Science, 246, 897

\bibitem[{{Giavalisco} {et~al.}(2004){Giavalisco}, {Ferguson}, {Koekemoer},
  {Dickinson}, {Alexander}, {Bauer}, {Bergeron}, {Biagetti}, {Brandt},
  {Casertano}, {Cesarsky}, {Chatzichristou}, {Conselice}, {Cristiani}, {Da
  Costa}, {Dahlen}, {de Mello}, {Eisenhardt}, {Erben}, {Fall}, {Fassnacht},
  {Fosbury}, {Fruchter}, {Gardner}, {Grogin}, {Hook}, {Hornschemeier}, {Idzi},
  {Jogee}, {Kretchmer}, {Laidler}, {Lee}, {Livio}, {Lucas}, {Madau},
  {Mobasher}, {Moustakas}, {Nonino}, {Padovani}, {Papovich}, {Park},
  {Ravindranath}, {Renzini}, {Richardson}, {Riess}, {Rosati}, {Schirmer},
  {Schreier}, {Somerville}, {Spinrad}, {Stern}, {Stiavelli}, {Strolger},
  {Urry}, {Vandame}, {Williams}, \& {Wolf}}]{Giavalisco2004}
{Giavalisco}, M., {Ferguson}, H.~C., {Koekemoer}, A.~M., {et~al.} 2004, \apjl,
  600, L93

\bibitem[{{Goh} {et~al.}(2019){Goh}, {Primack}, {Lee}, {Aragon-Calvo},
  {Hellinger}, {Behroozi}, {Rodriguez-Puebla}, {Eckholm}, \&
  {Johnston}}]{Goh2019}
{Goh}, T., {Primack}, J., {Lee}, C.~T., {et~al.} 2019, \mnras, 483, 2101

\bibitem[{{Gott} {et~al.}(1987){Gott}, {Weinberg}, \& {Melott}}]{Gott1987}
{Gott}, III, J.~R., {Weinberg}, D.~H., \& {Melott}, A.~L. 1987, \apj, 319, 1

\bibitem[{{Gunn} \& {Peterson}(1965)}]{Gunn1965}
{Gunn}, J.~E. \& {Peterson}, B.~A. 1965, \apj, 142, 1633

\bibitem[{{Guo} {et~al.}(2013){Guo}, {Ferguson}, {Giavalisco}, {Barro},
  {Willner}, {Ashby}, {Dahlen}, {Donley}, {Faber}, {Fontana}, {Galametz},
  {Grazian}, {Huang}, {Kocevski}, {Koekemoer}, {Koo}, {McGrath}, {Peth},
  {Salvato}, {Wuyts}, {Castellano}, {Cooray}, {Dickinson}, {Dunlop}, {Fazio},
  {Gardner}, {Gawiser}, {Grogin}, {Hathi}, {Hsu}, {Lee}, {Lucas}, {Mobasher},
  {Nandra}, {Newman}, \& {van der Wel}}]{Guo2013}
{Guo}, Y., {Ferguson}, H.~C., {Giavalisco}, M., {et~al.} 2013, \apjs, 207, 24

\bibitem[{{Haardt} \& {Madau}(1996)}]{Haardt1996}
{Haardt}, F. \& {Madau}, P. 1996, \apj, 461, 20

\bibitem[{{Hahn} {et~al.}(2009){Hahn}, {Porciani}, {Dekel}, \&
  {Carollo}}]{hahn2009}
{Hahn}, O., {Porciani}, C., {Dekel}, A., \& {Carollo}, C.~M. 2009, \mnras, 398,
  1742

\bibitem[{{Hamaus} {et~al.}(2016){Hamaus}, {Pisani}, {Sutter}, {Lavaux},
  {Escoffier}, {Wandelt}, \& {Weller}}]{hamaus16}
{Hamaus}, N., {Pisani}, A., {Sutter}, P.~M., {et~al.} 2016, Physical Review
  Letters, 117, 091302

\bibitem[{{Hirv} {et~al.}(2017){Hirv}, {Pelt}, {Saar}, {Tago}, {Tamm},
  {Tempel}, \& {Einasto}}]{hirv2017}
{Hirv}, A., {Pelt}, J., {Saar}, E., {et~al.} 2017, \aap, 599, A31

\bibitem[{{Hoeft} {et~al.}(2006){Hoeft}, {Yepes}, {Gottl{\"o}ber}, \&
  {Springel}}]{hoeft06}
{Hoeft}, M., {Yepes}, G., {Gottl{\"o}ber}, S., \& {Springel}, V. 2006, \mnras,
  371, 401

\bibitem[{{Horowitz} {et~al.}(2019){Horowitz}, {Lee}, {White}, {Krolewski}, \&
  {Ata}}]{horowitz19}
{Horowitz}, B., {Lee}, K.-G., {White}, M., {Krolewski}, A., \& {Ata}, M. 2019,
  arXiv e-prints, arXiv:1903.09049

\bibitem[{{Jagourel} {et~al.}(2018){Jagourel}, {Fitzsimons}, {Hammer}, {De
  Frondat}, {Puech}, {Evans}, {Sanchez}, {Guinouard}, {Chemla}, {Frotin},
  {Yang}, {Parr-Burman}, {Morris}, {Dubbeldam}, {Close}, {Middleton},
  {Rousset}, {Gendron}, {Kelz}, {Janssen}, {Pragt}, {Navarro}, {Larrieu}, {El
  Hadi}, {Dohlen}, {Dalton}, {Lewis}, {Rodrigues}, {Morris}, {Kaper}, {Barbuy},
  {Cuby}, \& {Le F{\`e}vre}}]{Jagourel2018}
{Jagourel}, P., {Fitzsimons}, E., {Hammer}, F., {et~al.} 2018, in Society of
  Photo-Optical Instrumentation Engineers (SPIE) Conference Series, Vol. 10702,
  Ground-based and Airborne Instrumentation for Astronomy VII, 10702A4

\bibitem[{{Jones} {et~al.}(2013){Jones}, {Noll}, {Kausch}, {Szyszka}, \&
  {Kimeswenger}}]{Jones2013}
{Jones}, A., {Noll}, S., {Kausch}, W., {Szyszka}, C., \& {Kimeswenger}, S.
  2013, \aap, 560, A91

\bibitem[{{Katz} {et~al.}(2019){Katz}, {Ramsoy}, {Rosdahl}, {Kimm}, {Blaizot},
  {Haehnelt}, {Michel-Dansac}, {Garel}, {Laigle}, \& {Devriendt}}]{katz2019}
{Katz}, H., {Ramsoy}, M., {Rosdahl}, J., {et~al.} 2019, arXiv e-prints,
  arXiv:1905.11414

\bibitem[{{Kaviraj} {et~al.}(2017){Kaviraj}, {Laigle}, {Kimm}, {Devriendt},
  {Dubois}, {Pichon}, {Slyz}, {Chisari}, \& {Peirani}}]{Kaviraj2017}
{Kaviraj}, S., {Laigle}, C., {Kimm}, T., {et~al.} 2017, \mnras, 467, 4739

\bibitem[{{Kim} {et~al.}(2001){Kim}, {Cristiani}, \& {D'Odorico}}]{Kim2001}
{Kim}, T.-S., {Cristiani}, S., \& {D'Odorico}, S. 2001, \aap, 373, 757

\bibitem[{{Komatsu} {et~al.}(2011){Komatsu}, {Smith}, {Dunkley}, {Bennett},
  {Gold}, {Hinshaw}, {Jarosik}, {Larson}, {Nolta}, {Page}, {Spergel},
  {Halpern}, {Hill}, {Kogut}, {Limon}, {Meyer}, {Odegard}, {Tucker}, {Weiland},
  {Wollack}, \& {Wright}}]{Komatsu2011}
{Komatsu}, E., {Smith}, K.~M., {Dunkley}, J., {et~al.} 2011, \apjs, 192, 18

\bibitem[{{Kraljic} {et~al.}(2018){Kraljic}, {Arnouts}, {Pichon}, {Laigle}, {de
  la Torre}, {Vibert}, {Cadiou}, {Dubois}, {Treyer}, {Schimd}, {Codis}, {de
  Lapparent}, {Devriendt}, {Hwang}, {Le Borgne}, {Malavasi}, {Milliard},
  {Musso}, {Pogosyan}, {Alpaslan}, {Bland-Hawthorn}, \& {Wright}}]{Kraljic2018}
{Kraljic}, K., {Arnouts}, S., {Pichon}, C., {et~al.} 2018, \mnras, 474, 547

\bibitem[{{Kraljic} {et~al.}(2019){Kraljic}, {Pichon}, {Dubois}, {Codis},
  {Cadiou}, {Devriendt}, {Musso}, {Welker}, {Arnouts}, {Hwang}, {Laigle},
  {Peirani}, {Slyz}, {Treyer}, \& {Vibert}}]{kraljic19}
{Kraljic}, K., {Pichon}, C., {Dubois}, Y., {et~al.} 2019, \mnras, 483, 3227

\bibitem[{{Krolewski} {et~al.}(2018){Krolewski}, {Lee}, {White}, {Hennawi},
  {Schlegel}, {Nugent}, {Luki{\'c}}, {Stark}, {Koekemoer}, {Le F{\`e}vre},
  {Lemaux}, {Maier}, {Rich}, {Salvato}, \& {Tasca}}]{krowleski18}
{Krolewski}, A., {Lee}, K.-G., {White}, M., {et~al.} 2018, \apj, 861, 60

\bibitem[{{Kuutma} {et~al.}(2017){Kuutma}, {Tamm}, \& {Tempel}}]{Kuutma2017}
{Kuutma}, T., {Tamm}, A., \& {Tempel}, E. 2017, \aap, 600, L6

\bibitem[{{Laigle} {et~al.}(2019){Laigle}, {Davidzon}, {Ilbert}, {Devriendt},
  {Kashino}, {Pichon}, {Capak}, {Arnouts}, {de la Torre}, \&
  {Dubois}}]{laigle19}
{Laigle}, C., {Davidzon}, I., {Ilbert}, O., {et~al.} 2019, \mnras, 486, 5104

\bibitem[{{Laigle} {et~al.}(2016){Laigle}, {McCracken}, {Ilbert}, {Hsieh},
  {Davidzon}, {Capak}, {Hasinger}, {Silverman}, {Pichon}, {Coupon}, {Aussel},
  {Le Borgne}, {Caputi}, {Cassata}, {Chang}, {Civano}, {Dunlop}, {Fynbo},
  {Kartaltepe}, {Koekemoer}, {Le F{\`e}vre}, {Le Floc'h}, {Leauthaud}, {Lilly},
  {Lin}, {Marchesi}, {Milvang-Jensen}, {Salvato}, {Sanders}, {Scoville},
  {Smolcic}, {Stockmann}, {Taniguchi}, {Tasca}, {Toft}, {Vaccari}, \&
  {Zabl}}]{Laigle2016}
{Laigle}, C., {McCracken}, H.~J., {Ilbert}, O., {et~al.} 2016, \apjs, 224, 24

\bibitem[{{Laigle} {et~al.}(2018){Laigle}, {Pichon}, {Arnouts}, {McCracken},
  {Dubois}, {Devriendt}, {Slyz}, {Le Borgne}, {Benoit-L{\'e}vy}, {Hwang},
  {Ilbert}, {Kraljic}, {Malavasi}, {Park}, \& {Vibert}}]{Laigle2018}
{Laigle}, C., {Pichon}, C., {Arnouts}, S., {et~al.} 2018, \mnras, 474, 5437

\bibitem[{{Laigle} {et~al.}(2015){Laigle}, {Pichon}, {Codis}, {Dubois}, {Le
  Borgne}, {Pogosyan}, {Devriendt}, {Peirani}, {Prunet}, {Rouberol}, {Slyz}, \&
  {Sousbie}}]{laigle15}
{Laigle}, C., {Pichon}, C., {Codis}, S., {et~al.} 2015, \mnras, 446, 2744

\bibitem[{{Laureijs} {et~al.}(2011){Laureijs}, {Amiaux}, {Arduini},
  {Augu{\`e}res}, {Brinchmann}, {Cole}, {Cropper}, {Dabin}, {Duvet}, {Ealet},
  {Garilli}, {Gondoin}, {Guzzo}, {Hoar}, {Hoekstra}, {Holmes}, {Kitching},
  {Maciaszek}, {Mellier}, {Pasian}, {Percival}, {Rhodes}, {Saavedra Criado},
  {Sauvage}, {Scaramella}, {Valenziano}, {Warren}, {Bender}, {Castander},
  {Cimatti}, {Le F{\`e}vre}, {Kurki-Suonio}, {Levi}, {Lilje}, {Meylan},
  {Nichol}, {Pedersen}, {Popa}, {Rebolo Lopez}, {Rix}, {Rottgering},
  {Zeilinger}, {Grupp}, {Hudelot}, {Massey}, {Meneghetti}, {Miller}, {Paltani},
  {Paulin-Henriksson}, {Pires}, {Saxton}, {Schrabback}, {Seidel}, {Walsh},
  {Aghanim}, {Amendola}, {Bartlett}, {Baccigalupi}, {Beaulieu}, {Benabed},
  {Cuby}, {Elbaz}, {Fosalba}, {Gavazzi}, {Helmi}, {Hook}, {Irwin}, {Kneib},
  {Kunz}, {Mannucci}, {Moscardini}, {Tao}, {Teyssier}, {Weller}, {Zamorani},
  {Zapatero Osorio}, {Boulade}, {Foumond}, {Di Giorgio}, {Guttridge}, {James},
  {Kemp}, {Martignac}, {Spencer}, {Walton}, {Bl{\"u}mchen}, {Bonoli},
  {Bortoletto}, {Cerna}, {Corcione}, {Fabron}, {Jahnke}, {Ligori}, {Madrid},
  {Martin}, {Morgante}, {Pamplona}, {Prieto}, {Riva}, {Toledo}, {Trifoglio},
  {Zerbi}, {Abdalla}, {Douspis}, {Grenet}, {Borgani}, {Bouwens}, {Courbin},
  {Delouis}, {Dubath}, {Fontana}, {Frailis}, {Grazian}, {Koppenh{\"o}fer},
  {Mansutti}, {Melchior}, {Mignoli}, {Mohr}, {Neissner}, {Noddle}, {Poncet},
  {Scodeggio}, {Serrano}, {Shane}, {Starck}, {Surace}, {Taylor},
  {Verdoes-Kleijn}, {Vuerli}, {Williams}, {Zacchei}, {Altieri}, {Escudero
  Sanz}, {Kohley}, {Oosterbroek}, {Astier}, {Bacon}, {Bardelli}, {Baugh},
  {Bellagamba}, {Benoist}, {Bianchi}, {Biviano}, {Branchini}, {Carbone},
  {Cardone}, {Clements}, {Colombi}, {Conselice}, {Cresci}, {Deacon}, {Dunlop},
  {Fedeli}, {Fontanot}, {Franzetti}, {Giocoli}, {Garcia-Bellido}, {Gow},
  {Heavens}, {Hewett}, {Heymans}, {Holland}, {Huang}, {Ilbert}, {Joachimi},
  {Jennins}, {Kerins}, {Kiessling}, {Kirk}, {Kotak}, {Krause}, {Lahav}, {van
  Leeuwen}, {Lesgourgues}, {Lombardi}, {Magliocchetti}, {Maguire}, {Majerotto},
  {Maoli}, {Marulli}, {Maurogordato}, {McCracken}, {McLure}, {Melchiorri},
  {Merson}, {Moresco}, {Nonino}, {Norberg}, {Peacock}, {Pello}, {Penny},
  {Pettorino}, {Di Porto}, {Pozzetti}, {Quercellini}, {Radovich}, {Rassat},
  {Roche}, {Ronayette}, {Rossetti}, {Sartoris}, {Schneider}, {Semboloni},
  {Serjeant}, {Simpson}, {Skordis}, {Smadja}, {Smartt}, {Spano}, {Spiro},
  {Sullivan}, {Tilquin}, {Trotta}, {Verde}, {Wang}, {Williger}, {Zhao},
  {Zoubian}, \& {Zucca}}]{Laureijs2011}
{Laureijs}, R., {Amiaux}, J., {Arduini}, S., {et~al.} 2011, arXiv e-prints,
  arXiv:1110.3193

\bibitem[{{Lavaux} \& {Wandelt}(2012)}]{lavaux12}
{Lavaux}, G. \& {Wandelt}, B.~D. 2012, \apj, 754, 109

\bibitem[{{Lee} {et~al.}(2014{\natexlab{a}}){Lee}, {Hennawi}, {Stark},
  {Prochaska}, {White}, {Schlegel}, {Eilers}, {Arinyo-i-Prats}, {Suzuki},
  {Croft}, {Caputi}, {Cassata}, {Ilbert}, {Garilli}, {Koekemoer}, {Le Brun},
  {Le F{\`e}vre}, {Maccagni}, {Nugent}, {Taniguchi}, {Tasca}, {Tresse},
  {Zamorani}, \& {Zucca}}]{Lee2014b}
{Lee}, K.-G., {Hennawi}, J.~F., {Stark}, C., {et~al.} 2014{\natexlab{a}},
  \apjl, 795, L12

\bibitem[{{Lee} {et~al.}(2014{\natexlab{b}}){Lee}, {Hennawi}, {White}, {Croft},
  \& {Ozbek}}]{Lee2014a}
{Lee}, K.-G., {Hennawi}, J.~F., {White}, M., {Croft}, R. A.~C., \& {Ozbek}, M.
  2014{\natexlab{b}}, \apj, 788, 49

\bibitem[{{Lee} {et~al.}(2016){Lee}, {Hennawi}, {White}, {Prochaska},
  {Font-Ribera}, {Schlegel}, {Rich}, {Suzuki}, {Stark}, {Le F{\`e}vre},
  {Nugent}, {Salvato}, \& {Zamorani}}]{Lee2016}
{Lee}, K.-G., {Hennawi}, J.~F., {White}, M., {et~al.} 2016, \apj, 817, 160

\bibitem[{{Lee} {et~al.}(2018){Lee}, {Krolewski}, {White}, {Schlegel},
  {Nugent}, {Hennawi}, {M{\"u}ller}, {Pan}, {Prochaska}, {Font-Ribera},
  {Suzuki}, {Glazebrook}, {Kacprzak}, {Kartaltepe}, {Koekemoer}, {Le
  F{\`e}vre}, {Lemaux}, {Maier}, {Nanayakkara}, {Rich}, {Sanders}, {Salvato},
  {Tasca}, \& {Tran}}]{Lee2018}
{Lee}, K.-G., {Krolewski}, A., {White}, M., {et~al.} 2018, \apjs, 237, 31

\bibitem[{{Libeskind} {et~al.}(2013){Libeskind}, {Hoffman}, {Steinmetz},
  {Gottl{\"o}ber}, {Knebe}, \& {Hess}}]{libeskind13}
{Libeskind}, N.~I., {Hoffman}, Y., {Steinmetz}, M., {et~al.} 2013, \apjl, 766,
  L15

\bibitem[{{Lindner} {et~al.}(1996){Lindner}, {Einasto}, {Einasto}, {Freudling},
  {Fricke}, {Lipovetsky}, {Pustilnik}, {Izotov}, \& {Richter}}]{lindner1996}
{Lindner}, U., {Einasto}, M., {Einasto}, J., {et~al.} 1996, \aap, 314, 1

\bibitem[{{LSST Science Collaboration} {et~al.}(2009){LSST Science
  Collaboration}, {Abell}, {Allison}, {Anderson}, {Andrew}, {Angel}, {Armus},
  {Arnett}, {Asztalos}, {Axelrod}, {Bailey}, {Ballantyne}, {Bankert},
  {Barkhouse}, {Barr}, {Barrientos}, {Barth}, {Bartlett}, {Becker}, {Becla},
  {Beers}, {Bernstein}, {Biswas}, {Blanton}, {Bloom}, {Bochanski}, {Boeshaar},
  {Borne}, {Bradac}, {Brandt}, {Bridge}, {Brown}, {Brunner}, {Bullock},
  {Burgasser}, {Burge}, {Burke}, {Cargile}, {Chand rasekharan}, {Chartas},
  {Chesley}, {Chu}, {Cinabro}, {Claire}, {Claver}, {Clowe}, {Connolly}, {Cook},
  {Cooke}, {Cooray}, {Covey}, {Culliton}, {de Jong}, {de Vries}, {Debattista},
  {Delgado}, {Dell'Antonio}, {Dhital}, {Di Stefano}, {Dickinson}, {Dilday},
  {Djorgovski}, {Dobler}, {Donalek}, {Dubois-Felsmann}, {Durech},
  {Eliasdottir}, {Eracleous}, {Eyer}, {Falco}, {Fan}, {Fassnacht}, {Ferguson},
  {Fernandez}, {Fields}, {Finkbeiner}, {Figueroa}, {Fox}, {Francke}, {Frank},
  {Frieman}, {Fromenteau}, {Furqan}, {Galaz}, {Gal-Yam}, {Garnavich},
  {Gawiser}, {Geary}, {Gee}, {Gibson}, {Gilmore}, {Grace}, {Green}, {Gressler},
  {Grillmair}, {Habib}, {Haggerty}, {Hamuy}, {Harris}, {Hawley}, {Heavens},
  {Hebb}, {Henry}, {Hileman}, {Hilton}, {Hoadley}, {Holberg}, {Holman},
  {Howell}, {Infante}, {Ivezic}, {Jacoby}, {Jain}, {R}, {Jedicke}, {Jee},
  {Garrett Jernigan}, {Jha}, {Johnston}, {Jones}, {Juric}, {Kaasalainen},
  {Styliani}, {Kafka}, {Kahn}, {Kaib}, {Kalirai}, {Kantor}, {Kasliwal},
  {Keeton}, {Kessler}, {Knezevic}, {Kowalski}, {Krabbendam}, {Krughoff},
  {Kulkarni}, {Kuhlman}, {Lacy}, {Lepine}, {Liang}, {Lien}, {Lira}, {Long},
  {Lorenz}, {Lotz}, {Lupton}, {Lutz}, {Macri}, {Mahabal}, {Mandelbaum},
  {Marshall}, {May}, {McGehee}, {Meadows}, {Meert}, {Milani}, {Miller},
  {Miller}, {Mills}, {Minniti}, {Monet}, {Mukadam}, {Nakar}, {Neill}, {Newman},
  {Nikolaev}, {Nordby}, {O'Connor}, {Oguri}, {Oliver}, {Olivier}, {Olsen},
  {Olsen}, {Olszewski}, {Oluseyi}, {Padilla}, {Parker}, {Pepper}, {Peterson},
  {Petry}, {Pinto}, {Pizagno}, {Popescu}, {Prsa}, {Radcka}, {Raddick},
  {Rasmussen}, {Rau}, {Rho}, {Rhoads}, {Richards}, {Ridgway}, {Robertson},
  {Roskar}, {Saha}, {Sarajedini}, {Scannapieco}, {Schalk}, {Schindler},
  {Schmidt}, {Schmidt}, {Schneider}, {Schumacher}, {Scranton}, {Sebag},
  {Seppala}, {Shemmer}, {Simon}, {Sivertz}, {Smith}, {Allyn Smith}, {Smith},
  {Spitz}, {Stanford}, {Stassun}, {Strader}, {Strauss}, {Stubbs}, {Sweeney},
  {Szalay}, {Szkody}, {Takada}, {Thorman}, {Trilling}, {Trimble}, {Tyson}, {Van
  Berg}, {Vand en Berk}, {VanderPlas}, {Verde}, {Vrsnak}, {Walkowicz}, {Wand
  elt}, {Wang}, {Wang}, {Warner}, {Wechsler}, {West}, {Wiecha}, {Williams},
  {Willman}, {Wittman}, {Wolff}, {Wood-Vasey}, {Wozniak}, {Young}, {Zentner},
  \& {Zhan}}]{LSST2009}
{LSST Science Collaboration}, {Abell}, P.~A., {Allison}, J., {et~al.} 2009,
  arXiv e-prints, arXiv:0912.0201

\bibitem[{{Malavasi} {et~al.}(2017){Malavasi}, {Arnouts}, {Vibert}, {de la
  Torre}, {Moutard}, {Pichon}, {Davidzon}, {Kraljic}, {Bolzonella}, {Guzzo},
  {Garilli}, {Scodeggio}, {Granett}, {Abbas}, {Adami}, {Bottini}, {Cappi},
  {Cucciati}, {Franzetti}, {Fritz}, {Iovino}, {Krywult}, {Le Brun}, {Le
  F{\`e}vre}, {Maccagni}, {Ma{\l}ek}, {Marulli}, {Polletta}, {Pollo}, {Tasca},
  {Tojeiro}, {Vergani}, {Zanichelli}, {Bel}, {Branchini}, {Coupon}, {De Lucia},
  {Dubois}, {Hawken}, {Ilbert}, {Laigle}, {Moscardini}, {Sousbie}, {Treyer}, \&
  {Zamorani}}]{Malavasi2017}
{Malavasi}, N., {Arnouts}, S., {Vibert}, D., {et~al.} 2017, \mnras, 465, 3817

\bibitem[{{Matsubara}(1995)}]{matsubara95}
{Matsubara}, T. 1995, in Large Scale Structure in the Universe, ed. J.~P.
  {M{\"u}cket}, S.~{Gottloeber}, \& V.~{M{\"u}ller}, 162

\bibitem[{{McQuinn}(2016)}]{McQuinn2016}
{McQuinn}, M. 2016, \araa, 54, 313

\bibitem[{{Melott} {et~al.}(1988){Melott}, {Weinberg}, \& {Gott}}]{Melott1988}
{Melott}, A.~L., {Weinberg}, D.~H., \& {Gott}, III, J.~R. 1988, \apj, 328, 50

\bibitem[{{Momcheva} {et~al.}(2016){Momcheva}, {Brammer}, {van Dokkum},
  {Skelton}, {Whitaker}, {Nelson}, {Fumagalli}, {Maseda}, {Leja}, {Franx},
  {Rix}, {Bezanson}, {Da Cunha}, {Dickey}, {F{\"o}rster Schreiber},
  {Illingworth}, {Kriek}, {Labb{\'e}}, {Ulf Lange}, {Lundgren}, {Magee},
  {Marchesini}, {Oesch}, {Pacifici}, {Patel}, {Price}, {Tal}, {Wake}, {van der
  Wel}, \& {Wuyts}}]{Momcheva2016}
{Momcheva}, I.~G., {Brammer}, G.~B., {van Dokkum}, P.~G., {et~al.} 2016, \apjs,
  225, 27

\bibitem[{{Moorman} {et~al.}(2016){Moorman}, {Moreno}, {White}, {Vogeley},
  {Hoyle}, {Giovanelli}, \& {Haynes}}]{Moorman16}
{Moorman}, C.~M., {Moreno}, J., {White}, A., {et~al.} 2016, \apj, 831, 118

\bibitem[{{Morris} {et~al.}(2018){Morris}, {Hammer}, {Jagourel}, {Evans},
  {Puech}, {Dalton}, {Rodrigues}, {Sanchez-Janssen}, {Fitzsimons}, {Barbuy},
  {Cuby}, {Kaper}, {Roth}, {Rousset}, {Myers}, {Le F{\`e}vre}, {Finogenov},
  {Kotilainen}, {Castilho}, {Ostlin}, {Feltzing}, {Korn}, {Gallego},
  {Castillo}, {Iglesias-P{\'a}ramo}, {Pentericci}, {Ziegler}, {Afonso},
  {Dubbledam}, {Close}, {Parr-Burman}, {Morris}, {Chemla}, {De Frondat},
  {Kelz}, {Guinouard}, {Lewis}, {Middleton}, {Navarro}, {Larrieu}, {Pragt},
  {Janssen}, {Dohlen}, {El Hadi}, {Gendron}, {Yang}, {Wells}, {Conan}, {Fusco},
  {Schaerer}, {Bergin}, {Taburet}, {Frotin}, \& {Berkourn}}]{Morris2018}
{Morris}, S., {Hammer}, F., {Jagourel}, P., {et~al.} 2018, in Society of
  Photo-Optical Instrumentation Engineers (SPIE) Conference Series, Vol. 10702,
  Society of Photo-Optical Instrumentation Engineers (SPIE) Conference Series,
  107021W

\bibitem[{{Musso} {et~al.}(2018){Musso}, {Cadiou}, {Pichon}, {Codis},
  {Kraljic}, \& {Dubois}}]{Musso2018}
{Musso}, M., {Cadiou}, C., {Pichon}, C., {et~al.} 2018, \mnras, 476, 4877

\bibitem[{{Noll} {et~al.}(2012){Noll}, {Kausch}, {Barden}, {Jones}, {Szyszka},
  {Kimeswenger}, \& {Vinther}}]{Noll2012}
{Noll}, S., {Kausch}, W., {Barden}, M., {et~al.} 2012, \aap, 543, A92

\bibitem[{{Ozbek} {et~al.}(2016){Ozbek}, {Croft}, \& {Khandai}}]{Ozbek2016}
{Ozbek}, M., {Croft}, R.~A.~C., \& {Khandai}, N. 2016, \mnras, 456, 3610

\bibitem[{{P{\^a}ris} {et~al.}(2012){P{\^a}ris}, {Petitjean}, {Aubourg},
  {Bailey}, {Ross}, {Myers}, {Strauss}, {Anderson}, {Arnau}, {Bautista},
  {Bizyaev}, {Bolton}, {Bovy}, {Brandt}, {Brewington}, {Browstein}, {Busca},
  {Capellupo}, {Carithers}, {Croft}, {Dawson}, {Delubac}, {Ebelke},
  {Eisenstein}, {Engelke}, {Fan}, {Filiz Ak}, {Finley}, {Font-Ribera}, {Ge},
  {Gibson}, {Hall}, {Hamann}, {Hennawi}, {Ho}, {Hogg}, {Ivezi{\'c}}, {Jiang},
  {Kimball}, {Kirkby}, {Kirkpatrick}, {Lee}, {Le Goff}, {Lundgren}, {MacLeod},
  {Malanushenko}, {Malanushenko}, {Maraston}, {McGreer}, {McMahon},
  {Miralda-Escud{\'e}}, {Muna}, {Noterdaeme}, {Oravetz},
  {Palanque-Delabrouille}, {Pan}, {Perez-Fournon}, {Pieri}, {Richards},
  {Rollinde}, {Sheldon}, {Schlegel}, {Schneider}, {Slosar}, {Shelden}, {Shen},
  {Simmons}, {Snedden}, {Suzuki}, {Tinker}, {Viel}, {Weaver}, {Weinberg},
  {White}, {Wood-Vasey}, \& {Y{\`e}che}}]{Paris2012}
{P{\^a}ris}, I., {Petitjean}, P., {Aubourg}, {\'E}., {et~al.} 2012, \aap, 548,
  A66

\bibitem[{{Park} {et~al.}(1992){Park}, {Gott}, {Melott}, \&
  {Karachentsev}}]{park92}
{Park}, C., {Gott}, III, J.~R., {Melott}, A.~L., \& {Karachentsev}, I.~D. 1992,
  \apj, 387, 1

\bibitem[{{Perrotta} {et~al.}(2016){Perrotta}, {D'Odorico}, {Prochaska},
  {Cristiani}, {Cupani}, {Ellison}, {L{\'o}pez}, {Becker}, {Berg},
  {Christensen}, {Denney}, {Hamann}, {P{\^a}ris}, {Vestergaard}, \&
  {Worseck}}]{Perrotta2016MNRAS}
{Perrotta}, S., {D'Odorico}, V., {Prochaska}, J.~X., {et~al.} 2016, \mnras,
  462, 3285

\bibitem[{{Petitjean} {et~al.}(1995){Petitjean}, {Mueket}, \&
  {Kates}}]{Petitjean1995}
{Petitjean}, P., {Mueket}, J.~P., \& {Kates}, R.~E. 1995, \aap, 295, L9

\bibitem[{{Pichon} {et~al.}(2001){Pichon}, {Vergely}, {Rollinde}, {Colombi}, \&
  {Petitjean}}]{Pichon2001}
{Pichon}, C., {Vergely}, J.~L., {Rollinde}, E., {Colombi}, S., \& {Petitjean},
  P. 2001, \mnras, 326, 597

\bibitem[{{Pieri} {et~al.}(2016){Pieri}, {Bonoli}, {Chaves-Montero},
  {P{\^a}ris}, {Fumagalli}, {Bolton}, {Viel}, {Noterdaeme},
  {Miralda-Escud{\'e}}, {Busca}, {Rahmani}, {Peroux}, {Font-Ribera}, \&
  {Trager}}]{Pieri2016}
{Pieri}, M.~M., {Bonoli}, S., {Chaves-Montero}, J., {et~al.} 2016, in
  SF2A-2016: Proceedings of the Annual meeting of the French Society of
  Astronomy and Astrophysics, ed. C.~{Reyl{\'e}}, J.~{Richard},
  L.~{Cambr{\'e}sy}, M.~{Deleuil}, E.~{P{\'e}contal}, L.~{Tresse}, \&
  I.~{Vauglin}, 259--266

\bibitem[{{Pontoppidan} {et~al.}(2016){Pontoppidan}, {Pickering}, {Laidler},
  {Gilbert}, {Sontag}, {Slocum}, {Sienkiewicz}, {Hanley}, {Earl}, {Pueyo},
  {Ravindranath}, {Karakla}, {Robberto}, {Noriega-Crespo}, \&
  {Barker}}]{Pontoppidan2016}
{Pontoppidan}, K.~M., {Pickering}, T.~E., {Laidler}, V.~G., {et~al.} 2016, in
  Society of Photo-Optical Instrumentation Engineers (SPIE) Conference Series,
  Vol. 9910, Observatory Operations: Strategies, Processes, and Systems VI,
  991016

\bibitem[{{Puech} {et~al.}(2018){Puech}, {Evans}, {Disseau}, {Japelj},
  {Ram{\'{\i}}rez-Agudelo}, {Rahmani}, {Trevisan}, {Wang}, {Rodrigues},
  {S{\'a}nchez-Janssen}, {Yang}, {Hammer}, {Kaper}, {Morris}, {Barbuy}, {Cuby},
  {Dalton}, {Fitzsimons}, \& {Jagourel}}]{Puech2018}
{Puech}, M., {Evans}, C.~J., {Disseau}, K., {et~al.} 2018, in Society of
  Photo-Optical Instrumentation Engineers (SPIE) Conference Series, Vol. 10702,
  Society of Photo-Optical Instrumentation Engineers (SPIE) Conference Series,
  107028R

\bibitem[{{Puech} {et~al.}(2012){Puech}, {Flores}, {Yang}, {Rodrigues}, {Gon{\c
  c}alves}, {Hammer}, \& {Disseau}}]{Puech2012}
{Puech}, M., {Flores}, H., {Yang}, Y.~B., {et~al.} 2012, in \procspie, Vol.
  8446, Ground-based and Airborne Instrumentation for Astronomy IV, 84467L

\bibitem[{{Puech} {et~al.}(2010){Puech}, {Rosati}, {Toft}, {Cimatti},
  {Neichel}, \& {Fusco}}]{Puech2010}
{Puech}, M., {Rosati}, P., {Toft}, S., {et~al.} 2010, \mnras, 402, 903

\bibitem[{{Puech} {et~al.}(2016){Puech}, {Yang}, {J{\'e}gouzo}, {Marchal},
  {Disseau}, {Paillous}, {Rodrigues}, {Taburet}, {Cl{\'e}net}, {Gratadour},
  {Flores}, \& {Hammer}}]{Puech2016}
{Puech}, M., {Yang}, Y., {J{\'e}gouzo}, I., {et~al.} 2016, in \procspie, Vol.
  9908, Ground-based and Airborne Instrumentation for Astronomy VI, 99089P

\bibitem[{{Reddy} \& {Steidel}(2009)}]{Reddy2009}
{Reddy}, N.~A. \& {Steidel}, C.~C. 2009, \apj, 692, 778

\bibitem[{{Rhodes} {et~al.}(2017){Rhodes}, {Nichol}, {Aubourg}, {Bean},
  {Boutigny}, {Bremer}, {Capak}, {Cardone}, {Carry}, {Conselice}, {Connolly},
  {Cuilland re}, {Hatch}, {Helou}, {Hemmati}, {Hildebrandt}, {Hlo{\v{z}}ek},
  {Jones}, {Kahn}, {Kiessling}, {Kitching}, {Lupton}, {Mand elbaum},
  {Markovic}, {Marshall}, {Massey}, {Maughan}, {Melchior}, {Mellier}, {Newman},
  {Robertson}, {Sauvage}, {Schrabback}, {Smith}, {Strauss}, {Taylor}, \& {Von
  Der Linden}}]{Rhodes2017}
{Rhodes}, J., {Nichol}, R.~C., {Aubourg}, {\'E}., {et~al.} 2017, \apjs, 233, 21

\bibitem[{{Rieke} {et~al.}(2015){Rieke}, {Wright}, {B{\"o}ker}, {Bouwman},
  {Colina}, {Glasse}, {Gordon}, {Greene}, {G{\"u}del}, {Henning}, {Justtanont},
  {Lagage}, {Meixner}, {N{\o}rgaard-Nielsen}, {Ray}, {Ressler}, {van Dishoeck},
  \& {Waelkens}}]{Rieke2015}
{Rieke}, G.~H., {Wright}, G.~S., {B{\"o}ker}, T., {et~al.} 2015, \pasp, 127,
  584

\bibitem[{{Rigby} {et~al.}(2018){Rigby}, {Bayliss}, {Chisholm}, {Bordoloi},
  {Sharon}, {Gladders}, {Johnson}, {Paterno-Mahler}, {Wuyts}, {Dahle}, \&
  {Acharyya}}]{Rigby18}
{Rigby}, J.~R., {Bayliss}, M.~B., {Chisholm}, J., {et~al.} 2018, \apj, 853, 87

\bibitem[{{Rodrigues} {et~al.}(2010){Rodrigues}, {Flores}, {Puech}, {Yang}, \&
  {Royer}}]{Rodrigues2010}
{Rodrigues}, M., {Flores}, H., {Puech}, M., {Yang}, Y., \& {Royer}, F. 2010, in
  \procspie, Vol. 7735, Ground-based and Airborne Instrumentation for Astronomy
  III, 77356L

\bibitem[{{Rojas} {et~al.}(2004){Rojas}, {Vogeley}, {Hoyle}, \&
  {Brinkmann}}]{Rojas2004}
{Rojas}, R.~R., {Vogeley}, M.~S., {Hoyle}, F., \& {Brinkmann}, J. 2004, \apj,
  617, 50

\bibitem[{{Rosales-Ortega} {et~al.}(2012){Rosales-Ortega}, {Arribas}, \&
  {Colina}}]{RosalesOrtega2012}
{Rosales-Ortega}, F.~F., {Arribas}, S., \& {Colina}, L. 2012, \aap, 539, A73

\bibitem[{{Schechter}(1976)}]{Schechter1976}
{Schechter}, P. 1976, \apj, 203, 297

\bibitem[{{Schmidt} {et~al.}(2019){Schmidt}, {Hennawi}, {Lee}, {Luki{\'c}},
  {O{\~n}orbe}, \& {White}}]{Schmidt2019}
{Schmidt}, T.~M., {Hennawi}, J.~F., {Lee}, K.-G., {et~al.} 2019, \apj, 882, 165

\bibitem[{{Scoville} {et~al.}(2007){Scoville}, {Aussel}, {Brusa}, {Capak},
  {Carollo}, {Elvis}, {Giavalisco}, {Guzzo}, {Hasinger}, {Impey}, {Kneib},
  {LeFevre}, {Lilly}, {Mobasher}, {Renzini}, {Rich}, {Sanders}, {Schinnerer},
  {Schminovich}, {Shopbell}, {Taniguchi}, \& {Tyson}}]{Scoville2007}
{Scoville}, N., {Aussel}, H., {Brusa}, M., {et~al.} 2007, \apjs, 172, 1

\bibitem[{{Shapley} {et~al.}(2003){Shapley}, {Steidel}, {Pettini}, \&
  {Adelberger}}]{Shapley2003}
{Shapley}, A.~E., {Steidel}, C.~C., {Pettini}, M., \& {Adelberger}, K.~L. 2003,
  \apj, 588, 65

\bibitem[{{Shibuya} {et~al.}(2015){Shibuya}, {Ouchi}, \&
  {Harikane}}]{Shibuya2015}
{Shibuya}, T., {Ouchi}, M., \& {Harikane}, Y. 2015, \apjs, 219, 15

\bibitem[{{Skelton} {et~al.}(2014){Skelton}, {Whitaker}, {Momcheva}, {Brammer},
  {van Dokkum}, {Labb{\'e}}, {Franx}, {van der Wel}, {Bezanson}, {Da Cunha},
  {Fumagalli}, {F{\"o}rster Schreiber}, {Kriek}, {Leja}, {Lundgren}, {Magee},
  {Marchesini}, {Maseda}, {Nelson}, {Oesch}, {Pacifici}, {Patel}, {Price},
  {Rix}, {Tal}, {Wake}, \& {Wuyts}}]{Skelton2014}
{Skelton}, R.~E., {Whitaker}, K.~E., {Momcheva}, I.~G., {et~al.} 2014, \apjs,
  214, 24

\bibitem[{{Sousbie}(2011)}]{Sousbie2011}
{Sousbie}, T. 2011, \mnras, 414, 350

\bibitem[{{Sousbie} {et~al.}(2011){Sousbie}, {Pichon}, \&
  {Kawahara}}]{sousbieetal2011}
{Sousbie}, T., {Pichon}, C., \& {Kawahara}, H. 2011, \mnras, 414, 384

\bibitem[{{Springel} {et~al.}(2006){Springel}, {Frenk}, \&
  {White}}]{Springel2006}
{Springel}, V., {Frenk}, C.~S., \& {White}, S.~D.~M. 2006, \nat, 440, 1137

\bibitem[{{Stark} {et~al.}(2015){Stark}, {Font-Ribera}, {White}, \&
  {Lee}}]{Stark2015b}
{Stark}, C.~W., {Font-Ribera}, A., {White}, M., \& {Lee}, K.-G. 2015, \mnras,
  453, 4311

\bibitem[{{Steidel} {et~al.}(1996){Steidel}, {Giavalisco}, {Pettini},
  {Dickinson}, \& {Adelberger}}]{Steidel1996}
{Steidel}, C.~C., {Giavalisco}, M., {Pettini}, M., {Dickinson}, M., \&
  {Adelberger}, K.~L. 1996, \apjl, 462, L17

\bibitem[{{Takada} {et~al.}(2014){Takada}, {Ellis}, {Chiba}, {Greene},
  {Aihara}, {Arimoto}, {Bundy}, {Cohen}, {Dor{\'e}}, {Graves}, {Gunn},
  {Heckman}, {Hirata}, {Ho}, {Kneib}, {Le F{\`e}vre}, {Lin}, {More},
  {Murayama}, {Nagao}, {Ouchi}, {Seiffert}, {Silverman}, {Sodr{\'e}},
  {Spergel}, {Strauss}, {Sugai}, {Suto}, {Takami}, \& {Wyse}}]{Takada2014}
{Takada}, M., {Ellis}, R.~S., {Chiba}, M., {et~al.} 2014, \pasj, 66, R1

\bibitem[{{Tegmark} {et~al.}(2004){Tegmark}, {Blanton}, {Strauss}, {Hoyle},
  {Schlegel}, {Scoccimarro}, {Vogeley}, {Weinberg}, {Zehavi}, {Berlind},
  {Budavari}, {Connolly}, {Eisenstein}, {Finkbeiner}, {Frieman}, {Gunn},
  {Hamilton}, {Hui}, {Jain}, {Johnston}, {Kent}, {Lin}, {Nakajima}, {Nichol},
  {Ostriker}, {Pope}, {Scranton}, {Seljak}, {Sheth}, {Stebbins}, {Szalay},
  {Szapudi}, {Verde}, {Xu}, {Annis}, {Bahcall}, {Brinkmann}, {Burles},
  {Castander}, {Csabai}, {Loveday}, {Doi}, {Fukugita}, {Gott}, {Hennessy},
  {Hogg}, {Ivezi{\'c}}, {Knapp}, {Lamb}, {Lee}, {Lupton}, {McKay}, {Kunszt},
  {Munn}, {O'Connell}, {Peoples}, {Pier}, {Richmond}, {Rockosi}, {Schneider},
  {Stoughton}, {Tucker}, {Vanden Berk}, {Yanny}, {York}, \& {SDSS
  Collaboration}}]{Tegmark2004}
{Tegmark}, M., {Blanton}, M.~R., {Strauss}, M.~A., {et~al.} 2004, \apj, 606,
  702

\bibitem[{{Tempel} \& {Libeskind}(2013)}]{Tempel2013}
{Tempel}, E. \& {Libeskind}, N.~I. 2013, \apjl, 775, L42

\bibitem[{{Turner} {et~al.}(2014){Turner}, {Schaye}, {Steidel}, {Rudie}, \&
  {Strom}}]{Turner2014}
{Turner}, M.~L., {Schaye}, J., {Steidel}, C.~C., {Rudie}, G.~C., \& {Strom},
  A.~L. 2014, \mnras, 445, 794

\bibitem[{{Tytler} {et~al.}(2009){Tytler}, {Gleed}, {Melis}, {Chapman},
  {Kirkman}, {Lubin}, {Paschos}, {Jena}, \& {Crotts}}]{Tytler2009}
{Tytler}, D., {Gleed}, M., {Melis}, C., {et~al.} 2009, \mnras, 392, 1539

\bibitem[{{Vernet} {et~al.}(2011){Vernet}, {Dekker}, {D'Odorico}, {Kaper},
  {Kjaergaard}, {Hammer}, {Randich}, {Zerbi}, {Groot}, {Hjorth}, {Guinouard},
  {Navarro}, {Adolfse}, {Albers}, {Amans}, {Andersen}, {Andersen}, {Binetruy},
  {Bristow}, {Castillo}, {Chemla}, {Christensen}, {Conconi}, {Conzelmann},
  {Dam}, {de Caprio}, {de Ugarte Postigo}, {Delabre}, {di Marcantonio},
  {Downing}, {Elswijk}, {Finger}, {Fischer}, {Flores}, {Fran{\c{c}}ois},
  {Goldoni}, {Guglielmi}, {Haigron}, {Hanenburg}, {Hendriks}, {Horrobin},
  {Horville}, {Jessen}, {Kerber}, {Kern}, {Kiekebusch}, {Kleszcz}, {Klougart},
  {Kragt}, {Larsen}, {Lizon}, {Lucuix}, {Mainieri}, {Manuputy}, {Martayan},
  {Mason}, {Mazzoleni}, {Michaelsen}, {Modigliani}, {Moehler}, {M{\o}ller},
  {Norup S{\o}rensen}, {N{\o}rregaard}, {P{\'e}roux}, {Patat}, {Pena}, {Pragt},
  {Reinero}, {Rigal}, {Riva}, {Roelfsema}, {Royer}, {Sacco}, {Santin},
  {Schoenmaker}, {Spano}, {Sweers}, {Ter Horst}, {Tintori}, {Tromp}, {van
  Dael}, {van der Vliet}, {Venema}, {Vidali}, {Vinther}, {Vola}, {Winters},
  {Wistisen}, {Wulterkens}, \& {Zacchei}}]{Vernet2011}
{Vernet}, J., {Dekker}, H., {D'Odorico}, S., {et~al.} 2011, \aap, 536, A105

\bibitem[{{Viel} {et~al.}(2004){Viel}, {Haehnelt}, \& {Springel}}]{Viel2004}
{Viel}, M., {Haehnelt}, M.~G., \& {Springel}, V. 2004, \mnras, 354, 684

\bibitem[{{Visbal} \& {Croft}(2008)}]{Visbal2008}
{Visbal}, E. \& {Croft}, R. A.~C. 2008, \apj, 674, 660

\bibitem[{{Wang} {et~al.}(2012){Wang}, {Chen}, \& {Park}}]{Wang2012}
{Wang}, X., {Chen}, X., \& {Park}, C. 2012, \apj, 747, 48

\bibitem[{{Weinberg} {et~al.}(1987){Weinberg}, {Gott}, \&
  {Melott}}]{Weinberg1987}
{Weinberg}, D.~H., {Gott}, III, J.~R., \& {Melott}, A.~L. 1987, \apj, 321, 2

\bibitem[{{Wolfe} {et~al.}(2005){Wolfe}, {Gawiser}, \& {Prochaska}}]{Wolfe2005}
{Wolfe}, A.~M., {Gawiser}, E., \& {Prochaska}, J.~X. 2005, \araa, 43, 861

\bibitem[{{Yang} {et~al.}(2012){Yang}, {Puech}, {Flores}, {Hammer},
  {Rodrigues}, \& {Disseau}}]{Yang2012}
{Yang}, Y.~B., {Puech}, M., {Flores}, H., {et~al.} 2012, in \procspie, Vol.
  8446, Ground-based and Airborne Instrumentation for Astronomy IV, 84467Q

\bibitem[{{Zel'dovich}(1970)}]{Zeldovich1970}
{Zel'dovich}, Y.~B. 1970, \aap, 5, 84

\bibitem[{{Zhang} {et~al.}(2015){Zhang}, {Yang}, {Wang}, {Wang}, {Luo}, {Mo},
  \& {van den Bosch}}]{zhang2015}
{Zhang}, Y., {Yang}, X., {Wang}, H., {et~al.} 2015, \apj, 798, 17

\bibitem[{{Zunckel} {et~al.}(2011){Zunckel}, {Gott}, \& {Lunnan}}]{Zunckel2011}
{Zunckel}, C., {Gott}, J.~R., \& {Lunnan}, R. 2011, \mnras, 412, 1401

\end{thebibliography}

\appendix

\section{Stacked spectra}
\label{stacked}
\renewcommand{\thefigure}{A.\arabic{figure}}
\setcounter{figure}{0}

\begin{figure*}[!t]
\begin{center}
\includegraphics[scale=0.58]{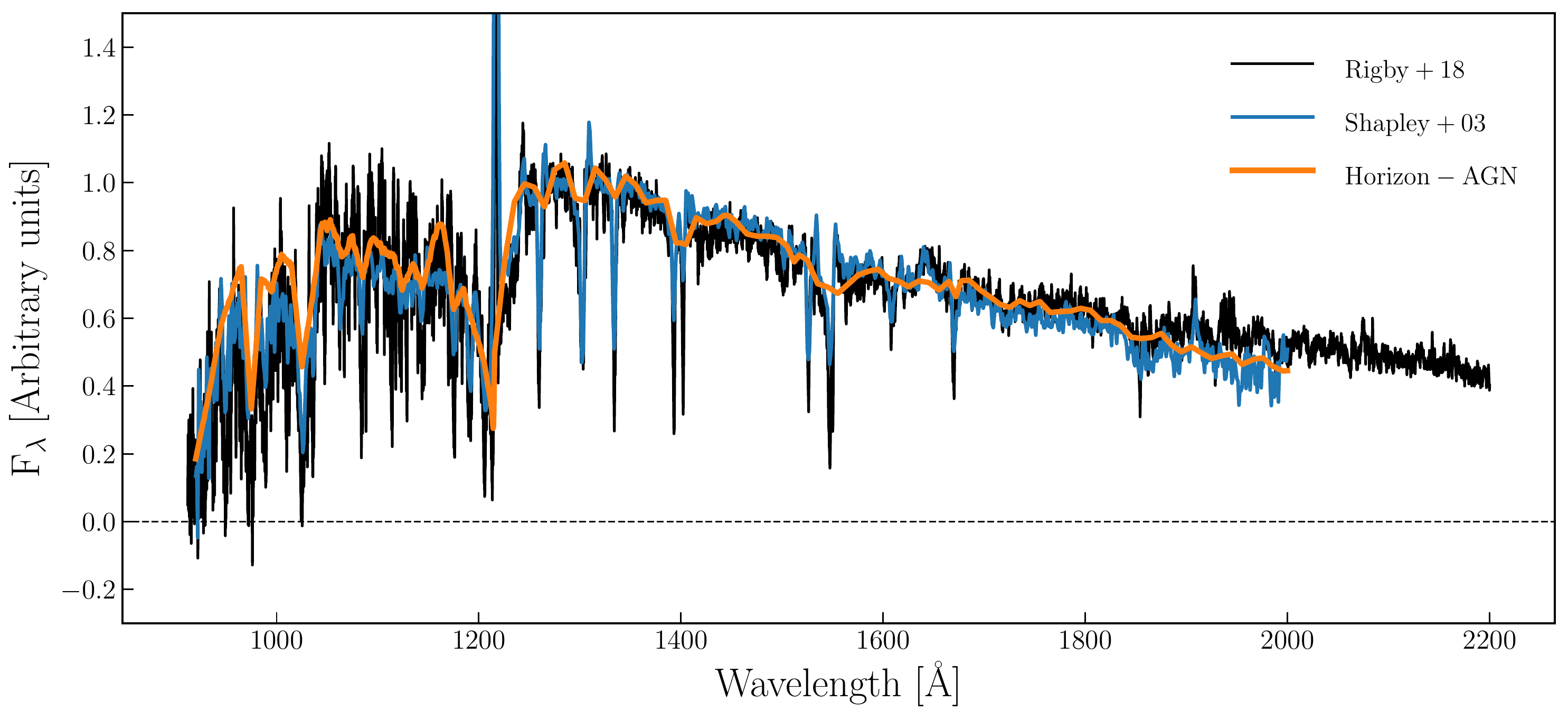}
\caption{Stacked rest-frame spectra of galaxies generated in the Horizon-AGN simulation compared to the stacked spectra of 811 galaxies at $2.4 < z < 3.5$ of \citet{Shapley2003} and to 14 galaxies at $1.6 < z < 3.6$ of \citet{Rigby18}.}
\label{figB1}
\end{center}
\end{figure*}

The stacking of the Horizon-AGN galaxy spectra was made in the following way. First, the spectra were smoothed to the desired resolution in the observer frame. We considered three different resolutions: R = 5000 (corresponding to the MOSAIC resolution), 3300 (the average resolution in the \citet{Rigby18} sample), and 560 (the average resolution of the \citet{Shapley2003} sample). The stacking was made following the procedure of \citet{Rigby18}. We shifted the spectra to their rest-frame wavelengths, re-sampled them to a common wavelength grid, normalized them at $\lambda \approx 1270~\mathrm{\AA,}$ and stacked them. For the purpose of comparison, we stacked only a subsample of $z \sim 3 - 3.2$ galaxies, in order to have similar mean redshift as \citet{Shapley2003}. Including the galaxies at higher redshifts results in a lower average transmission in the region below  \lya\,due to the progressively increased number density of Lyman-alpha absorbers with redshift \citep[e.g.][]{Kim2001}. Because there are so many absorbers, the average spectrum does not change regardless of whether we include noise in the simulated spectra. The resolution does not change the overall picture either, therefore we only worked with the R = 5000 average spectrum.

The average spectrum of the Horizon-AGN galaxies is compared to the stacked spectra of \citet{Shapley2003} and \citet{Rigby18} in Fig. \ref{figB1}. The Horizon-AGN spectrum compares very well with that reported in \citet{Shapley2003}. The galaxy ISM absorption lines are lacking because they were not included in the generation of the Horizon-AGN spectra.

\section{Details on the skeleton extraction}
\label{App:SkelExt}

\renewcommand{\thefigure}{B.\arabic{figure}}
\setcounter{figure}{0}

\subsection{Choosing the persistence threshold}
\begin{figure}
\begin{center}
\includegraphics[scale=0.58]{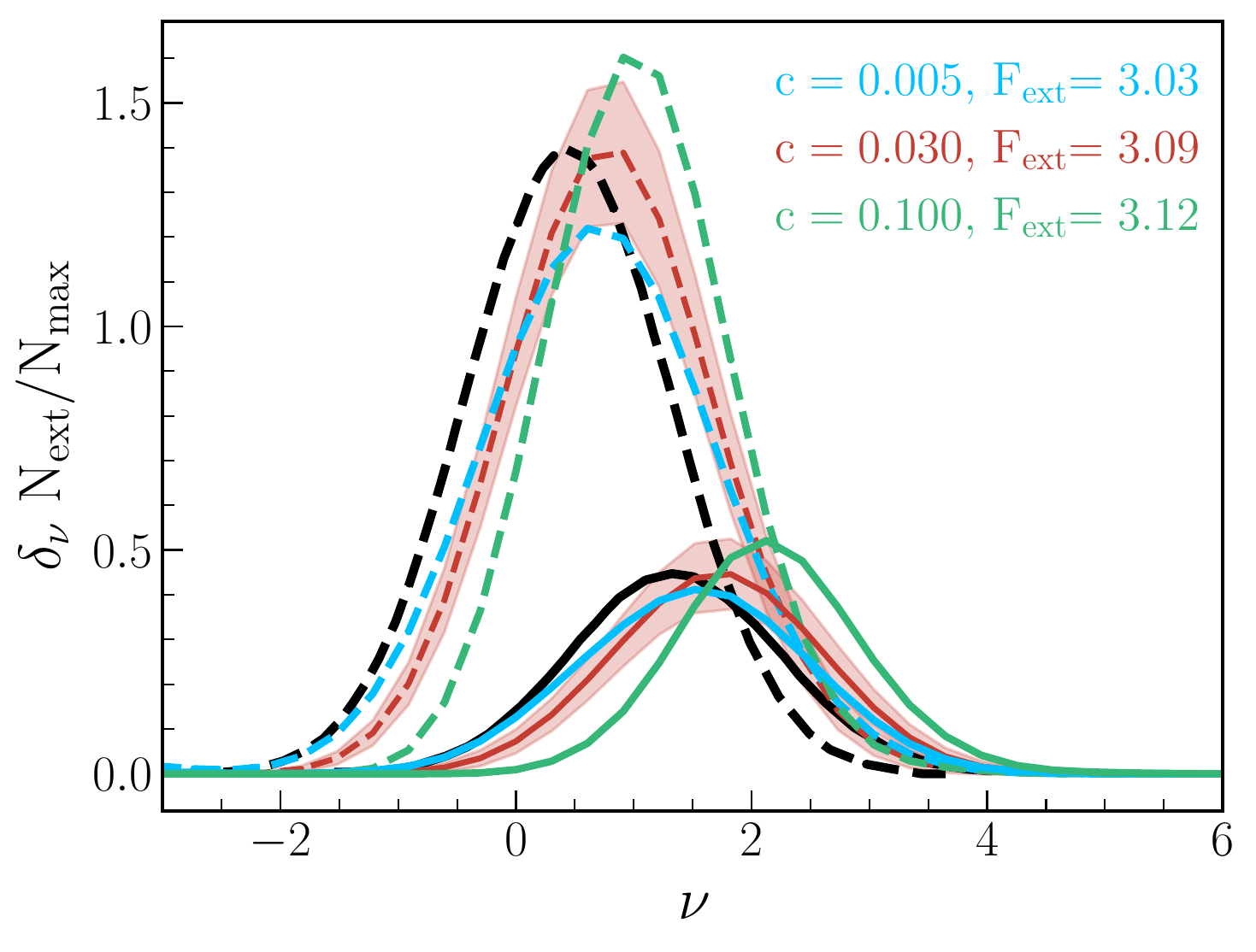}
\caption{Counts of saddle points (dashed line) and maxima (solid line) in the original field for different persistence cuts of the skeleton. The black lines correspond to the theoretical prediction for a Gaussian field with spectral index $n_s=-2$.}
\label{figC0}
\end{center}
\end{figure}
In order to define the most appropriate persistence threshold for the skeleton extraction, we compared the count of saddle points  and peaks in the original field at a given smoothing scale to the counts that are expected from a Gaussian field (which are known exactly) at the same smoothing scale. These counts are commonly expressed as a function of the height $\nu=\rho/\sigma,$ where $\sigma$ is the rms of the field. In order to compare our critical point counts with those in a Gaussian field, we define  $\nu_{f}$, the  threshold in optical depth (where the optical depth is taken as a proxy for density, in the original or the reconstructed field) such that the excursion set has the same volume fraction as a Gaussian random field \citep[e.g.][]{Gott1987,Weinberg1987,Melott1988,Appleby18},
\begin{equation}
f=\dfrac{1}{2\pi}\int_{\nu_{\rm f}}^{\infty} \exp\left(-t^2/2\right) dt\,,\end{equation}
where $f$ is the filling factor of the excursion set. This parametrization allows us to minimize the impact of non-Gaussianities in the density field. An example of the counts of saddle points and maxima $\delta_{\nu} {N_{\rm ext}}/{N_{\rm max}}$ in the original field for different persistence cuts of the skeleton is presented in Fig.~\ref{figC0} for the configuration $C_2$, and it is compared to the theoretical prediction for a Gaussian field with a power-law spectrum with spectral index $n_s=-2$. The count is normalized by $N_{\rm max}$, the total number of maxima, that is, it corresponds to the number density of critical points in a sphere of radius $R_{p}$ (where $R_{p}$ is defined as the radius of the sphere that on average contains exactly one peak). We also show the fraction ${\rm F}_{\rm ext}=N_{\rm sad}/N_{\rm max}$ , which is 3.055 for a Gaussian random field in 3D. 
The small volume of the simulated lightcone only allows a qualitative comparison, and we found that  persistence cuts $c=0.03$ and $c=0.005$ in the C2 and C1 configurations, respectively, better match the predictions (where $c$ is given in units of $\tau$). Varying this value by a factor of a few does not really affect our result. 
In a given configuration, the same persistence cut was adopted for all tested $S/N$.
For the purpose of the analysis presented in Section~\ref{sec:mapgal}, only the 50\% densest filaments (in terms of $\tau$) were used in each field. 

\subsection{Filament alignment}
\begin{figure}
\begin{center}
\includegraphics[scale=0.58]{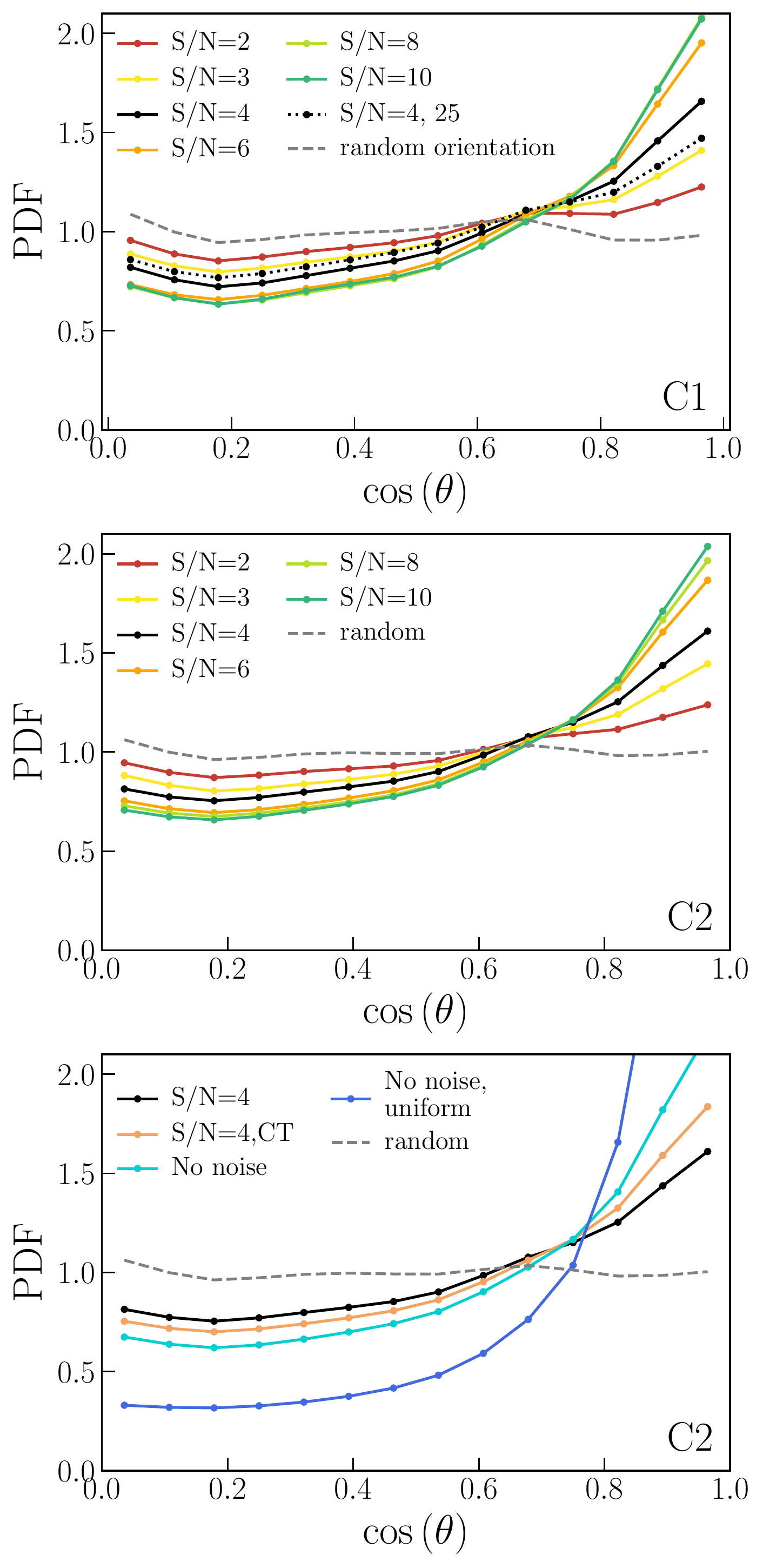}
\caption{Probability density function of $\cos \theta$, where $ \theta$ ($0\leq\theta\leq90$) is defined for each segment in the reconstructed skeleton as the angle between this segment and its closest neighbour in the original skeleton. The dashed line shows the signal for random segment orientations. }
\label{figC6}
\end{center}
\end{figure}

A complementary test to quantify how well cosmic filaments can be extracted from the reconstructed fields is measuring their relative orientations with respect to the filament orientations in the original fields. 
For all segments in the reconstructed field in a given configuration and at a given $S/N$, we measured $\cos \theta$, where $\theta$ is the angle between this segment and the closest one in the original skeleton ($0\leq\theta\leq90$). The PDF of $\cos \theta$ is displayed in Fig.~\ref{figC6}, and the random signal (which was obtained by reshuffling the orientation of the segments in the original skeleton)  is shown as the dashed line.  For all $S/N$, there is an excess of probability for $\cos \theta>0.5$, that is, for the two segments to be aligned. This signal increases with increasing $S/N$, although there is not much improvement from $S/N=6$.

\subsection{Galaxies in filaments}
\label{Ap:galinfil}
\begin{figure*}[!t]
\begin{center}
\includegraphics[scale=0.58]{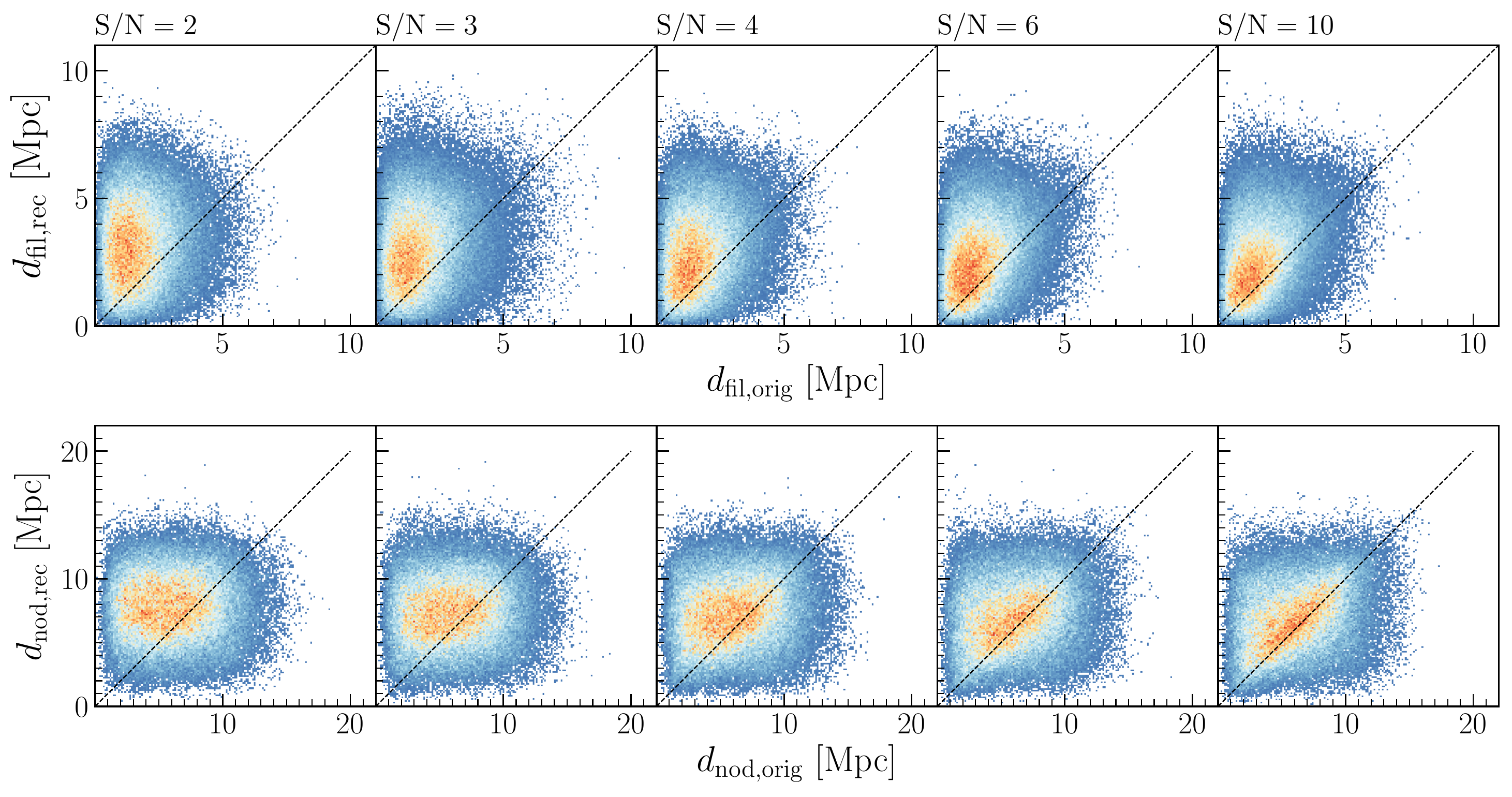}
\caption{Comparison of the distances of galaxies to the nearest filament (top) and to the nearest node (bottom) measured from the skeleton computed at the original density map ({\it x}-axis) and the reconstructed density map ({\it y}-axis). The skeleton was applied to configuration 1 ($R=1000$ and $L_{\rm T}=4$ Mpc). The dashed line indicates the one-to-one correspondence.}
\label{figC1}
\end{center}
\end{figure*}

\begin{figure*}[!t]
\begin{center}
\includegraphics[scale=0.58]{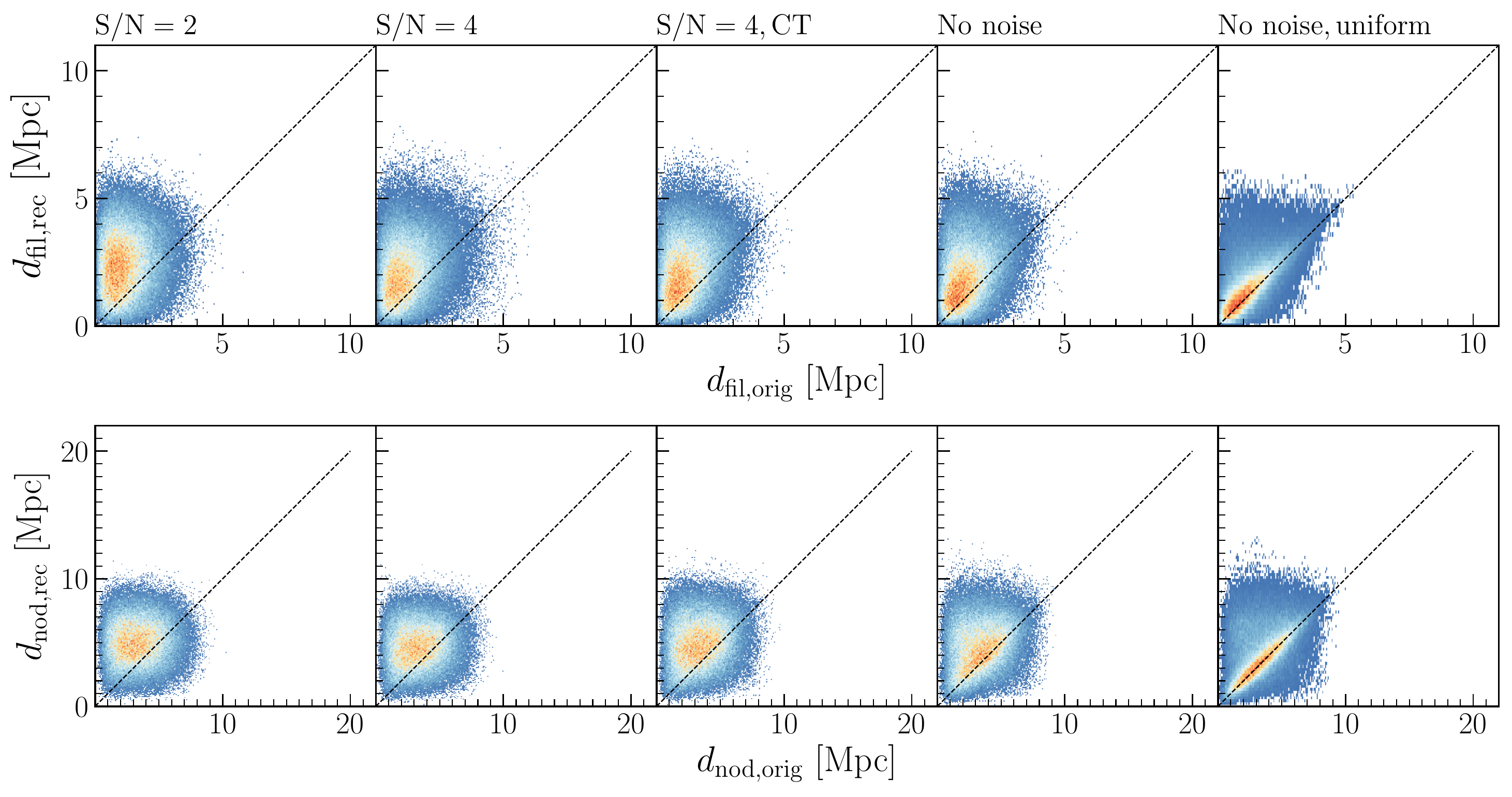}
\caption{Same as Fig.~\ref{figC1}, but for the second configuration ($R=2000$ and $L_{\rm T}=2.5$ Mpc).}
\label{figC2}
\end{center}
\end{figure*}

As a complement of the main text analysis (Fig.~\ref{fig12}), the comparison of galaxy distances to filaments and to nodes in the reconstructed skeletons and in the original skeletons is displayed in Figs.~\ref{figC1} and~\ref{figC2}. Because the smoothing scale is larger in configuration C1, fewer filaments are extracted and therefore the distances to nodes and filaments are larger than in C2. In both cases, the distances to nodes are in general better recovered than distances to filaments.

\begin{figure*}[!t]
\begin{center}
\includegraphics[width=\textwidth]{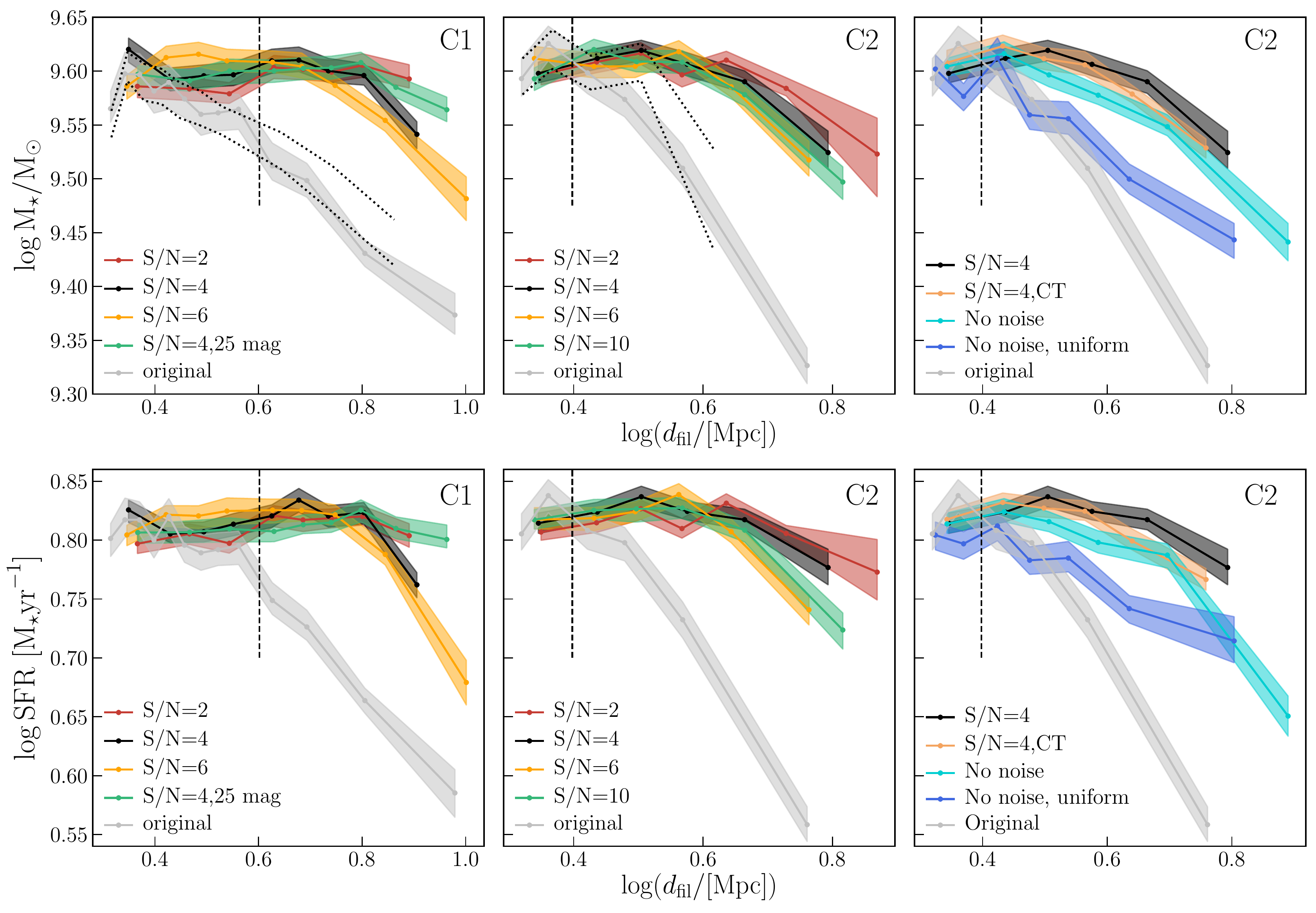}
\caption{Distributions of the mean galaxy stellar mass (top) and SFR (bottom) as a function of the distance to the filaments ($d_{\mathrm{fil}}$) for galaxies with $\log M_{\star}/M_{\odot} > 9.0$. Results are shown for the C1 ($R=1000$) and the C2 ($R=2000$) configurations. The vertical dashed line indicates the reconstruction scales of 4 and 2.5 Mpc for the C1 and C2 configuration, respectively. The dotted black lines show the mass gradient in the original field when all the filaments (not only the densest 50\%) are kept in the analysis.}
\label{figC3}
\end{center}
\end{figure*}

\renewcommand{\thefigure}{C.\arabic{figure}}
\setcounter{figure}{0}   

\begin{figure*}
\begin{center}
\includegraphics[scale=0.52]{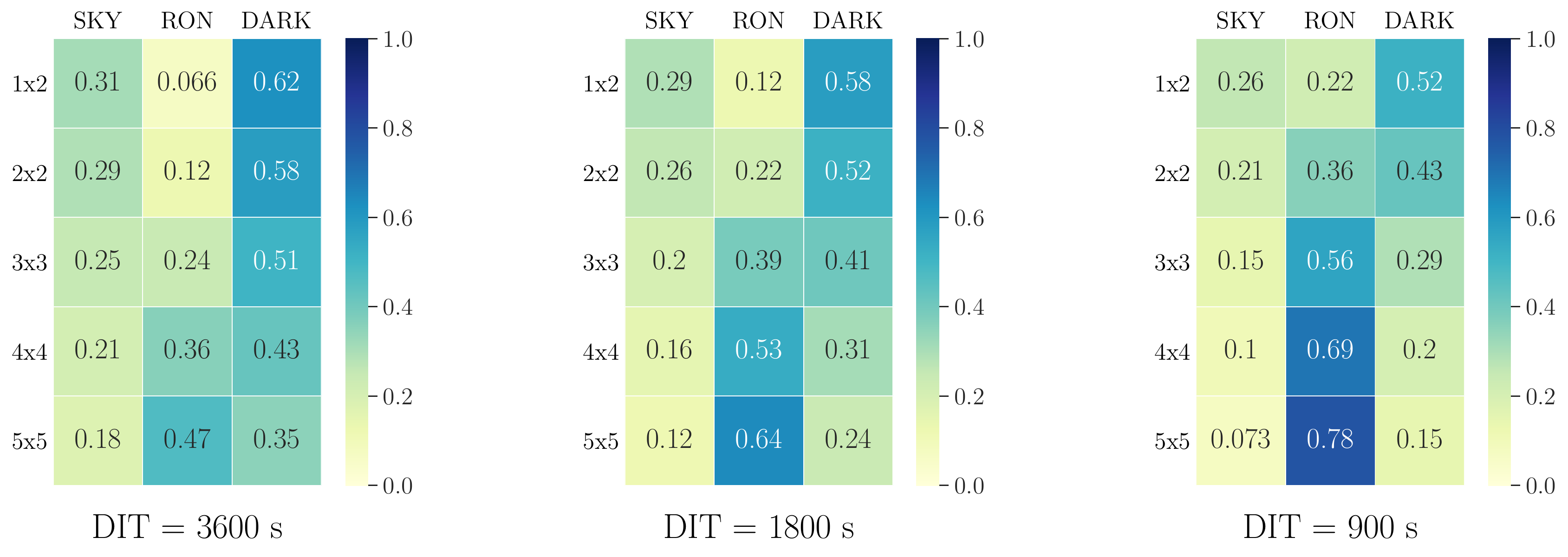}
\caption{Relative contributions of the sky background, RON, and dark current to the total RMS noise as a function of on-the-CCD binning ($S_{\mathrm{s}}\times S_{\lambda}$). Colours and numbers indicate the fraction of the noise contribution of each source in each line. The three plots correspond to different exposure times (DIT).}
\label{figA00}
\end{center}
\end{figure*}

\begin{figure}
\begin{center}
\includegraphics[scale=0.58]{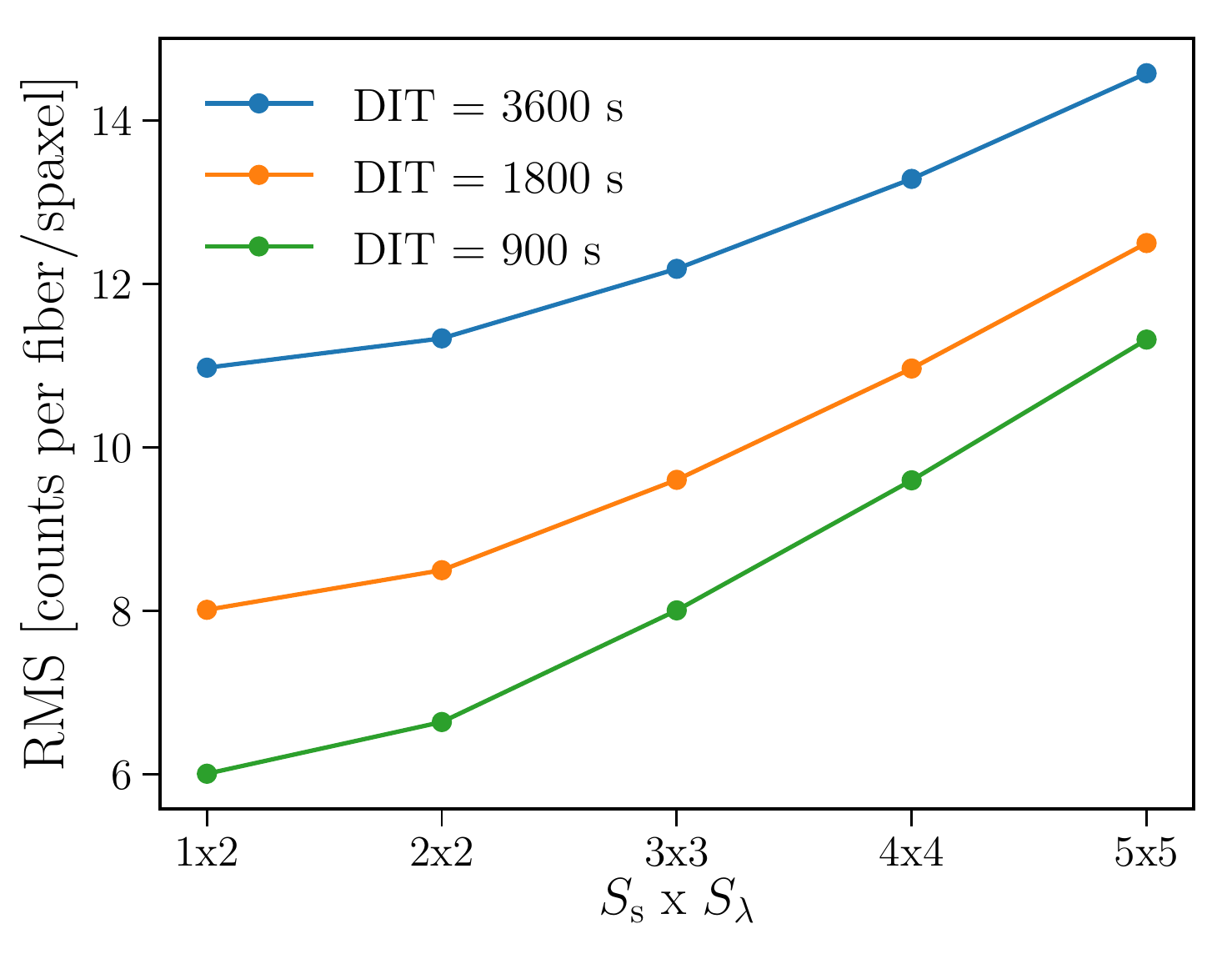}
\caption{Total noise per fiber as a function of on-the-CCD binning ($S_{\mathrm{s}}\times S_{\lambda}$) for three exposure times (DIT).}
\label{figA01}
\end{center}
\end{figure}

\begin{figure*}
\begin{center}
\includegraphics[scale=0.42]{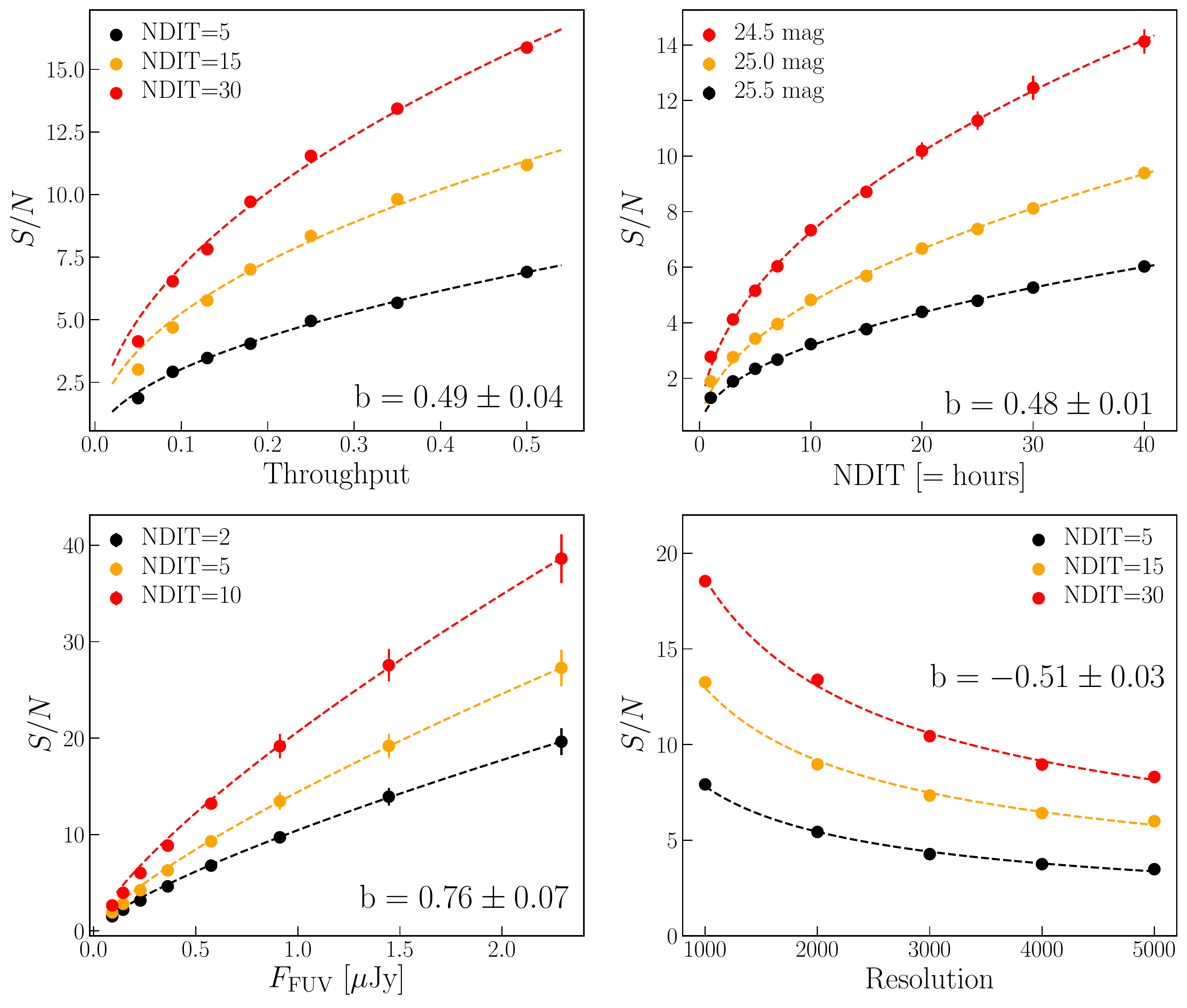}
\caption{Representative plots to show how $S/N$ scales (for the HMM-VIS mode) with the number of exposures (exposure time), resolution, total throughput, and brightness of the source. Scalings were obtained by fitting a power-law function $\left( x/x_{0}\right)^{b}$ to each $S/N$ curve, where $x$ represents one of the four quantities, $x_{0}$ is the  normalization (see text and Equation \ref{scalingIGM}), and $b$ is a power-law index.}
\label{figA0}
\end{center}
\end{figure*}

\begin{figure*}
\begin{center}
\begin{tabular}{cc}
\includegraphics[scale=0.54]{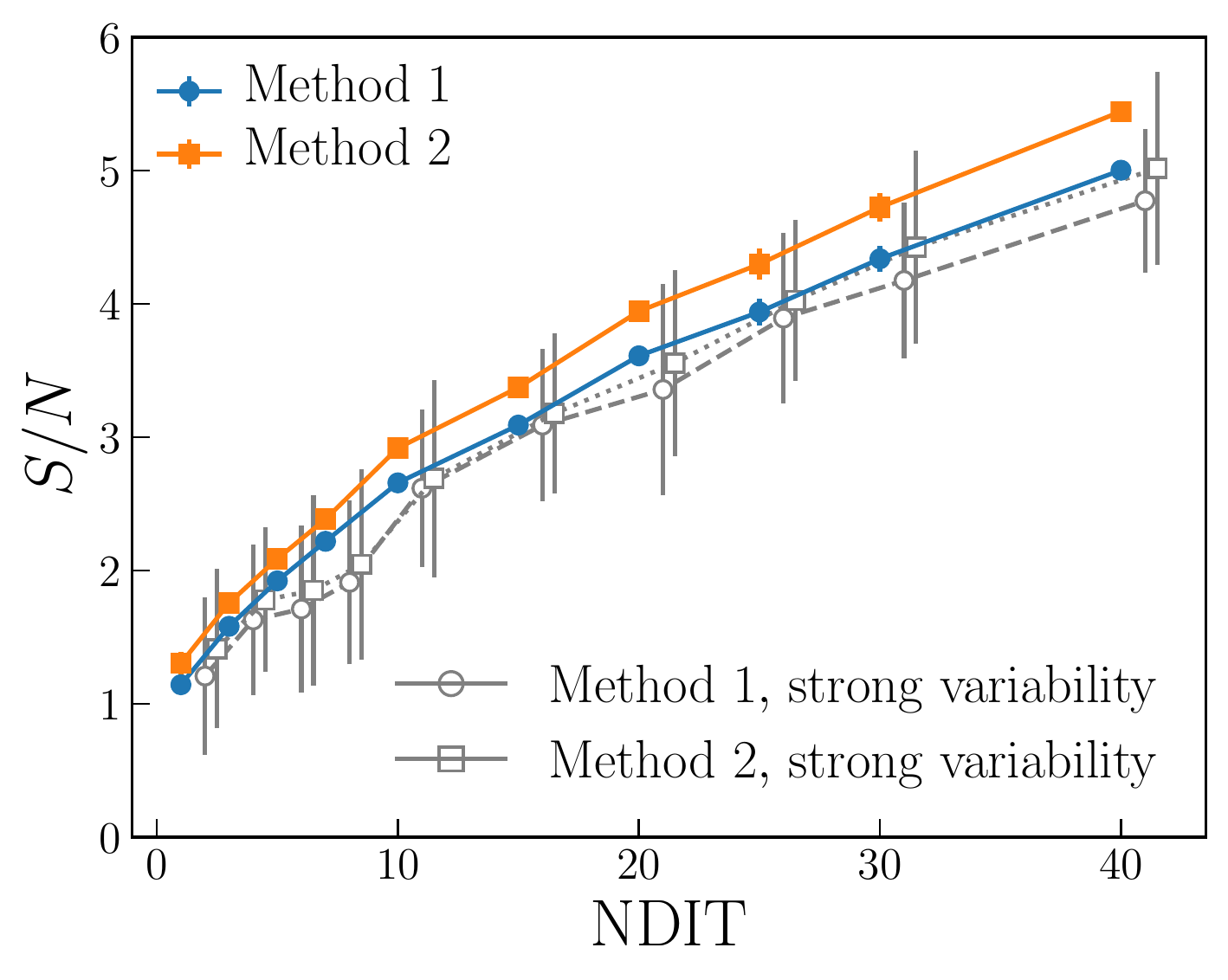}&
\includegraphics[scale=0.54]{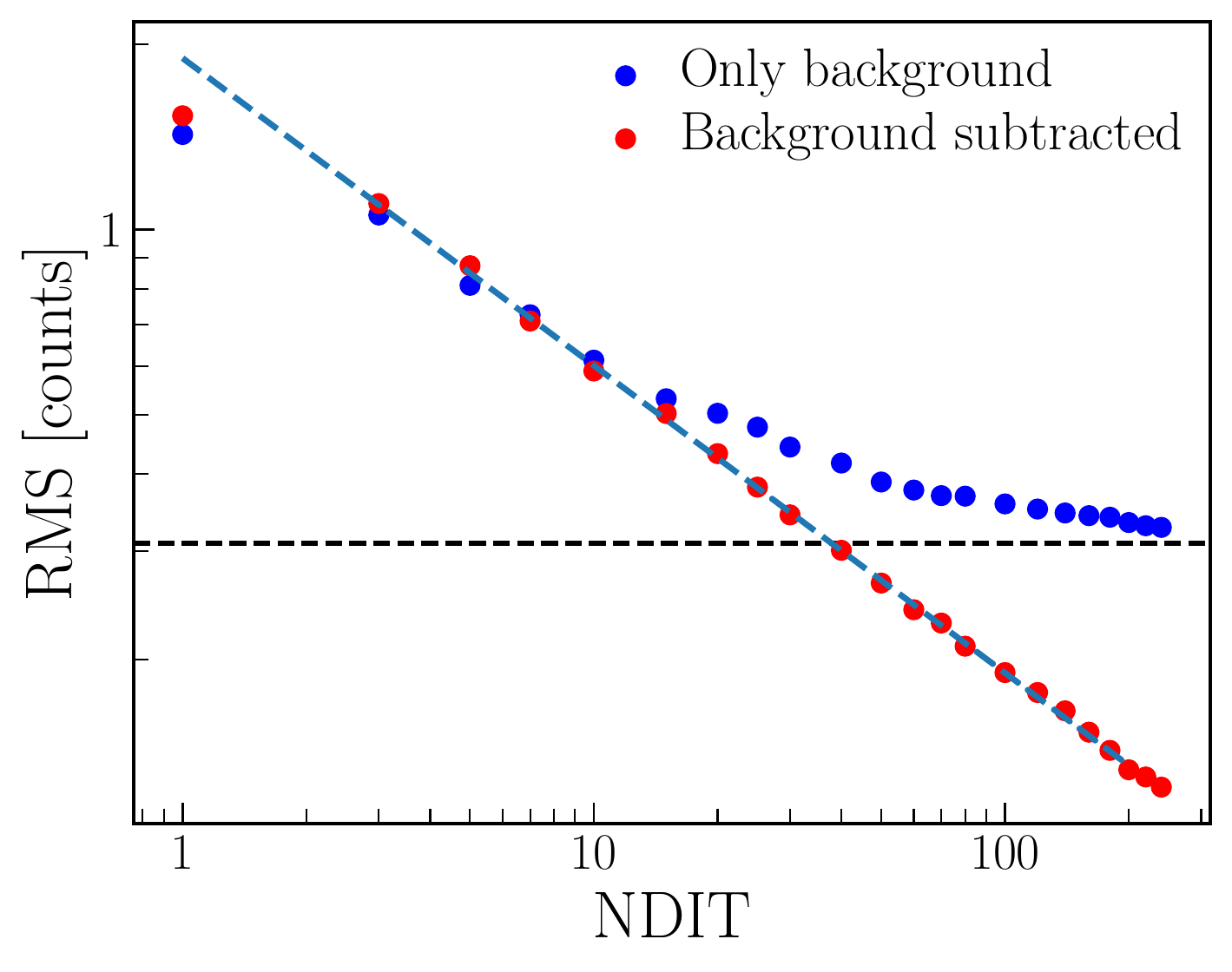}
\end{tabular}
\caption{Towards understanding the effect of sky variability on the $S/N$ in the case of HMM observations. {\it Left}:  $\text{Signal-to-noise ratio}$ as a function of NDIT for two different methods of background subtraction (see text). The coloured points correspond to the simulation with an average sky variability \citep{Puech2012}, and the grey points correspond to the simulation with enhanced amplitudes of sky variability. The latter points are shifted in the {\it x}-axis direction for the purpose of visualization. {\it Right}: RMS as a function of NDIT for one simulation in the case of very bad sky conditions, i.e. when the amplitude of sky variations is ten times stronger than the typical variations. The blue curve corresponds to the RMS of the sky alone (i.e. equivalent to the observation of the sky), while the red curve corresponds to the RMS of the source spectrum after the background is subtracted. The values are given with respect to the normalized (=1) continuum. The red curve has been scaled in the {\it y}-axis in order to match the blue curve for small NDIT. The dashed black line shows the amplitude of the sky variations. The dashed blue line indicates the decline expected in the case of Poissonian noise (NDIT$^{-0.5}$).}
\label{figA1}
\end{center}
\end{figure*}

\section{Details on the simulations of the MOSAIC performance}

\subsection{Signal-to-noise ratio regimes}
\label{noiseregime}

The spectroscopic targets considered in this work are faint, which means that the noise will be dominated by the sky background, RON, or dark current. The total noise per spaxel (or fiber in the HMM-VIS case) is 
\begin{equation}
{\rm RMS} = \sqrt{n_{\rm s}n_{\lambda}N_{\rm sky}^{\rm px}t_{\rm exp} + n_{\rm s}n_{\lambda}{\rm RON}^2 + n_{\rm s}n_{\lambda}N_{\rm dark}^{\rm px}t_{\rm exp}},
\end{equation}
where $N_{\rm sky}^{\rm px}$ and $N_{\rm dark}^{\rm px}$ are photon counts per pixel per second of the sky and dark current, $n_{\rm s}$ and $n_{\lambda}$ are the number of pixels in the spatial and dispersion direction over which a spaxel (fiber) is imaged on the detector, and $t_{\rm exp} = {\rm DIT}$ is the exposure time. The default imaging of a spaxel (fiber) on the detector is $n_{\rm s}\times n_{\lambda} = 5\times 5$. In our simulation we assumed RON=2 and $N_{\rm dark}^{\rm px} = 3~{\rm counts} ~{\rm h^{-1}} ~{\rm px}^{-1}$. $N_{\rm sky}^{\rm px}$ was determined by adopting the Paranal sky model (see Sect. \ref{atmospheric}). These values correspond to the optical detectors, and the sky was computed at $\lambda = 4900~\mathrm{\AA}$. 

The spatial and spectral sampling on the detector can be binned during the readout. Binning will only affect the RON noise component, that is,

\begin{equation}
{\rm RMS} = \sqrt{N_{\rm sky}^{\rm sp}t_{\rm exp} + S_{\rm s}S_{\lambda}{\rm RON}^2 + N_{\rm dark}^{\rm sp}t_{\rm exp}},
\end{equation}

\noindent where $N_{\rm sky}^{\rm sp} = n_{\rm s}n_{\lambda}N_{\rm sky}^{\rm px} = 25N_{\rm sky}^{\rm px}$ and $N_{\rm dark}^{\rm sp} = 25N_{\rm dark}^{\rm px}$ are the integrated photon counts per spaxel (or fiber) per second of the sky and dark current. $S_{\rm s}$ and $S_{\lambda}$ represent the number of bins in the spatial and dispersion direction after rebinning. For example, an extreme binning of 5 px both in the spatial and dispersion direction would give $S_{\rm s} \times S_{\lambda} = 1\times1$ (while $n_{\rm s}\times n_{\lambda} = 5\times5$).

The relative fraction of the noise contributed by each of the three noise sources is shown in Fig.~\ref{figA00} for different on-CCD binning and three exposure times: 3600, 1800, and 900 s. The RON noise becomes less important as the DIT increases. At a constant DIT, the rebinning also reduces the relative importance of the RON noise. The dark current dominates the noise at strong binning. The change of the total noise, where all three noise components are combined, with binning and exposure time is plotted in Fig.~\ref{figA01}.

These results suggest that some degree of binning is to be applied to the spaxel (fiber) image when observing at optical wavelengths. For simplicity, and to form an impression of the best possible results, we assumed a binning of 1x2 in this work. We note that the difference in the total noise from a more realistic sampling of 2x2 is small (see Fig.~\ref{figA01}); given all the uncertainties (Sec.~\ref{inputs} and \ref{reqtime}), the S/N scaling relations and the estimates of the number of nights required to carry out the IGM tomography survey would not change significantly if the 2x2 sampling were adopted.

\subsection{Scaling relations for the HMM-VIS mode}

Figure ~\ref{figA0} shows the scaling relations for the simulated HMM-VIS mode of the MOSAIC instrument.

\subsection{Sky variability}
\label{skyvar}

The sky variability could have a significant effect on the quality of the reduced data coming from the HMM mode observations. For this reason, the modelling of the sky variability was added to the simulator. By default, the variability is described by the average values of the amplitude and scale length of the variable sky at $\lambda \sim 9000~\mathrm{\AA}$ as measured by \citet{Puech2012} (see also \citealt{Rodrigues2010,Yang2012}). We determined the effect of the sky variability by simulating four fiber bundles around each science fiber. We used two different sky subtraction methods.
\begin{itemize}
\item Method 1: Only one bundle was assumed for the background estimate. This was motivated by the fact that no more than one bundle is allowed for a sky observation. The average spectrum in this bundle was subtracted from all the fibers in the science bundle. This was done for each DIT separately.
\item Method 2: All four bundles were assumed for background subtraction. At each wavelength, the sky counts of the four bundles were averaged, and this averaged spectrum was then subtracted from the science bundle. This was done for each DIT separately.
\end{itemize}
\noindent The bundles were positioned close (e.g. $1\arcsec$) to the source. The variability, as measured by \citet{Puech2012}, is characterized by several contributions with different spatial scales and amplitudes. The amplitude of the mode with the smallest spatial scale ($\sim 0.8\arcsec$) also is an order of magnitude stronger than that of the other modes with (much) larger spatial scales. The conclusions therefore do not change when the sky bundles are placed at larger distances from the source.

The analysis was conducted for the $m_{\rm rest,UV}$ = 25.5 mag source. The results are shown in Fig. \ref{figA1} (left). First we considered the default sky-variability parameters. The $S/N$ curves obtained from different bundles (i.e. different positions in the sky around the science source) is essentially the same. The noise in this case is Poissonian. As expected, averaging the sky measured from different positions in the sky (i.e. method 2 subtraction) slightly improves the resulting $S/N$. Inspecting the scatter around the continuum as a function of NDIT, we find that the scatter values are very high compared to the amplitude of sky variations. In these conditions, it would take NDIT $\sim 6000$ to reach the regime in which the sky variability would begin to affect the $S/N$ appreciably.

We also tested the effect of very bad sky conditions, for which we assumed that the amplitude of sky variations is ten times stronger than in the default variations. As illustrated in Fig. \ref{figA1} (left), in this case, the variability has a strong effect on the performance and the scatter is dominated by the sky variability. 
Even under these extreme conditions, the saturation regime is not yet reached. This is shown in Fig. \ref{figA1} (right). In the shot-noise regime, the noise is expected to decrease with time as $\sqrt{T}$. For long integration times, the noise saturates because of systematics; in our case, these were due to the sky variation within the HMM bundle (see blue curve). When we subtracted the sky, we removed the systematic floor and then the noise indeed decreased as $\sqrt{T}$ (the red curve in Fig. \ref{figA1}). It would still take NDIT $\sim 300$ to approach the saturation regime. These are extreme conditions, however. We therefore conclude that under normal circumstances, the effects of sky variability on the performance of the instrument in the blue band are negligible. This means that in practice, at least for observations of $< 26$ mag sources, we do not need paired fibers for sky subtraction, that is, more than half of the available fibers can be used for observations of sources.

\end{document}